\documentclass[10pt,a4paper]{article}
\pdfoutput=1
\usepackage{jheppub}

\usepackage{amsmath}
\usepackage{amssymb}
\usepackage{graphicx,color}

\usepackage{amsmath,amsthm}
\usepackage{amssymb}

\usepackage{graphicx}
\usepackage{bm}

\newcommand{\Ref}[1]{(\ref{#1})}

\newtheorem{Theorem}{Theorem}[section]
\newtheorem{Definition}{Definition}[section]
\newtheorem{Lemma}[Theorem]{Lemma}

\newtheorem{Proposition}[Theorem]{Proposition}
\newcommand{\startproof}{\textbf{Proof:\ \ }}
\newcommand{\finishproof}{\hfill $\Box$ \\}

\newcommand{\half}{\frac{1}{2}}

\newcommand{\Slc}{\mathrm{SL}(2,\mathbb{C})}
\newcommand{\slc}{\fs\fl_2\mathbb{C}}

\def\be{\begin{eqnarray}}
\def\ee{\end{eqnarray}}

\newcommand{\ca}{\mathcal A}

\newcommand{\ch}{\mathcal H}

\newcommand{\cj}{\mathcal J}
\newcommand{\ck}{\mathcal K}
\newcommand{\cl}{\mathcal L}
\newcommand{\cm}{\mathcal M}

\newcommand{\calr}{\mathcal R}

\newcommand{\fl}{\mathfrak{l}}

\newcommand{\fs}{\mathfrak{s}}

\renewcommand{\a}{\alpha}
\renewcommand{\b}{\beta}
\newcommand{\g}{\gamma}
\newcommand{\G}{\Gamma}

\newcommand{\eps}{\varepsilon}
\newcommand{\varth}{\vartheta}

\newcommand{\sig}{\sigma}

\renewcommand{\l}{\lambda}

\renewcommand{\o}{\omega}
\renewcommand{\O}{\Omega}

\newcommand{\rmd}{\mathrm d}

\newcommand{\lt}{\left}
\newcommand{\rt}{\right}

\newcommand{\lag}{\left\langle}
\newcommand{\rag}{\right\rangle}

\newcommand{\tr}{\mathrm{tr}}
\newcommand{\bbc}{\mathbb{C}}

\newcommand{\bp}{\mathbf{P}}

\title{Asymptotics of Spinfoam Amplitude on Simplicial Manifold: Lorentzian Theory}

\author[]{Muxin Han,  \ \  Mingyi Zhang}

\affiliation[]{Centre de Physique Th\'eorique%
\footnote{Unit\'e mixte de recherche (UMR 6207) du CNRS et des Universit\'es
de Provence (Aix-Marseille I), de la Meditarran\'ee (Aix-Marseille II) et du Sud (Toulon-Var); laboratoire affili\'e \`a la FRUMAM (FR 2291).}, CNRS-Luminy, Case 907, 13288 Marseille, France}

\emailAdd{Muxin.Han(AT)cpt.univ-mrs.fr} \emailAdd{Mingyi.Zhang(AT)cpt.univ-mrs.fr} 

\abstract{The present paper studies the large-j asymptotics of the Lorentzian EPRL spinfoam amplitude on a 4d simplicial complex with an arbitrary number of simplices. The asymptotics of the spinfoam amplitude is determined by the critical configurations. Here we show that, given a critical configuration in general, there exists a partition of the simplicial complex into three type of regions $\calr_{\text{Nondeg}},\calr_{\text{Deg-A}},\calr_{\text{Deg-B}}$, where the three regions are simplicial sub-complexes with boundaries. The critical configuration implies different types of geometries in different types of regions, i.e. (1) the critical configuration restricted into $\calr_{\text{Nondeg}}$ implies a nondegenerate discrete Lorentzian geometry, (2) the critical configuration restricted into $\calr_{\text{Deg-A}}$ is degenerate of type-A in our definition of degeneracy, but implies a nondegenerate discrete Euclidean geometry in $\calr_{\text{Deg-A}}$, (3) the critical configuration restricted into $\calr_{\text{Deg-B}}$ is degenerate of type-B, and implies a vector geometry in $\calr_{\text{Deg-B}}$.

With the critical configuration, we further make a subdivision of the regions $\calr_{\text{Nondeg}}$ and $\calr_{\text{Deg-A}}$ into sub-complexes (with boundary) according to their Lorentzian/Euclidean oriented 4-volume $V_4(v)$ of the 4-simplices, such that $\mathrm{sgn}(V_4(v))$ is a constant sign on each sub-complex. Then in the each sub-complex, the spinfoam amplitude at the critical configuration gives the Regge action in Lorentzian or Euclidean signature respectively in $\calr_{\text{Nondeg}}$ or $\calr_{\text{Deg-A}}$. The Regge action reproduced here contains a sign prefactor $\mathrm{sgn}(V_4(v))$ related to the oriented 4-volume of the 4-simplices. Therefore the Regge action reproduced here can be viewed a discretized Palatini action with on-shell connection.

Finally the asymptotic formula of the spinfoam amplitude is given by a sum of the amplitudes evaluated at all possible critical configurations, which are the products of the amplitudes associated to different type of geometries. }


\arxivnumber{}

\begin{document}

\maketitle

\section{Introduction}

Loop Quantum Gravity (LQG) is an attempt to make a background independent, non-perturbative
quantization of 4-dimensional General Relativity (GR) -- for reviews, see \cite{book,rev,sfrevs}. It is
inspired by the classical formulation of GR as a dynamical theory of connections. Starting
from this formulation, the kinematics of LQG is well-studied and results in a successful kinematical
framework (see the corresponding chapters in the books \cite{book}), which is also unique in a
certain sense. However, the framework of the dynamics in LQG is still largely open so
far. There are two main approaches to the dynamics of LQG, they are (1) the Operator formalism of
LQG, which follows the spirit of Dirac quantization or reduced phase space quantization of constrained dynamical system, and performs a canonical quantization of GR \cite{QSD}; (2) the covariant formulation of LQG, which is currently understood in terms of the Spinfoam Models \cite{sfrevs,BC,EPRL,FK,HT}. The
relation between these two approaches is well-understood in the case of 3d quantum gravity
\cite{perez}, while in 4d the situation is much more complicated and there are
some recent attempts \cite{links} for relating these two approaches.

The present article is concerning the spinfoam approach of LQG. The current spinfoam models for quantum gravity are mostly inspired by the 4-dimensional Plebanski formulation of GR \cite{plebanski} (or Plebanski-Holst formulation by including the Barbero-Immirzi parameter $\g$), which is a BF theory constrained by the condition that the $B$ field should be ``simple'' i.e. there is a tetrad field $e^I$ such that $B=*(e\wedge e)$. Currently one of the successful spinfoam models is the EPRL model defined in \cite{EPRL}, whose implementation of simplicity constraint is understood in the sense of \cite{DingYou}. The EPRL vertex amplitude is shown to reproduce the classical discrete GR in the large-j asymptotics \cite{semiclassical}. Recently, The fermion coupling is included in the framework of EPRL spinfoam model \cite{SFfermion}, and a q-deformed EPRL spinfoam model is defined and gives discrete GR with cosmological constant in the large-j asymptotics \cite{QSF,QSFasymptotics}.

The semiclassical behavior of spinfoam model is currently understood in terms of the \emph{large-j asymptotics} of the spinfoam amplitude, i.e. if we consider a spinfoam model as a state-sum
\be
A(\ck)=\sum_{j_f}\mu(j_f)A_{j_f}(\ck)
\ee
where $\mu(j_f)$ is a measure, we are investigating the asymptotic behavior of the (partial-)amplitude $A_{j_f}$ as all the spins $j_f$ are taken to be large uniformly. The area spectrum in LQG is given approximately by $A_f=\g j_f\ell_p^2$, so the semiclassical limit of spinfoam models is argued to be achieved by taking $\ell_p^2\to0$ while keeping the area $A_f$ fixed, which results in $j_f\to\infty$ uniformly as $\g$ is a fixed Barbero-Immirzi parameter. There is another argument relating the large-j asymptotics of the spinfoam amplitude to the semiclassical limit, by imposing the semiclassical boundary state to the vertex amplitude \cite{holomorphic}. Mathematically the asymptotic problem is posed by making a uniform scaling for the spins $j_f\mapsto\l j_f$, and studying the asymptotic behavior of the amplitude $A_{\l j_f}(\ck)$ as $\l\to\infty$.

There were various investigations for the large-j asymptotics of the spinfoam models. The asymptotics of the Barrett-Crane vertex amplitude (10j-symbol) was studied in \cite{10j}, which showed that the degenerate configurations in Barrett-Crane model were nonÐoscillatory, but dominant. The large-j asymptotics of the FK model was studied in \cite{CF}, concerning the nondegenerate Riemanian geometry, in the case of a simplicial manifold without boundary. The large-j asymptotics of the EPRL model was initially investigated in \cite{semiclassical} for both Euclidean and Lorentzian cases, where the analysis concerned a single 4-simplex amplitude (EPRL vertex amplitude). It was shown that the asymptotics of the vertex amplitude is mainly a Cosine of the Regge action in a 4-simplex if the boundary data admits a nondegenerate 4-simplex geometry, and the asymptotics is non-oscillatory if the boundary data doesn't admit a nondegenerate 4-simplex geometry. There were also recent works to find the Regge gravity from the Euclidean/Lorentzian spinfoam amplitude on a simplicial complex via a certain ``double scaling limit'' \cite{claudio}.

The present work analyzes the large-j asymptotic analysis of the Lorentzian EPRL spinfoam amplitude to the general situation of a 4d simplicial manifold with or without boundary, with an arbitrary number of simplices. The analysis for the Euclidean EPRL model is presented in \cite{HZ}. The asymptotics of the spinfoam amplitude is determined by the critical configurations of the ``spinfoam action'', and is given by a sum of the amplitudes evaluated at the critical configurations. Therefore the large-j asymptotics is clarified once we find all the critical configurations and clarify their geometrical implications. Here for the Lorentzian EPRL spinfoam amplitude, a critical configuration in general is given by the data $(j_f, g_{ve},\xi_{ef},z_{vf})$ that solves the critical point equations, where $j_f$ is an SU(2) spin assigned to each triangle, $g_{ve}$ is an $\Slc$ group variable, and $\xi_{ef},z_{vf}$ are two types of spinors. Here in this work we show that given a general critical configuration, there exists a partition of the simplicial complex $\ck$ into three types of regions $\calr_{\text{Nondeg}},\calr_{\text{Deg-A}},\calr_{\text{Deg-B}}$, where the three regions are simplicial sub-complexes with boundaries, and they may be disconnected regions. The critical configuration implies different types of geometries in different types of regions:
\begin{itemize}
\item The critical configuration restricted into $\calr_{\text{Nondeg}}$ is nondegenerate in our definition of degeneracy. It implies a nondegenerate discrete Lorentzian geometry on the simplicial sub-complex $\calr_{\text{Nondeg}}$.

\item The critical configuration restricted into $\calr_{\text{Deg-A}}$ is degenerate of type-A in our definition of degeneracy. However, it implies a nondegenerate discrete Euclidean geometry on the simplicial sub-complex $\calr_{\text{Deg-A}}$

\item The critical configuration restricted into $\calr_{\text{Deg-B}}$ is degenerate of type-B in our definition of degeneracy. It implies a vector geometry on the simplicial sub-complex $\calr_{\text{Deg-B}}$
\end{itemize}

With the critical configuration, we further make a subdivision of the regions $\calr_{\text{Nondeg}}$ and $\calr_{\text{Deg-A}}$ into sub-complexes (with boundary) $\ck_{1}(\calr_{*}),\cdots,\ck_n(\calr_{*})$ ($*$=Nondeg,Deg-A) according to their Lorentzian/Euclidean oriented 4-volume $V_4(v)$ of the 4-simplices, such that $\mathrm{sgn}(V_4(v))$ is a constant sign on each $\ck_i(\calr_{*})$. Then in the each sub-complex $\ck_i(\calr_{\text{Nondeg}})$ or $\ck_i(\calr_{\text{Deg-A}})$, the spinfoam amplitude at the critical configuration gives an exponential of Regge action in Lorentzian or Euclidean signature respectively. However we emphasize that the Regge action reproduced here contains a sign factor $\mathrm{sgn}(V_4(v))$ related to the oriented 4-volume of the 4-simplices, i.e.
\be
S=\mathrm{sgn}(V_4)\sum_{\text{Internal}\ f} A_f\Theta_f+\mathrm{sgn}(V_4)\sum_{\text{Boundary}\ f} A_f\Theta_f^B
\ee
where $A_f$ is the area of the triangle $f$ and $\Theta_f,\Theta_f^B$ are deficit angle and dihedral angle respectively. Recall that the Regge action without $\mathrm{sgn}(V_4)$ is a discretization of Einstein-Hilbert action of GR. Therefore the Regge action reproduced here is actually a discretized Palatini action with the on-shell connection (compatible with the tetrad).

The asymptotic formula of the spinfoam amplitude is given by a sum of the amplitudes evaluated at all possible critical configurations, which are the products of the amplitudes associated to different type of geometries. 

Additionally, we also show in Section \ref{parity} that given a spinfoam amplitude $A_{j_f}(\ck)$ with the spin configuration $j_f$, any pair of the non-degenerate critical configurations associated with $j_f$ are related each other by a \emph{local} parity transformation. The parity transformation is the one studied in \cite{semiclassical} in the case of a single 4-simplex. A similar result holds for any pair of the degenerate configuration of type-A associated with $j_f$, since it implies a nondegenerate Euclidean geometry.

\section{Lorentzian Spinfoam Amplitude}

Given a simplicial complex $\ck$ (with or without boundary), the Lorentzian spinfoam amplitude on $\ck$ can be expressed in the coherent state representation:  
\be
A(\ck) &=&\sum_{j_{f}}\prod_{f}\mu \left( j_{f}\right)\prod_{\left(
v,e\right) } \int_{\Slc} \rmd g_{ve}\prod_{\left( e,f\right)
} \int_{S^2} \rmd \hat{n}_{ef}\,\prod_{v\in f}\left\langle j_{f},\xi_{ef}\left\vert Y^{\dag
}g_{ev}g_{ve^{\prime }}Y\right\vert j_{f},\xi_{e^{\prime }f}\right\rangle \label{start}
\ee
Here  $\mu(j_f)$ is the face amplitude of the spinfoam, given by $\mu(j_f)=(2j_f+1)$. $\lt\vert j_{f},\xi_{e^{\prime }f}\right\rangle$ is an SU(2) coherent state in the Spin-j representation. The coherent state is labeled by the spin $j$ and a normalized 2-component spinor $|\xi_{ef}\rangle=g(\xi_{ef})|\half,\half\rangle$ ($n_{ef}\in\text{SU(2)}$), while $\hat{n}_{ef}:=g(\xi_{ef})\rhd\hat{z}$ is a unit 3-vector. $Y$ is an embedding map from the Spin-j irrep $\ch^j$ of SU(2) to the unitary irrep $\ch^{(j,\g j)}$ of $\Slc$ with $(k,p)=(j,\g j)$. The embedding $Y$ identify $\ch^j$ with the lowest level in the decomposition $\ch^{(j,\g j)}=\oplus_{k=j}^\infty\ch^k$. Therefore we define an $\Slc$ coherent state by the embedding
\be
\lt|(j_f,\g j_f);j_f,\xi_{ef}\rag:=Y\lt\vert j_{f},\xi_{e^{\prime }f}\right\rangle=\Pi^{(j_f,\g j_f)}\Big(g(\xi_{ef})\Big)\lt|(j_f,\g j_f);j_f,j_f\rag.
\ee 

In order to write the $\ch^{(j_f,\g j_f)}$ inner product in Eq.\Ref{start} explicitly, we express the $\Slc$ coherent state in terms of the canonical basis \cite{ruhl}. The Hilbert space $\ch^{(k,p)}$ can be represented as a space of homogeneous functions of two complex variables $(z^0,z^1)$ with degree $(-1+ip+k;-1+ip-k)$, i.e.
\be
f(\l z^\a)=\l^{-1+ip+k}\bar{\l}^{-1+ip-k}f(z^\a)
\ee 
Given a normalized 2-component spinor $z^\a$ ($\a=0,1$) with $\lag z,z\rag:=\delta_{\a\dot{\a}}\bar{z}^{\dot{\a}}z^\a=1$, we construct the SU(2) matrix
\be
g(z)=\left(\begin{array}{cc}z^0 & -\bar{z}^1 \\ z^1 & \bar{z}^0 \end{array}\right)\equiv (z,Jz)
\ee
where $J(z^0,z^1)^t:=(-\bar{z}^1,\bar{z}^0 )^t$. The canonical basis $f^j_m(z)^{(k,p)}=\lt|(k,p);j,m\rag$ in the $\Slc$ unitary irrep $\ch^{(k,p)}$ is given by the following when restricted to the normalized spinors
\be
f^{j}_m(z)^{(k,p)}=\sqrt{\frac{\dim(j)}{\pi}}D^j_{mk}\Big(g(z)\Big)
\ee
where $D_{mk}^j(g)$ is the SU(2) representation matrix. The canonical basis $f^{j}_m(z)^{(k,p)}$ evaluated on the non-normalized spinor $z^\a$ is then given by the homogeneity 
\be
f^{j}_m(z)^{(k,p)}=\sqrt{\frac{\dim(j)}{\pi}}\lag z,z\rag^{ip-1-j}D^j_{mk}\Big(g(z)\Big)
\ee
while here $D^j_{mk}\Big(g(z)\Big)$ is a analytic continuation of the SU(2) representation matrix. Thus we can write down explicitly the highest weight state in the $j$-representation and in the case of $(k,p)=(j,\g j)$
\be
f^{j}_j(z)^{(j,\g j)}=\sqrt{\frac{\dim(j)}{\pi}}\lag z,z\rag^{i\g j-1-j}(z^0)^{2j}
\ee
Therefore the coherent state is given explicitly by
\be
\lt|(j,\g j);j,\xi\rag=f^{j}_\xi(z)^{(j,\g j)}=f^{j}_j\Big(g(\xi)^tz\Big)^{(j,\g j)}=\sqrt{\frac{\dim(j)}{\pi}}\lag z,z\rag^{i\g j-1-j}\lag\bar{z},\xi \rag^{2j}
\ee
As a result we can write down explicitly the inner product in Eq.\Ref{start} in terms of a $L^2$ inner product on $\mathbb{CP}^1$ between the coherent states $f^{j}_\xi(z)^{(j,\g j)}$
\be
\left\langle j_{f},\xi_{ef}\left\vert Y^{\dag
}g_{ev}g_{ve^{\prime }}Y\right\vert j_{f},\xi_{e^{\prime }f}\right\rangle
&=&\lag(j_f,\g j_f);j_f,\xi_{ef}\big|g_{ev}g_{ve^{\prime }}\big|(j_f,\g j_f);j_f,\xi_{e'f}\rag\nonumber\\
&=&\int_{\mathbb{CP}^{1}}\Omega _{z_{vf}}\overline{f_{\xi_{ef}}^{(
j_{f},\g j_{f}) }\Big( g_{ve}^{t}z_{vf}\Big) }%
f_{\xi_{e^{\prime }f}}^{( j_{f},\g j_{f})
}\Big( g_{ve^{\prime }}^{t}z_{vf}\Big) \label{CP1inner}
\ee
where $\Omega _{z}=\frac{i}{2}\left( z_{0}dz_{1}-z_{1}dz_{0}\right) \wedge \left(
\bar{z}_{0}d\bar{z}_{1}-\bar{z}_{1}d\bar{z}_{0}\right)$ is a homogeneous measure on $\mathbb{C}^2$. 

We insert the result Eq.\Ref{CP1inner} back into Eq.\Ref{start} and define a new spinor variable $Z_{vef}$ and a measure on $\mathbb{CP}^1$ (a scaling invariant measure)
\begin{eqnarray}
Z_{vef} &:= &g_{ve}^{\dag }z_{vf} \nonumber\\
\Omega _{vf} &:= &\frac{\Omega _{z_{vf}}}{\left\langle
Z_{vef},Z_{vef}\right\rangle \left\langle Z_{ve^{\prime }f},Z_{ve^{\prime
}f}\right\rangle }
\end{eqnarray}%
Then the spinfoam amplitude $A(\ck)$ can be written as
\be
A(\ck)=\sum_{j_{f}}\prod_{f}\mu \left( j_{f}\right) \prod_{\left( v,e\right)}\int_{\Slc}
\rmd g_{ve} \prod_{\left( e,f\right)}\int_{S^2} \rmd \hat{n}_{ef}\prod_{v\in \partial f}\int_{\mathbb{CP}^{1}}\left( \frac{\dim(j_f)}{\pi }\Omega
_{vf}\right) e^{S}
\ee%
where we have a ``spinfoam action'' $S=\sum_f S_f$ and 
\begin{equation}
S_{f}=\sum_{v\in f}S_{vf}=\sum_{v\in f}\left( j_{f}\ln \frac{\left\langle
\xi_{ef},Z_{vef}\right\rangle ^{2}\left\langle Z_{ve^{\prime }f},\xi_{e^{\prime
}f}\right\rangle ^{2}}{\left\langle Z_{vef},Z_{vef}\right\rangle
\left\langle Z_{ve^{\prime }f},Z_{ve^{\prime }f}\right\rangle }+i\gamma
j_{f} \ln \frac{\left\langle Z_{ve^{\prime }f},Z_{ve^{\prime
}f}\right\rangle }{\left\langle Z_{vef},Z_{vef}\right\rangle }\right).
\label{eq:actionL}
\end{equation}

In this paper we consider the large-j regime of the spinfoam amplitude $A(\ck)$. Concretely, we define the partial-amplitude
\be
A_{j_f}(\ck)&:=&\prod_{\left( v,e\right)}\int_{\Slc}
\rmd g_{ve} \prod_{\left( e,f\right)}\int_{S^2} \rmd \hat{n}_{ef}\prod_{v\in \partial f}\int_{\mathbb{CP}^{1}}\left( \frac{\dim(j_f)}{\pi }\Omega
_{vf}\right) e^{S}  \label{Aj}\\  
A(\ck)&=&\sum_{j_f}\prod_{f}\mu \left( j_{f}\right)A_j(\ck)\nonumber
\ee 
and consider the regime in the sum $\sum_{j_f}$ where all the spins $j_f$ are large. In this regime, the spinfoam amplitude is a sum over the asymptotics of partial amplitude $A_j(\ck)$ with large spins $j_f$. In the following, we study the large-j asymptotics of the partial-amplitudes $A_{j_f}(\ck)$ by making the uniform scaling $j_f\mapsto\l j_f$ and taking the limit $\l\to\infty$. Each face action $S_f\mapsto\l S_f$ scales linearly with $\l$, so we can use the generalized stationary phase approximation \cite{stationaryphase} to study the asymptotical behavior of $A_{j_f}(\ck)$ in large-j regime.

Before coming to the asymptotic analysis, we note that in all the following discussions, we only consider the spin configurations such that $\sum_{f\subset t_e}\epsilon_f j_f\neq 0$ with $\epsilon_f=\pm1$ for all $e$. Therefore the geometric tetrahedron with the oriented area $j_f \hat{n}_{ef}$, $f\subset t_e$ is  always assumed to be nondegenerate.

\subsection{Derivation of Critical Point Equations}

We use the generalized stationary phase method to study the large-j asymptotics of the above spinfoam amplitude. The spinfoam amplitude have been reduced to the following type of integral: 
\be
f(\l)=\int_D\rmd x\ a(x)\ e^{\l S(x)}
\ee
where $D$ is a closed manifold, $S(x)$ and $a(x)$ are smooth, complex valued functions, and $\mathrm{Re}S\leq 0$ (this will be shown in the following for the spinfoam amplitude). For large parameter $\l$ the dominant contributions for the above integral comes from the \emph{critical points} $x_c$, which are the stationary point of $S(x)$ and satisfy $\mathrm{Re}S(x_c)=0$. The asymptotic behavior of the above integral for large $\l$ is given by
\be
f(\l)=\sum_{x_c}a(x_c)\lt(\frac{2\pi}{\l}\rt)^{\frac{r(x_c)}{2}}\frac{e^{i\mathrm{Ind}H'(x_c)}}{\sqrt{|\det_r H'(x_c)|}}e^{\l S(x_c)}\lt[1+o(\frac{1}{\l})\rt]
\ee
for isolated critical points, where $r(x_c)$ is the rank of the Hessian matrix $H_{ij}(x_c)=\partial_i\partial_j S(x_c)$ at a critical point, and $H'(x_c)$ is the invertible restriction on $\mathrm{ker}H(x_c)^\perp$. When the critical points are not isolated, the above $\sum_{x_c}$ is replaced by a integral over a submanifold of critical points. If the $S(x)$ doesn't have any critical point $f(\l)$ decreases faster than any power of $\l^{-1}$. From the above asymptotic formula, we see that the asymptotics of the spinfoam amplitude is clarified by finding all the critical points of the action and evaluating the integrand at each critical point.  
                     
In order to find the critical points of the spinfoam action, first of all, we show that the spinfoam action $S$ satisfies $\mathrm{Re}S\leq0$. For each $S_{vf}$, by using the Cauchy-Schwarz inequality 
\be
\mathrm{Re}S_{vf}=j_{f}\ln \frac{\left\vert \left\langle
\xi_{ef},Z_{vef}\right\rangle \right\vert ^{2}\left\vert \left\langle
Z_{ve^{\prime }f},\xi_{e^{\prime }f}\right\rangle \right\vert ^{2}}{%
\left\langle Z_{vef},Z_{vef}\right\rangle \left\langle Z_{ve^{\prime
}f},Z_{ve^{\prime }f}\right\rangle }
\leq j_{f}\ln \frac{\left\langle\xi_{ef},\xi_{ef}\rag\lag Z_{vef},Z_{vef}\right\rangle  \left\langle\xi_{e'f},\xi_{e'f}\rag\lag Z_{ve'f},Z_{ve'f}\right\rangle}{%
\left\langle Z_{vef},Z_{vef}\right\rangle \left\langle Z_{ve^{\prime
}f},Z_{ve^{\prime }f}\right\rangle }
\leq 0
\ee
Therefore 
\be
\mathrm{Re}S=\sum_{f,v}\mathrm{Re}S_{vf} \leq0
\ee
From $\mathrm{Re}S=0$, we obtain the following equations
\begin{equation}
\xi_{ef}=\frac{e^{i\phi _{ev}}}{\left\Vert Z_{vef}\right\Vert }Z_{vef},\quad
\text{and}\quad \xi_{e^{\prime }f}=\frac{e^{i\phi _{e^{\prime }v}}}{\left\Vert
Z_{ve^{\prime }f}\right\Vert }Z_{ve^{\prime }f}  \label{eq:ReS}
\end{equation}%
where $\left\Vert Z_{vef}\right\Vert \equiv \left\vert \left\langle
Z_{vef},Z_{vef}\right\rangle \right\vert ^{1/2}$. If we define $\phi
_{eve^{\prime }}=\phi _{ev}-\phi _{e^{\prime }v}$, the above equation results in that
\begin{equation}
\left( g_{ve}^{\dag }\right) ^{-1}\xi_{ef}=\frac{\left\Vert Z_{ve^{\prime
}f}\right\Vert }{\left\Vert Z_{vef}\right\Vert }e^{i\phi _{eve^{\prime
}}}\left( g_{ve^{\prime }}^{\dag }\right) ^{-1}\xi_{e^{\prime }f}
\label{eq:PreGlu}
\end{equation}%
Here we use the property of anti-linear map $J$
\begin{eqnarray}
JgJ^{-1} &=&\left( g^{\dag }\right) ^{-1} 
\end{eqnarray}%
to Eq.(\ref{eq:PreGlu}), we find%
\begin{equation}
g_{ve}\left( J\xi_{ef}\right) =\frac{\left\Vert Z_{ve^{\prime }f}\right\Vert }{%
\left\Vert Z_{vef}\right\Vert }e^{-i\phi _{eve^{\prime }}}g_{ve^{\prime
}}\left( J\xi_{e^{\prime }f}\right)   \label{eq:GluingJn}
\end{equation}

Now we compute the derivative of the action $S$ on the variables $z_{vf},\xi_{ef},g_{ve}$ to find the staionary point of $S$. We first consider the derivative with respect to the $\mathbb{CP}^1$ variable $z_{vf}$. Given a spinor $z^\a=(z_0,z_1)^t$, $z^\a$ and $(Jz)^\a=(-\bar{z}_1,\bar{z}_0)$ is a basis of the space $\bbc^2$ of 2-component spinors. The following variation can be written in general by
\be
\delta z^\a=\eps (Jz)^\a+\o z^\a
\ee 
where $\eps,\o$ are complex number. Since $z\in\mathbb{CP}^1$, we can choose a partial gauge fixing that $\lag z,z\rag=1$, which gives $\lag \delta z,z\rag=-\lag z,\delta z\rag$. Thus we obtain $\o=i\eta$ with a real number $\eta$. Moreover if we choose the variation with $\eps=0$, it leads to $\delta z^\a=i\eta z^\a$, which gives $\eta=0$ for $z\in\mathbb{CP}^1$. Using the variation $\delta z_{vf}^\a=\eps_{vf} (Jz_{vf})^\a$ and $\delta \bar{z}_{vf}^\a=\bar{\eps}_{vf} (J\bar{z}_{vf})^\a$, we obtain that
\begin{eqnarray}
0&=&\delta _{z_{vf}}S_{vf} \nonumber\\
&=&j_{f}\left( 2\frac{\delta _{z_{vf}}\left\langle
\xi_{ef},Z_{vef}\right\rangle }{\left\langle \xi_{ef},Z_{vef}\right\rangle }+2%
\frac{\delta _{z_{vf}}\left\langle Z_{ve^{\prime }f},\xi_{e^{\prime
}f}\right\rangle }{\left\langle Z_{ve^{\prime }f},\xi_{e^{\prime
}f}\right\rangle }-\frac{\delta _{z_{vf}}\left\langle
Z_{vef},Z_{vef}\right\rangle }{\left\langle Z_{vef},Z_{vef}\right\rangle }-%
\frac{\delta _{z_{vf}}\left\langle Z_{ve^{\prime }f},Z_{ve^{\prime
}f}\right\rangle }{\left\langle Z_{ve^{\prime }f},Z_{ve^{\prime
}f}\right\rangle }\right)  \nonumber\\
&&+i\gamma j_{f} \left( \frac{\delta _{z_{vf}}\left\langle
Z_{ve^{\prime }f},Z_{ve^{\prime }f}\right\rangle }{\left\langle
Z_{ve^{\prime }f},Z_{ve^{\prime }f}\right\rangle }-\frac{\delta
_{z_{vf}}\left\langle Z_{vef},Z_{vef}\right\rangle }{\left\langle
Z_{vef},Z_{vef}\right\rangle }\right) \nonumber \\
&=&j_{f}\left( \frac{\varepsilon_{vf} \left\langle \xi_{ef},g_{ve}^{\dag
}Jz_{vf}\right\rangle }{\left\langle \xi_{ef},Z_{vef}\right\rangle }+\frac{%
\bar{\varepsilon}_{vf}\big\langle g_{ve^{\prime }}^{\dag }Jz_{vf},\xi_{e^{\prime
}f}\big\rangle }{\left\langle Z_{ve^{\prime }f},\xi_{e^{\prime
}f}\right\rangle }-\frac{\bar{\varepsilon}_{vf}\left\langle g_{ve}^{\dag
}Jz_{vf},Z_{vef}\right\rangle }{\left\langle Z_{vef},Z_{vef}\right\rangle }-%
\frac{\varepsilon_{vf} \big\langle Z_{ve^{\prime }f},g_{ve^{\prime }}^{\dag
}Jz_{vf}\big\rangle }{\left\langle Z_{ve^{\prime }f},Z_{ve^{\prime
}f}\right\rangle }\right) \nonumber \\
&&+i\gamma  j_{f} \left( \frac{\varepsilon_{vf} \big\langle
Z_{ve^{\prime }f},g_{ve^{\prime }}^{\dag }Jz_{vf}\big\rangle +\bar{%
\varepsilon}_{vf}\big\langle g_{ve^{\prime }}^{\dag }Jz_{vf},Z_{ve^{\prime
}f}\big\rangle }{\left\langle Z_{ve^{\prime }f},Z_{ve^{\prime
}f}\right\rangle }-\frac{\varepsilon_{vf} \left\langle Z_{vef},g_{ve}^{\dag
}Jz_{vf}\right\rangle +\bar{\varepsilon}_{vf}\left\langle g_{ve}^{\dag
}Jz_{vf},Z_{vef}\right\rangle }{\left\langle Z_{vef},Z_{vef}\right\rangle }%
\right)
\end{eqnarray}%
Using Eq.(\ref{eq:ReS}), we obtain the following equation
\begin{equation}
\big\langle Jz_{vf},g_{ve}\xi_{ef}\big\rangle=\frac{\left\Vert Z_{vef}\right\Vert }{\left\Vert Z_{ve^{\prime
}f}\right\Vert }e^{i\phi _{eve^{\prime }}}\big\langle Jz_{vf},g_{ve^{\prime }}\xi_{e^{\prime }f}\big\rangle
\end{equation}%
Also from Eq.\Ref{eq:ReS}, because of $\lag\xi_{ef},\xi_{ef}\rag=\lag\xi_{e'f},\xi_{e'f}\rag=1$
\be
\lag z_{vf},g_{ve}\xi_{ef}\rag=\frac{\left\Vert Z_{vef}\right\Vert }{\left\Vert Z_{ve'f}\right\Vert }e^{i\phi _{eve'}}\lag z_{vf},g_{ve'}\xi_{e'f}\rag
\ee
Therefore since $z^\a$ and $(Jz)^\a$ is a basis of the space $\bbc^2$ of 2-component spinors,
\be
g_{ve}\xi_{ef}=\frac{\left\Vert Z_{vef}\right\Vert }{\left\Vert Z_{ve'f}\right\Vert }e^{i\phi _{eve'}}g_{ve'}\xi_{e'f}\label{eq:Gluingn}
\ee

We consider the variation with respect to $\xi_{ef}$. Since the spinor $\xi_{ef}$ is normalized, we should use $\delta\xi_{ef}^\a=\o_{ef}( J\xi_{ef})^\a+i\eta_{ef}\xi^\a_{ef}$ for complex infinitesimal parameter $\o\in\bbc$ and $\eta\in\mathbb{R}$. The variation of the action vanishes automatically 
\begin{eqnarray}
\delta _{\xi_{ef}}S &=&j_{f}\left( 2\frac{\delta _{\xi_{ef}}\left\langle
\xi_{ef},Z_{vef}\right\rangle }{\left\langle \xi_{ef},Z_{vef}\right\rangle }+2%
\frac{\delta _{\xi_{ef}}\left\langle Z_{v^{\prime }ef},\xi_{ef}\right\rangle }{%
\left\langle Z_{v^{\prime }ef},\xi_{ef}\right\rangle }\right)=j_{f}\left( 2\frac{\bar{\o}_{ef}\left\langle
J\xi_{ef},Z_{vef}\right\rangle }{\left\langle \xi_{ef},Z_{vef}\right\rangle }+2%
\frac{\o_{ef} \left\langle Z_{v^{\prime }ef},J\xi_{ef}\right\rangle }{%
\left\langle Z_{v^{\prime }ef},\xi_{ef}\right\rangle }\right) \nonumber \\
&=&0\label{deltaxi}
\end{eqnarray}%
by using Eq.(\ref{eq:ReS}) and the identity $\left\langle J\xi_{ef},\xi_{ef}\right\rangle =0$.

Finally we consider the stationary point for the group variables $g_{ve}$. We parameterize
the group with the parameter $\theta _{IJ}$ around a saddle point ${g
}_{ve}$, i.e. $g'_{ve}={g}_{ve}e^{-i\theta _{IJ}^{ve}\cj^{IJ}}$, where $\cj^{IJ}$ is the generator of the Lie algebra $\slc$. Then we have%
\begin{eqnarray}
0&=&\frac{\partial S_{vf}}{\partial \theta _{IJ}^{ve}}\Big|_{\theta ^{ve}=0}\nonumber\\
&=&\sum_{f\ \text{incoming}\ e}\Bigg[j_{f}\left( 2\frac{\big\langle \xi_{ef},i\cj^{IJ\dagger}Z_{vef}\big\rangle }{\left\langle
\xi_{ef},Z_{vef}\right\rangle }-\frac{\big\langle i\cj^{IJ\dagger} Z_{vef},Z_{vef}\big\rangle +\big\langle Z_{vef},i\cj^{IJ\dagger}Z_{vef}\big\rangle }{\left\langle
Z_{vef},Z_{vef}\right\rangle }\right) \nonumber \\
&&+i\gamma j_{f} \left( -\frac{\big\langle i\cj^{IJ\dagger}Z_{vef},Z_{vef}\big\rangle +\big\langle Z_{vef},i\cj^{IJ\dagger} Z_{vef}\big\rangle }{\left\langle
Z_{vef},Z_{vef}\right\rangle }\right) \Bigg]\nonumber \\
&&+\sum_{f\ \text{outgoing}\ e}\Bigg[j_{f}\left( 2\frac{\big\langle i\cj^{IJ\dagger} Z_{vef},\xi_{ef}\big\rangle }{\left\langle
Z_{vef},\xi_{ef}\right\rangle }-\frac{\big\langle i\cj^{IJ\dagger} Z_{vef},Z_{vef}\big\rangle +\big\langle Z_{vef},i\cj^{IJ\dagger} Z_{vef}\big\rangle }{\left\langle
Z_{vef},Z_{vef}\right\rangle }\right)  \nonumber\\
&&+i\gamma j_{f} \left( \frac{\big\langle i\cj^{IJ\dagger} Z_{vef},Z_{vef}\big\rangle +\big\langle Z_{vef},i\cj^{IJ\dagger} Z_{vef}\big\rangle }{\left\langle
Z_{vef},Z_{vef}\right\rangle }\right) \Bigg]
\end{eqnarray}%
We again apply Eq.(\ref{eq:ReS}) and find
\begin{eqnarray}
0&=&\sum_{f\ \text{incoming}\ e}\lt[j_{f}\left( \big\langle \xi_{ef},i\cj^{IJ\dagger}\xi_{ef}\big\rangle +\big\langle
\xi_{ef},i\cj^{IJ}\xi_{ef}\big\rangle \right) +i\gamma  j_{f} \left( \big\langle
\xi_{ef},i\cj^{IJ}\xi_{ef}\big\rangle -\big\langle \xi_{ef},i\cj^{IJ\dagger}\xi_{ef}\big\rangle \right) \rt]  \nonumber \\
&&+\sum_{f\ \text{outgoing}\ e}\lt[j_{f}\left( -\big\langle
\xi_{ef},i\cj^{IJ}\xi_{ef}\big\rangle -\big\langle \xi_{ef},i\cj^{IJ\dagger}\xi_{ef}\big\rangle \right)  +i\gamma j_{f} \left( -\big\langle\xi_{ef},i\cj^{IJ}\xi_{ef}\big\rangle +\big\langle \xi_{ef},i\cj^{IJ\dagger}\xi_{ef}\big\rangle \right) \rt]  \nonumber \\
&=&\sum_{f\in t_e}^{4}\eps_{ef}(v)\lt[j_{f}\left( \big\langle \xi_{ef},i\cj^{IJ\dagger}\xi_{ef}\big\rangle +\big\langle
\xi_{ef},i\cj^{IJ}\xi_{ef}\big\rangle \right) +i\gamma  j_{f} \left( \big\langle
\xi_{ef},i\cj^{IJ}\xi_{ef}\big\rangle -\big\langle \xi_{ef},i\cj^{IJ\dagger}\xi_{ef}\big\rangle \right) \rt] \label{eq:dSdg}
\end{eqnarray}%
where $\eps_{ef}(v)=\pm1$ is determined (up to a global sign) by the following relations
\be
\eps_{ef}(v)=-\eps_{e'f}(v)\ \ \ \ \text{and}\ \ \ \ \eps_{ef}(v)=-\eps_{ef}(v')
\ee
for the triangle $f$ shared by the tetrahedra $t_e$ and $t_{e'}$ in the 4-simplex $\sig_v$, and the dual edge $e=(v,v')$. As usual we can rewrite Lorentz Lie algebra generator $\cj^{IJ}$ in terms of rotation part $\vec{J}$ and boost part $\vec{K}$ where where $J_{i}=\frac{i}{2}\epsilon
_{0ijk}\cj^{jk}$, $K_{i}=-i\cj^{0i}$. In the Spin-$\half$ representation, the rotation generators $\vec{J}=\frac{i}{2}\vec{\sigma}$ and the boost generators $\vec{K}=\frac{1}{2}\vec{\sigma}$. Recall that
\be
\left\langle \xi\left\vert \vec{\sigma}\right\vert \xi\right\rangle =\hat{n}_\xi\ \ \ \ \text{with}\ \ \ \ \
\hat{n}_\xi=(\xi^0\bar{\xi}^1+\xi^1\bar{\xi}^0)\hat{\mathbf{x}}-i(\xi^0\bar{\xi}^1-\xi^1\bar{\xi}^0)\hat{\mathbf{y}}+(\xi^0\bar{\xi}^0-\xi^1\bar{\xi}^1)\hat{\mathbf{z}}
\ee
we have%
\begin{eqnarray}
\left\langle \xi_{ef},\vec{J}\xi_{ef}\right\rangle  &=&-\left\langle
\xi_{ef},\vec{J}^{\dag }\xi_{ef}\right\rangle =\frac{i}{2}\hat{n}_{ef} \\
\left\langle\xi_{ef},\vec{K}\xi_{ef}\right\rangle  &=&\left\langle
\xi_{ef},\vec{K}^{\dag }\xi_{ef}\right\rangle =\frac{1}{2}\hat{n}_{ef}
\end{eqnarray}
Using the above relations, Eq.(\ref{eq:dSdg}) results in the closure condition
\be
\sum_{f\subset t_e}^{4}\eps_{ef}(v)j_{f}\hat{n}_{ef}=0.
\ee%
Thus we finish the derivation of all the critical point equations.

\subsection{Analysis of Critical Point Equations}

We summarize the critical point equations for a spinfoam configuration $(j_f,g_{ev},\xi_{ef},z_{vf})$
\begin{eqnarray}
g_{ve}\left( J\xi_{ef}\right)  &=&\frac{\left\Vert Z_{ve^{\prime
}f}\right\Vert }{\left\Vert Z_{vef}\right\Vert }e^{-i\phi _{eve^{\prime
}}}g_{ve^{\prime }}\left( J\xi_{e^{\prime }f}\right)   \label{gluingJ}
\\
g_{ve}\xi_{ef} &=&\frac{\left\Vert Z_{vef}\right\Vert }{\left\Vert
Z_{ve^{\prime }f}\right\Vert }e^{i\phi _{eve^{\prime }}}g_{ve^{\prime
}}\xi_{e^{\prime }f}  \label{gluing} \\
0&=&\sum_{f\subset t_e}^{4}\eps_{ef}(v)j_{f}\hat{n}_{ef}  \label{closure}
\end{eqnarray}%
where Eq.(\ref{closure}) stands for the closure condition for each tetrahedron. $\eps_{ef}(v)$ is the sign factor coming from the variation with respect to $g_{ev}$. It is determined (up to a global sign) by the following relations
\be
\eps_{ef}(v)=-\eps_{e'f}(v)\ \ \ \ \text{and}\ \ \ \ \eps_{ef}(v)=-\eps_{ef}(v')
\ee
for the triangle $f$ shared by the tetrahedra $t_e$ and $t_{e'}$ in the 4-simplex $\sig_v$, and the dual edge $e=(v,v')$.

In the following, we show that Eqs.(\ref{gluingJ}) and (\ref{gluing}) give the parallel transportation condition of the bivectors. Given a spinor $\xi^\a$, it naturally constructs a null vector $\xi^\a\bar{\xi}^{\dot{\a}}=\iota(\xi)^I\sig_I^{\a\dot{\a}}$ where $\sig_I=(1,\vec{\sig})$. It is straight-forward to check that
\be
\xi\bar{\xi}=\half(1+\vec{\sig}\cdot\hat{n}_\xi)\ \ \ \ \
\text{with}\ \ \ \ \
\hat{n}_\xi=(\xi^0\bar{\xi}^1+\xi^1\bar{\xi}^0)\hat{\mathbf{x}}-i(\xi^0\bar{\xi}^1-\xi^1\bar{\xi}^0)\hat{\mathbf{y}}+(\xi^0\bar{\xi}^0-\xi^1\bar{\xi}^1)\hat{\mathbf{z}}
\ee
$\hat{n}_\xi$ is a unit 3-vector since $\xi$ is a normalized spinor. Thus we obtain that
\be
\iota({\xi})^I=\half(1,\hat{n}_\xi)
\ee
Similarly for the spinor $J\xi$, we define the null vector $J\xi^\a\overline{J\xi}{}^{\dot{\a}}=\iota(J\xi)^I\sig_I^{\a\dot{\a}}$ and obtain that
\be
\iota(J{\xi})^I=\half(1,-\hat{n}_\xi)
\ee
We can write Eqs.\Ref{gluingJ} and \Ref{gluing} in their Spin-1 representation
\be
\hat{g}_{ve}\ \iota( J\xi_{ef})  =\frac{\left\Vert Z_{ve^{\prime
}f}\right\Vert^2 }{\left\Vert Z_{vef}\right\Vert^2 }\hat{g}_{ve^{\prime }}\ \iota( J\xi_{e^{\prime }f})
\ \ \ \ \text{and}\ \ \ \
\hat{g}_{ve}\ \iota(\xi_{ef}) =\frac{\left\Vert Z_{vef}\right\Vert ^2}{\left\Vert
Z_{ve^{\prime }f}\right\Vert^2 }\hat{g}_{ve^{\prime
}}\ \iota(\xi_{e^{\prime }f})
\ee
It is obvious that if we construct a bivector\footnote{the pre-factor is a convention for simplifying the notation in the following discussion.}
\be
X_{ef}^{IJ}=-4\g j_f\lt[\iota(\xi_{ef})\wedge\iota(J\xi_{ef})\rt]^{IJ}\label{Xef0}
\ee
$X_{ef}$ satisfies the parallel transportation condition within a 4-simplex
\be
(\hat{g}_{ve})^{I}_{\ K}(\hat{g}_{ve})^{J}_{\ L}X^{KL}_{ef}=(\hat{g}_{ve'})^{I}_{\ K}(\hat{g}_{ve'})^{J}_{\ L}X^{KL}_{e'f}.
\ee
We define the bivector $X^{IJ}_f$ located at each vertex $v$ of the dual face $f$ by the parallel transportation
\be
X_f^{IJ}(v):=(\hat{g}_{ve})^{I}_{\ K}(\hat{g}_{ve})^{J}_{\ L}X^{KL}_{ef}.
\ee
which is independent of the choice of $e$ by the above parallel transportation condition. Then we have the parallel transportation relation of $X^{IJ}_f(v)$
\be
X_f^{IJ}(v)=(\hat{g}_{vv'})^{I}_{\ K}(\hat{g}_{vv'})^{J}_{\ L}X^{KL}_{f}(v')
\ee
because the spinor $\xi_{ef}$ belonging to the tetrahedron $t_e$ is shared as the boundary data by two neighboring 4-simplex.

On the other hand, we can write the bivector $X^{IJ}_{ef}$ as a matrix:
\be
X_{ef}^{IJ}&=&2\g j_f
\left(\begin{array}{cccc}0 &\ \hat{n}^1_{ef} &\ \hat{n}^2_{ef} &\ \hat{n}^3_{ef} \\-\hat{n}^1_{ef} &\ 0 &\ 0 &\ 0 \\-\hat{n}_{ef}^2 &\ 0 &\ 0 &\ 0 \\ -\hat{n}^3_{ef} &\ 0 &\ 0 &\ 0\end{array}\right)\nonumber\\
\lt|X_{ef}^{IJ}\rt|&=&\sqrt{\lt|\half X_{ef}^{IJ}X^{ef}_{IJ}\rt|}\ =\ 2\g j_f\label{Xef1}
\ee
However the matrix $(X_{ef})^{I}_{\ J}=X_{ef}^{IK}\eta_{KJ}$ read
\be
X_{ef}\equiv (X_{ef})^{I}_{\ J}=2\g j_f
\left(\begin{array}{cccc}0 &\ \hat{n}^1_{ef} &\ \hat{n}^2_{ef} &\ \hat{n}^3_{ef} \\ \hat{n}^1_{ef} &\ 0 &\ 0 &\ 0 \\ \hat{n}_{ef}^2 &\ 0 &\ 0 &\ 0 \\  \hat{n}^3_{ef} &\ 0 &\ 0 &\ 0\end{array}\right)=2\g j_f\hat{n}_{ef}\cdot\vec{K}\label{Xef}
\ee
where $\vec{K}$ denotes the boost generator of Lorentz Lie algebra $\slc$ in the Spin-1 representation. The rotation generator in $\slc$ is denoted by $\vec{J}$. The generators in $\slc$ satisfies the commutation relations $[J^i,J^j]=-\eps^{ijk}J^k,\ [J^i,K^j]=-\eps^{ijk}K^k,\ [K^i,K^j]=\eps^{ijk}J^k$. The relation $X_{ef}=2\g j_f\hat{n}_{ef}\cdot\vec{K}$ gives a representation of the bivector in terms of the $\slc$ lie algebra generators. Moreover it is not difficult to verify that in the Spin-$\half$ representation $\vec{J}=\frac{i}{2}\vec{\sig}$ and $\vec{K}=\frac{1}{2}\vec{\sig}$. Thus in Spin-$\half$ representation
\be
X_{ef}= \g j_f\vec{\sig}\cdot\hat{n}_{ef}
\ee
For this $\slc$ Lie algebra representation of the bivector $X_{ef}$, the parallel transportation is represented by the adjoint action of the Lie group on its Lie algebra. Therefore we have
\be
g_{ve}X_{ef}g_{ev}=g_{ve'}X_{e'f}g_{e'v},\ \ \ \ X_{f}(v):=g_{ve}X_{ef}g_{ev},\ \ \ \ X_{f}(v):=g_{vv'}X_{f}(v')g_{v'v}
\ee
where $g_{ve}=g_{ev}^{-1}, g_{v'v}=g_{vv'}^{-1}$. We note that the above equations are valid for all the representations of $\Slc$.

There is the duality map acting on $\slc$ by $*\vec{J}=-\vec{K}$, $*\vec{K}=\vec{J}$. For self-dual/anti-self-dual bivector $\vec{T}_\pm:=\half(\vec{J}\pm i\vec{K})$, One can verify that $*\vec{T}_\pm=\pm i\vec{T}_\pm$. In the Spin-1 representation (bivector representation), the duality map is represented by $*X^{IJ}=\half\eps^{IJKL}X_{KL}$. In the Spin-$\half$ representation, the duality map is represented by $*X=iX$ since $\vec{J}=\frac{i}{2}\vec{\sig}$ and $\vec{K}=\frac{1}{2}\vec{\sig}$ in the Spin-$\half$ representation. From Eq.\Ref{Xef}, we see that
\be
X_{ef}=-*(2\g j_f\hat{n}_{ef}\cdot\vec{J})
\ee
From its bivector representation one can see that
\be
\eta_{IJ}u^I*\!X_{ef}^{JK}=0,\ \ \ \ \ u^I=(1,0,0,0).
\ee
It motivates us to define a unit vector at each vertex $v$ for each tetrahedron $t_e$ by
\be
N^I_e(v):=(\hat{g}_{ve})^{I}_{\ J}u^J\label{Ne}
\ee
Then for all triangles $f$ in the tetrahedron $t_e$, $N^I_e(v)$ is orthogonal to all the bivectors $*\!X_f(v)$ with $f$ belonging to $t_e$.
\be
\eta_{IJ}N^I_e(v)*\!X^{JK}_f(v)=0.
\ee

In addition, from the closure constraint Eq.\Ref{closure}, we obtain for each tetrahedron $t_e$
\be
\sum_{f\subset t_e}\eps_{ef}(v)X_f(v)=0.
\ee

We summarize the above analysis of the critical point equations Eqs.\Ref{gluingJ}, \Ref{gluing}, and \Ref{closure} into the following proposition:

\begin{Proposition}\label{fromcritical}

Given the data $(j_f,g_{ev},\xi_{ef},z_{vf})$ be a spinfoam configuration that solves the critical point equations Eqs.\Ref{gluingJ}, \Ref{gluing}, and \Ref{closure}, we construct the bivector variables (in the $\slc$ Lie algebra representation) for the spinfoam amplitude $X_{ef}=-*\!(2\g j_f\hat{n}_{ef}\cdot\vec{J})$ and $X_{ef}(v):=g_{ve}X_{ef}g_{ev}$, where $|X_{ef}(v)|=\sqrt{\half\tr\lt(X_{ef}(v)X_{ef}(v)\rt)}=2\g j_f$. The critical point equations implies the following equations for the bivector variables
\be
&&X_{ef}(v)=X_{e'f}(v)\equiv X_f(v),\ \ \ \ X_{f}(v):=g_{vv'}X_{f}(v')g_{v'v}, \nonumber\\
&&\eta_{IJ}N^I_e(v)*\!X^{JK}_f(v)=0,\ \ \ \ \ \ \ \sum_{f\subset t_e}\eps_{ef}(v)X_f(v)=0.\label{summarize}
\ee
where $t_e$ and $t_{e'}$ are two different tetrahedra of a 4-simplex dual to $v$, $f$ is a triangle shared by the two tetrahedra $t_e$ and $t_{e'}$, and $N^I_e(v)=(\hat{g}_{ve})^{I}_{\ J}u^J$ with $u^J=(1,0,0,0)$ is a unit vector associated with the tetrahedron $t_e$. $\eps_{ef}(v)$ is a sign factor determined (up to a global sign) by the following relations
\be
\eps_{ef}(v)=-\eps_{e'f}(v)\ \ \ \ \text{and}\ \ \ \ \eps_{ef}(v)=-\eps_{ef}(v')
\ee
for the triangle $f$ shared by the tetrahedra $t_e$ and $t_{e'}$ in the 4-simplex $\sig_v$, and the dual edge $e=(v,v')$.
\end{Proposition}


\section{Nondegenerate Geometry on a Simplicial Complex}\label{geometry}

\subsection{Discrete Bulk Geometry}

In order to relate the spinfoam configurations solving the critical point equations with the a discrete Regge geometry, here we introduce the \emph{classical} geometric variables for the discrete Lorentzian geometry on a 4-manifold \cite{CF,deficit}.

Given a simplicial complex $\ck$ triangulating the 4-manifold $\cm$ with Lorentzian metric $g_{\mu\nu}$, we associate each 4-simplex $\sig_v$ (dual to the vertex $v$) a reference frame. In this reference frame the vertices $[p_1(v),\cdots,p_5(v)]$ of the 4-simplex $\sig_v$ have the coordinates 
\be
p_i(v)=\{x^I_i(v)\}_{i=1,\cdot,5}
\ee
Consider another 4-simplex $\sig_{v'}$ neighboring $\sig_v$, there is an edge $e$ connecting $v$ and $v'$, and there is a tetrahedron $t_e$ shared by $\sig_v,\sig_{v'}$ with vertices $[p_2(v),\cdots,p_5(v)]=[p_2(v'),\cdots,p_5(v')]$. Then it is possible to associate the edge $e=(v,v')$ uniquely an element of Poincar\'e group $\lt\{(\O_{e})^I_{\ J},(\O_e)^I\rt\}$, such that for the vertices $p_2,\cdots, p_5$ of $t_e$
\be
(\O_{e})^I_{\ J}x_i^J(v')+(\O_{e})^I=x_i^I(v)\ \ \ \ \ i=2,\cdots,5
\ee
Here the matrix $(\O_{e})^I_{\ J}$ describes the change of the reference frames in $\sig_v$ and $\sig_{v'}$, while $(\O_{e})^I$ describes the transportation of the frame origins from $\sig_v$ to $\sig_{v'}$. We assume the triangulation is orientable, and we choose the reference frames in $\sig_v,\sig_{v'}$ in such a way that $\O_{e}\in\text{SO(1,3)}$. 

We focus on a 4-simplex $\sig_v$ whose center is the vertex $v$. For each oriented edge $\ell=[p_i(v),p_{j}(v)]$ in the 4-simplex, we associate an edge vector $E^I_\ell(v)=x_i^I(v)-x^I_j(v)$. Thus under the change of reference frame from $\sig_v$ to $\sig_{v'}$
\be
(\O_e)^I_JE^I_\ell(v')=E_\ell(v)\ \ \ \ \forall\ \ell\subset t_e
\ee
In this paper we assume all the edge vectors $E^I_{\ell}(v)$ are \emph{spatial} in the sense of the flat metric $\eta_{IJ}=\mathrm{diag}(-1,1,1,1)$. It is straight-forward to check from the definition that the edge vectors $E_{\ell}^I(v)$ satisfies:
\begin{itemize}
\item if we reverse the orientation of $\ell$, then
\be
E^I_{-\ell}(v)=-E^I_\ell(v),\label{E}
\ee

\item for all triangle $f$ in the simplex $\sig_v$ with edge $\ell_1,\ell_2,\ell_3$, the vectors $E^I_\ell(v)$ close, i.e.
\be
E_{\ell_1}^I(v)+E_{\ell_2}^I(v)+E_{\ell_3}^I(v)=0\label{edgeclose}
\ee
The set of $E_\ell^I(v)$ at $v$ satisfying Eqs.(\ref{E}) and (\ref{edgeclose}) is called a co-frame at the vertex $v$.

\item Moreover given a tetrahedron $t$ shared by two 4-simplices $\sig_v,\sig_{v'}$, for all pair of edges $\ell_1,\ell_2$ of the tetrahedron, we further require that
\be
\eta_{IJ}E_{\ell_1}^I(v)E_{\ell_2}^J(v)=\eta_{IJ}E_{\ell_1}^I(v')E_{\ell_2}^J(v')\label{EE}
\ee

\end{itemize}

\begin{Definition}
The collection of the vectors $E_\ell(v)$ satisfying Eqs.\Ref{E}, \Ref{edgeclose}, and \Ref{EE} at all the vertices is called a co-frame on the simplicial complex $\ck$. The discrete (spatial) metric on the each tetrahedron $t$ induced from $g_{\mu\nu}$ is given by
\be
g_{\ell_1\ell_2}(v)=\eta_{IJ}E_{\ell_1}^I(v)E_{\ell_2}^J(v)
\ee
which is actually independent of $v$ because of Eq.\Ref{EE}. 

\end{Definition}

We assume the co-frame $E_\ell^I(v)$ is nondegenerate, i.e. for each 4-simplex $\sig_v$, the set of $E_\ell^I(v)$ with $\ell\subset\partial\sig_v$ spans a 4-dimensional vector space.

An edge $\ell$ can be denoted by its end-points, say $p_1,p_2$, i.e. $\ell=[p_1,p_2]$. There are 5 vertices $p_i,i=1,\cdots,5$ for a 4-simplex $\sig_v$. Then each $p_i$ is one-to-one corresponding to a tetrahedron $t_{e_i}$ of the 4-simplex $\sig_v$. Therefore we can denote the edge $\ell=[p_1,p_2]$ also by $\ell=(e_1,e_2)$, once a 4-simplex $\sig_v$ is specified. Thus we can also write the co-frame $E^I_\ell(v)$ at the vertex $v$ by $E^I_{e_1e_2}(v)$. In this notation, for example Eqs.\Ref{E} and \Ref{edgeclose} become
\be
E^I_{e_1e_2}(v)=-E^I_{e_2e_1}(v),\ \ \ \ E_{e_1e_2}^I(v)+E_{e_2e_3}^I(v)+E_{e_3e_1}^I(v)=0.\label{coframe}
\ee
In the following we use both of the notations, according to the convenience by the context.

\begin{Lemma}\label{connection}
Given a co-frame $E_{\ell}^I(v)$ on the triangulation, it determines uniquely an SO(1,3) matrix $(\O_e)^I_{\ J}$ associated to each edge $e=(v,v')$ such that for all the edge of the tetrahedron $t_e$ shared by $\sig_v$ and $\sig_{v'}$
\be
(\O_e)^{I}_{\ J}E^J_{\ell}(v')=E^I_{\ell}(v) \ \ \ \ \forall\ \ell\subset t_e
\ee
We can associate a reference frame in each 4-simplex such that SO(1,3) matrix $(\O_e)^I_{\ J}$ changing the frame from $\sig_v$ to $\sig_{v'}$.

\end{Lemma}

\startproof Given a tetrahedron $t_e$ shared by two 4-simplices $\sig_v,\sig_{v'}$, we consider the relation between the co-frame vectors $E^I_\ell(v)$ at the vertex $v$ and $E^I_\ell(v')$ at $v'$, for all 6 edges $\ell$ of the tetrahedron $t_e$. The spatial vectors $E^I_\ell(v)$ $\ell\subset t_e$ spans a 3-dimensional subspace, and the same holds for $E^I_\ell(v')$. We choose the time-like unit normal vectors $\hat{U}(v)$ and $\hat{U}(v')$ orthogonal to $E^I_\ell(v)$ and $E^I_\ell(v')$ respectively, and require that
\be
\mathrm{sgn}\det\lt(E_{\ell_1}(v),E_{\ell_2}(v),E_{\ell_3}(v),\hat{U}(v)\rt)=\mathrm{sgn}\det\lt(E_{\ell_1}(v'),E_{\ell_2}(v'),E_{\ell_3}(v'),\hat{U}(v')\rt)\label{sgndet}
\ee
where $E_{\ell_1}(v),E_{\ell_2}(v),E_{\ell_3}(v)$ form a basis in the 3-dimensional subspace spanned by $E^I_\ell(v)$ $\ell\subset t_e$. From Eq.\Ref{sgndet}, Eq.(\ref{EE}) and $E_{\ell_i}(v)\cdot\hat{U}(v)=E_{\ell_i}(v')\cdot\hat{U}(v')=0$, $i=1,2,3$, an SO(1,3) matrix $\O_{e}$ is determined by
\be
(\O_e)^{I}_{\ J}E^J_{\ell_i}(v')=E^I_{\ell_i}(v)\ \ \ \ \ (\O_e)^{I}_{\ J}\hat{U}^J(v')=\hat{U}^I(v).\label{Omega}
\ee
Suppose there are two SO(1,3) matrices $\O_{e},\O'_{e}$ satisfying
\be
(\O_e)^{I}_{\ J}E^J_{\ell_i}(v')=E^I_{\ell_i}(v)\ \ \ \ \ (\O'_e)^{I}_{\ J}E^J_{\ell_i}(v')=E^I_{\ell_i}(v)
\ee
we then have $\O_{e}=\O'_{e}$. 

We choose a numbering $[p_1,\cdots,p_5]$ of the vertices of $\sig_v,\sig_{v'}$ such that $[p_2(v),\cdots,p_5(v)]=[p_2(v'),\cdots,p_5(v')]$ are the vertices of the tetrahedron $t_e$. Two reference frame in the 4-simplices $\sig_v,\sig_{v'}$ are specified by the coordinates $\lt\{E^I_{e_2e_1}(v),E^I_{e_3e_1}(v),E^I_{e_4e_1}(v),E^I_{e_5e_1}(v)\rt\}$ and $\lt\{E^I_{e_2e_1}(v'),E^I_{e_3e_1}(v'),E^I_{e_4e_1}(v'),E^I_{e_5e_1}(v')\rt\}$ by defining $x^I_j(v):=E^I_{e_je_1}(v)$ and similar for $x^I_j(v')$. Since
\be
E_{e_2e_1}=E_{e_2e_5}-E_{e_1e_5},\ \ \ \ E_{e_3e_1}=E_{e_3e_5}-E_{e_1e_5},\ \ \ \ E_{e_4e_1}=E_{e_4e_5}-E_{e_1e_5}
\ee
and there exists a unique $(\O_e)^I_{\ J}\in\text{SO(1,3)}$ that $E^I_{e_ie_j}(v)=(\O_e)^I_{\ J}E^I_{e_ie_j}(v')$, $i,j=2,\cdots,5$, we can relate the coordinates $\lt\{E^I_{e_2e_1}(v),E^I_{e_3e_1}(v),E^I_{e_4e_1}(v),E^I_{e_5e_1}(v)\rt\}$ and $\lt\{E^I_{e_2e_1}(v'),E^I_{e_3e_1}(v'),E^I_{e_4e_1}(v'),E^I_{e_5e_1}(v')\rt\}$ in two different 4-simplices by
\be
E^I_{e_2e_1}(v)&=&(\O_e)^I_{\ J}E^I_{e_2e_1}(v')+(\O_e)^I_{\ J}E_{e_1e_5}^J(v')-E_{e_1e_5}^I(v)\nonumber\\
E^I_{e_3e_1}(v)&=&(\O_e)^I_{\ J}E^I_{e_3e_1}(v')+(\O_e)^I_{\ J}E_{e_1e_5}^J(v')-E_{e_1e_5}^I(v)\nonumber\\
E^I_{e_4e_1}(v)&=&(\O_e)^I_{\ J}E^I_{e_4e_1}(v')+(\O_e)^I_{\ J}E_{e_1e_5}^J(v')-E_{e_1e_5}^I(v)\nonumber\\
E^I_{e_5e_1}(v)&=&(\O_e)^I_{\ J}E^I_{e_5e_1}(v')+(\O_e)^I_{\ J}E_{e_1e_5}^J(v')-E_{e_1e_5}^I(v)
\ee
The coordinates of $p_2,\cdots,p_5$ are given by $x^I_j(v):=E^I_{e_je_1}(v)$ with respective the reference frame in $\sig_v$, thus the Poincar\'e transformation relating two reference frames are given by an SO(1,3) matrix and a translation $\lt\{(\O_e)^I_{\ J},(\O_e)^I\rt\}$, where the translation vector $(\O_e)^I$ is given by
\be
(\O_e)^I:=(\O_e)^I_{\ J}E_{e_1e_5}^J(v')-E_{e_1e_5}^I(v)
\ee 
\finishproof

The orientation of a 4-simplex $\sig_v$ is represented by an ordering of its 5 vertices, i.e. a tuple $[p_1,\cdots,p_5]$. Two orientations are opposite to each other if the two orderings are related by an odd permutation, e.g. $[p_1,p_2,\cdots,p_5]=-[p_2,p_1\cdots,p_5]$. We say that two neighboring 4-simplices $\sig,\sig'$ are consistently oriented, if the orientation of their shared tetrahedron $t$ induced from $\sig$ is opposite to the orientation induced from $\sig'$. For example, $\sig=[p_1,p_2,\cdots,p_5]$ and $\sig'=-[p'_1,p_2,\cdots,p_5]$ are consistently oriented since the opposite orientations $t=\pm[p_2,\cdots,p_5]$ are induced respectively from $\sig$ and $\sig'$. The simplicial complex $\ck$ is said to be orientable if it is possible to orient consistently all pair of neighboring 4-simplices. Such a choice of consistent 4-simplex orientations is called a global orientation. We assume we define a global orientation of the triangulation $\ck$. Then for each 4-simplex $\sig_v=[p_1,p_2,\cdots,p_5]$, we define an oriented volume (assumed to be nonvanishing as the nondegeneracy)
\be
V_4(v):=\det\Big(E_{e_2e_1}(v),E_{e_3e_1}(v),E_{e_4e_1}(v),E_{e_5e_1}(v)\Big)
\ee
In general the oriented 4-volume $V_4(v)$ can be positive or negative for different 4-simplices.

\begin{Definition}\label{connection1}

Given two neighboring 4-simplices $\sig_v$ and $\sig_{v'}$, if their oriented volumes are both positive or both negative, i.e. $\mathrm{sgn}(V_4(v))=\mathrm{sgn}(V_4(v'))$. The SO(1,3) matrix $(\O_e)^I_{\ J}$, $e=(v,v')$ is the discrete spin connection compatible with $E_\ell(v)^I$.

\end{Definition}

For each vertex $v$ and a dual edge $e$ connecting $v$, we define a time-like vector $U_e(v)$ at the vertex $v$ by (choosing any $j\neq k$, the definition is independent of the choice of $j$ by Eqs.(\ref{E}) and \Ref{edgeclose})
\be
U^{e_k}_I(v):=\frac{1}{3!V_4(v)}\sum_{l,m,n}\eps^{jklmn}\eps_{I JKL}E^J_{e_l e_j}(v)E^K_{e_m e_j}(v)E^L_{e_n e_j}(v)
\ee
In total there are 5 vectors $U_e(v)$ at each vertex $v$. Using Eq.\Ref{E} and \Ref{edgeclose}, one can show that
\be
U_J^{e_j}(v)E^J_{e_k e_l}(v)=\delta_{jk}-\delta_{jl}\label{frame}
\ee
Thus we call the collection of $U_e(v)$ a discrete frame since $E_{e_1e_2}(v)$ is called a discrete co-frame. Moreover from this equation we see that $U^J_e(v)$ is a vector at $v$ normal to the tetrahedron $t_e$. If we sum over all 5 frame vectors $U_e(v)$ at $v$ in Eq.\Ref{frame}
\be
\sum_{j=1}^5U_J^{e_j}(v)E^J_{e_k e_l}(v)=\sum_{j=1}^5\delta_{jk}-\sum_{j=1}^5\delta_{jl}=0\ \ \ \ \forall\ e_k,e_l
\ee
which shows the closure of $U_e(v)$ at each vertex $v$, i.e.
\be
\sum_{e=1}^5U_e(v)=0\label{Uclose}
\ee
by the nondegenercy of $E_{ee'}(v)$. Eq.(\ref{Uclose}) shows that the 5 vectors $U_e(v)$ are all out-pointing or all in-pointing normal vectors to the tetrahedra. Also following from Eq.\Ref{frame} (fix $l=1$ and let $j=2,3,4,5$), we have that the $4\times 4$ matrix $\lt(U^{e_2}(v),U^{e_3}(v),U^{e_4}(v),U^{e_5}(v)\rt)^t$ is the inverse of the matrix $\lt(E_{e_2e_1}(v),E_{e_3e_1}(v),E_{e_4e_1}(v),E_{e_5e_1}(v)\rt)$. Therefore
\be
\frac{1}{V_4(v)}=\det\Big(U^{e_2}(v),U^{e_3}(v),U^{e_4}(v),U^{e_5}(v)\Big).
\ee
It implies ($i,j,k,l=2,3,4,5$)
\be
&&V_4(v)\eps^{IJKL}U^{e_i}_I(v)U^{e_j}_J(v)U^{e_k}_K(v)U^{e_l}_L(v)=\eps^{ijkl}\nonumber\\
&&V_4(v)\eps_{ijkl}U^{e_i}_I(v)U^{e_j}_J(v)U^{e_k}_K(v)U^{e_l}_L(v)=\eps_{IJKL}
\ee
where the above $\eps_{ijkl}=\eps^{ijkl}$, $\eps_{IJKL}=\eps^{IJKL}$ are all Levi-Civita symbols. Then using the fact that the matrix $U^{e_i}_I(v)$ is the inverse of $ E^I_{e_ie_1}(v)$, we can verify that
\be
&&E_{e_ke_j}^I(v)=\frac{V_4(v)}{3!}\sum_{l,m,n}\eps_{jklmn}\eps^{I JKL}U_J^{e_l}(v)U_K^{e_m}(v)U_L^{e_n }(v)\nonumber\\
&&V_4(v)U^{e_i}_{[I}(v) U^{e_j}_{J]}(v)=\half\sum_{m,n}\eps^{kijmn}\eps_{IJKL}E^K_{e_m e_k}(v)E^L_{e_n e_k}(v)
\ee
where the last equation is a relation for the area bivector $E_\ell(v)\wedge E_{\ell'}(v)$ of each triangle $f$. For example, given a triangle $f$ shared by $t_{e_4}$ and $t_{e_5}$ in a 4-simplex $\sig_v$. one has
\be
*\!\lt[E_{e_1 e_2}(v)\wedge E_{e_2 e_3}(v)\rt]=V_4(v)\lt[U^{e_4}(v)\wedge U^{e_5}(v)\rt]\label{EU}
\ee
where $*[E_1\wedge E_2]\equiv\eps_{IJKL}E^K_1E^L_2$.

\subsection{Discrete Boundary Geometry}

\begin{figure}[h]
\begin{center}
\includegraphics[width=7cm]{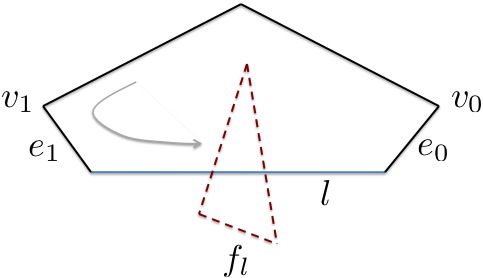}
\caption{The face dual to a boundary triangle $f_l$ shared by two tetrahedra $t_{e_0},t_{e_1}$.}
\label{boundary}
\end{center}
\end{figure}

All the above discussions are considering the discrete geometry in the bulk of the triangulation, where all the co-frame vectors $E_\ell(v)$ and frame vectors $U_e(v)$ are located at internal vertices $v$. Now we consider a triangulation with boundary, where the boundary is a simplical complex $\partial\ck$ built by tetrahedra triangulating a boundary 3-manifold. On the boundary $\partial\ck$, each triangle is shared by precisely two boundary tetrahedra. This triangle is dual to a unique boundary link $l$, connecting the centers of the two boundary tetrahedra sharing the triangle. We denote this triangle $f_l$. On the other hand, from the viewpoint of the whole triangulation $\ck$, there is a unique face dual to the triangle $f_l$, where two edges $e_0,e_1$ of this dual face are dual to the two boundary tetrahedra $t_{e_0},t_{e_1}$ sharing $f_l$. This dual face intersects the boundary uniquely by the link $l$\footnote{If the dual face intersects the boundary by more than one link, then it means that the triangle $f_l$ is shared by more than two tetrahedra, which is impossible for a 3-dimensional triangulation.}. Thus we denote this dual face also by $f_l$ because of the one-to-one correspondence of the duality for $\ck$. See FIG.\ref{boundary} for an example of a face dual to a boundary triangle.

The end-points $s(l),t(l)$ of the boundary link $l$ are centers of the tetrahedra $t_{e_0},t_{e_1}$ respectively. For each edge $\ell$ of the tetrahedron $t_{e_i}$ ($i=0,1$), we associate a spatial vector $E_\ell(e_i)$ at the center of $t_{e_i}$, satisfying the following requirement:

\begin{itemize}
\item Given the time-like unit vector $u^I=(1,0,0,0)$, all the vectors $E_\ell(e_i)$ ($i=0,1$) are orthogonal to $u^I$, i.e.
\be
u_IE^I_\ell(e_i)=0\ \ \ \ \forall\ \ell\in t_{e_i}.\label{uE}
\ee

\item If we reverse the orientation of $\ell$, then
\be
E_{-\ell}(e_i)=-E_\ell(e_i)\ \ \ \ \forall\ \ell\in t_{e_i}.\label{EB}
\ee

\item For all triangle $f$ of the boundary tetrahedron $t_{e_i}$ with edge $\ell_1,\ell_2,\ell_3$, the vectors $E_\ell(e_i)$ close, i.e.
\be
E_{\ell_1}(e_i)+E_{\ell_2}(e_i)+E_{\ell_3}(e_i)=0.\label{edgecloseB}
\ee

\item There is a internal vertex $v_i$ as one of the end-points of the dual edge $e_i$ ($i=0,1$), i.e. the boundary tetrahedron $t_{e_i}$ belongs to the boundary of the 4-simplex $\sig_{v_i}$. Then we require that
\be
\eta_{IJ}E_{\ell_1}^I(e_i)E_{\ell_2}^J(e_i)=\eta_{IJ}E_{\ell_1}^I(v_i)E_{\ell_2}^J(v_i)\ \ \ \ \forall\ \ell_1,\ell_2\in t_{e_i}.\label{EEB}
\ee

\end{itemize}
The set of $E_\ell^I(e_i)$ ($i=0,1$) at the center of $t_{e_i}$ satisfying the above requirements is called a boundary (3-dimensional) co-frame at the center of $t_{e_i}$ (at the node $s(l)$). The discrete metric
\be
g_{\ell_1\ell_2}(e_i):=\eta_{IJ}E_{\ell_1}^I(e_i)E_{\ell_2}^J(e_i)
\ee
is the induced metric on the boundary $\partial\ck$.

Consider a boundary tetrahedron $t_{e_i}$ belonging to a 4-simplex $\sig_{v_i}$, then the edge $e_i$ dual to $t_{e_i}$ connects to a boundary node (the center of $t_{e_i}$). We choose 3 linearly independent co-frame vectors $E_{\ell_1}(e_i),E_{\ell_2}(e_i),E_{\ell_3}(e_i)$ at the center of $t_{e_i}$ associated with 3 edges $\ell_1,\ell_2,\ell_3$, and also choose 3 linearly independent co-frame vectors $E_{\ell_1}(v_i),E_{\ell_2}(v_i),E_{\ell_3}(v_i)$ at the vertex $v_i$ associated with the same set of edges. Given a unit vector $\hat{U}(v_i)$ orthogonal to $E_{\ell_1}(v_i),E_{\ell_2}(v_i),E_{\ell_3}(v_i)$ such that
\be
\mathrm{sgn}\det\lt(E_{\ell_1}(v_i),E_{\ell_2}(v_i),E_{\ell_3}(v_i),\hat{U}(v_i)\rt)=\mathrm{sgn}\det\Big(E_{\ell_1}(e_i),E_{\ell_2}(e_i),E_{\ell_3}(e_i),u\Big)\label{sgndetB}
\ee
by the requirement Eq.\Ref{EEB}, there exist a unique SO(1,3) matrix $\O_{e_i}$ such that
\be
(\O_{e_i})^{I}_{\ J}E^J_{\ell_j}({e_i})=E^I_{\ell_j}(v_i)\ \ \ \ \ (\O_{e_i})^{I}_{\ J}u^J=\hat{U}^I(v_i).\label{OmegaB}
\ee
Thus $\O_{e_i}$ is identify as the spin connection compatible with $E_\ell(v_i),E_\ell(e_i)$.

Consider a dual face bounded by a boundary link $l$ (see, e.g. FIG.\ref{boundary}), by using the defining requirement of the co-frames in the bulk and on the boundary, i.e. Eqs.\Ref{EE} and \Ref{EEB}, we have
\be
\eta_{IJ}E_{\ell_j}^I(e_0)E_{\ell_k}^J(e_0)=\eta_{IJ}E_{\ell_j}^I(e_1)E_{\ell_k}^J(e_1)
\ee
where $\ell_j,\ell_k$ are two of the three edges of the triangle $f_l$ dual to the face. Therefore we obtain the \emph{shape-matching} condition between the triangle geometries of $f_l$ viewed in the frame of $t_{e_0}$ and $t_{e_1}$. More precisely, there exists an SO(3) matrix $\hat{g}_l$ such that for all the three $\ell$'s forming the boundary of the triangle $f_l$
\be
(\hat{g}_l)^I_{\ J} E_{\ell}^J(e_0)=E_{\ell}^I(e_1)
\ee
by the fact that both $E_{\ell}(e_0)$ and $E_{\ell}(e_1)$ are orthogonal to $u^I=(1,0,0,0)$.

Now we consider a single boundary tetrahedron $t_e$ dual to an edge $e$ connecting to the boundary. Since all the boundary co-frame vectors $E_\ell(e)$ at the center of $t_e$ are orthogonal to the time-like unit vector $u^I=(1,0,0,0)$, we now only consider the 3-dimensional spatial subspace orthogonal to $u^I=(1,0,0,0)$. We further assume the boundary tetrahedral geometry is nondegenerate, i.e. the (oriented) 3-volume of the tetrahedron
\be
V_3(e)=\det\Big(E_{\ell_1}(e),E_{\ell_2}(e),E_{\ell_3}(e)\Big)
\ee
is nonvanishing, where $\ell_1,\ell_2,\ell_3$ are the three edges of $t_e$ connecting to a vertex $p$ of $t_e$. Since there are 4 vertices of $t_e$ and an edge $\ell$ is determined by its end-points $p_i,p_j$, we denote $E_{\ell}(e)$ by $E_{p_i p_j}(e)$. Choose a vertex $p_1$ and construct the nondegenerate $3\times 3$ matrix
\be
\Big(E_{p_2p_1 }(e),E_{p_3p_1}(e),E_{p_4p_1}(e)\Big)
\ee
we construct is inverse
\be
\Big(n_{p_2}(e),n_{p_3}(e),n_{p_4}(e)\Big)^t
\ee
with $n_{p_i}(e)\cdot E_{p_j p_1}(e)=\delta_{ij}$. Repeat the same construction for all the other 3 verices $p_2,p_3,p_4$, we obtain four 3-vector $n_{p_i}(e)$ such that
\be
n_{p_i}(e)\cdot E_{p_j p_k}(e)=\delta_{ij}-\delta_{ik}.
\ee
From this relation, one can verify that: (i) The 3-vector $n_{p_i}(e)$ is orthogonal to the triangle $(p_{j},p_k,p_l)$ spanned by $E_{p_j p_k}(e),E_{p_j p_l}(e),E_{p_l p_k}(e)$ with $i\neq j,k,l$. Therefore we denote $n_{p}(e)$ by $n_{ef}$ where $f$ is the triangle determined by the 3 vertices other than $p$. (ii) the four $n_{ef}$ satisfy the closure condition
\be
\sum_{f=1}^4 n_{ef}=0.
\ee
We call the set of $n_{ef}$ a 3-dimensional frame at the center of $t_e$. Explicitly, the vector $n_{ef}$ is given by
\be
n_{ef}=V_3(e)^{-1}E_{\ell_1}(e)\times E_{\ell_2}(e)\ \ \ \ \text{or}\ \ \ \ n_{p_1}(e)=V_3(e)^{-1}E_{p_2p_3}(e)\times E_{p_3p_4}(e)
\ee
The norm $|n_{ef}|=2A_f/|V_3(e)|$ is proportional to the area of the triangle $A_f=\half\lt|E_{\ell_1}(e)\times E_{\ell_2}(e)\rt|$.

\section{Geometric Interpretation of Nondegenerate Critical Configuration}\label{solution}

\subsection{Classical Geometry from Spinfoam Critical Configuration}

Now we come back to the discussion of the critical point of spinfoam amplitude. The purpose of this section is to make a relation between the solution of the critical point equations Eqs.\Ref{gluingJ}, \Ref{gluing}, and \Ref{closure} and a (Lorentzian) discrete geometry described in Section \ref{geometry}.

Given a spinfoam configuration $(j_f,g_{ev},\xi_{ef},z_{vf})$ that solves the critical point equations, let's recall Proposition \ref{fromcritical} and consider a triangle $f$ shared by two tetrahedra $t_e$ and $t_{e'}$ of a 4-simplex $\sig_v$. In Eq.\Ref{summarize}, there are the simplicity conditions $N^e_I(v) *\!X^{IJ}_{ef}(v)=0$ and $N^{e'}_I(v) *\!X^{IJ}_{e'f}(v)=0$ from the viewpoint of the two tetrahedra $t_e$ and $t_{e'}$. The two simplicity conditions implie that there exists two 4-vectors $M_{ef}^I(v)$ and $M_{ef}^I(v)$ such that $X_{ef}(v)=N_e(v)\wedge M_{ef}(v)$ and $X_{e'f}(v)=N_{e'}(v)\wedge M_{e'f}(v)$. However we have in Eq.\Ref{summarize} the gluing condition $X_{ef}(v)=X_{e'f}(v)=X_{f}(v)$, which implies that $N_{e'}(v)$ belongs to the plane spanned by $N_e(v), M_{ef}(v)$, i.e. $N_{e'}(v)=a_{ef}M_{ef}(v)+b_{ef}N_e(v)$. If we assume the following nondegeneracy condition\footnote{Note that the nondegenerate here is purely a condition for the group variables $g_{ve}$ since $N_{e}(v)=g_{ve}(1,0,0,0)^t$.}:
\be
\prod_{e_1,e_2,e_3,e_4=1}^5\det\Big(N_{e_1}(v),N_{e_2}(v),N_{e_3}(v),N_{e_4}(v)\Big)\neq0\label{proddet}
\ee
then $N_e(v),N_{e'}(v)$ cannot be parallel with each other, for all pairs of $e,e'$, which excludes the case of vanishing $a_{ef}$ in the above. Denoting $\a_{ee'}=a_{ef}^{-1}$, we obtain that $M_{ef}(v)=\a_{ee'}N_{e'}(v)-\a_{ee'}b_{ef}N_e(v)$. Therefore
\be
X_f(v)=\a_{ee'}(v)\lt[N_{e}(v)\wedge N_{e'}(v)\rt]\label{NN}
\ee
for all $f$ shared by $t_e$ and $t_{e'}$. Note that within a simplex $\sig_v$ there is a one-to-one correspondence between a pair of tetrahedra $t_e$ and $t_{e'}$ and a triangle $f$ shared by them. Thus we can write the bivector $X_{f}(v)\equiv X_{ee'}(v)=\a_{ee'}(v)\lt[N_{e}(v)\wedge N_{e'}(v)\rt]$.

We label the 5 tetrahedra of $\sig_v$ by $t_{e_i}$, $i=1,\cdots,5$. Then Eq.\Ref{NN} reads
\be
X_{e_i e_j}(v)=\a_{ij}(v)\lt[N_{e_i}(v)\wedge N_{e_j}(v)\rt]
\ee
Then the closure condition $\sum_{j=1}^4\eps_{e_i e_j}(v)X_{e_i e_j}(v)=0$\footnote{Here $\eps_{e_ie_j}(v)=-\eps_{e_je_i}(v)$ and $X_{e_i e_j}(v)=X_{e_j e_i}(v)$.} gives that $\forall\ i=1,\cdots,5$
\be
0=\sum_{j=1}^4\eps_{e_i e_j}(v)\a_{ij}(v)\lt[N_{e_i}(v)\wedge N_{e_j}(v)\rt]=N_{e_i}(v)\wedge \sum_{j=1}^4\eps_{e_i e_j}(v)\a_{ij}(v)N_{e_j}(v)
\ee
which implies that for a choice of diagonal element $\b_{ii}(v)$,
\be
\sum_{j=1}^5\b_{ij}(v)N_{e_j}(v)=0\label{sumN}
\ee
where we denote $\b_{ij}(v):=\eps_{e_i e_j}(v)\a_{ij}(v)$. Here $\b_{ii}(v)$ must be chosen as nonzero, because if $\b_{ii}(v)=0$, Eq.\Ref{sumN} would reduce to $\sum_{j\neq i}\b_{ij}(v)N_{e_j}(v)=0$, which gives all the coefficients $\b_{ij}(v)=0$ by linearly independence of any four $N_e(v)$ (from the nondegeneracy Eq.\Ref{proddet}).

We consider
\be
0 =\b_{km}(v)\sum_{j=1}^5\b_{lj}(v)N_{e_j}(v)-\b_{lm}(v)\sum_{j=1}^5\b_{kj}(v)N_{e_j}(v)=\sum_{j\neq m}\Big[\b_{km}(v)\b_{lj}(v)-\b_{lm}(v)\b_{kj}(v)\Big]N_{e_j}(v)
\ee
Since we assume the nondegeneray condition Eq.\Ref{proddet}, any four of the five $N_{e}(v)$ are linearly independent. Thus
\be
\b_{km}(v)\b_{lj}(v)=\b_{lm}(v)\b_{kj}(v)
\ee
Let us pick one $j_0$ for each 4-simplex, and ask $l=j=j_0$ we obtain
\be
\b_{km}(v)=\frac{\b_{kj_0}(v)\b_{mj_0}(v)}{\b_{j_0j_0}(v)}.
\ee
Therefore we have the factorization of $\b_{ij}(v)$
\be
\b_{ij}(v)=\mathrm{sgn}(\b_{j_0j_0}(v))\b_i(v)\b_j(v)
\ee
where $\b_j(v)={\b_{jj_0}(v)}\big/\sqrt{\lt|\b_{j_0j_0}(v)\rt|}$. We denote $\mathrm{sgn}(\b_{j_0j_0}(v))=\tilde{\eps}(v)$ which is a constant within a 4-simplex $\sig_v$. Thus we have the following expression of the bivector $\eps_{e_i e_j}(v)X_{e_i e_j}(v)$
\be
\eps_{e_i e_j}(v)X_{e_i e_j}(v)=\tilde{\eps}(v)\Big(\b_i(v)N_{e_i}(v)\Big)\wedge \Big(\b_j(v)N_{e_j}(v)\Big)
\ee
The Eq.\Ref{sumN} takes the form
\be
\sum_{j=1}^5\b_{j}(v)N_{e_j}(v)=0.
\ee

Now we construct the frame vectors $U_{e_i}(v)$ for a classical discrete geometry at each vertex $v$\footnote{We denote the dual vector $N^e_I$ by $N^e$ and the vector $N_e^I$ by $N_e$, and the same convention holds for $U_e$ and $U^e$. }:
\be
U^{e_i}_I(v):=\pm\frac{\b_i(v)N^{e_i}_I(v)}{\sqrt{|V_4(v)|}}\ \ \ \ \text{with}\ \ \ \ V_4(v):=\det\Big(\b_2(v)N^{e_2}(v),\b_3(v)N^{e_2}(v),\b_4(v)N^{e_2}(v),\b_5(v)N^{e_2}(v)\Big)\label{U}
\ee
where $U^I_{e_i}(v)$ are time-like 4-vectors by Eq.(\ref{Ne}), and any four of the five frame vectors  $U_{e_i}(v)$ span a 4-dimensional vector space by the assumption of nondegeneracy. Moreover the frame vectors satisfy the closure condition
\be
\sum_{j=1}^5U_{e_j}(v)=0
\ee
and
\be
\frac{1}{V_4(v)}=\det\Big(U^{e_2}(v),U^{e_3}(v), U^{e_4}(v),U^{e_5}(v)\Big)
\ee
and
\be
\eps_{e_i e_j}(v)X^{e_i e_j}_{IJ}(v)=\tilde{\eps}(v)|V_4(v)|\lt[ U_{e_i}(v)\wedge U_{e_j}(v)\rt]_{IJ}=\eps(v)V_4(v)\lt[ U_{e_i}(v)\wedge U_{e_j}(v)\rt]_{IJ}.
\ee
where $\eps(v)=\tilde{\eps}(v)\mathrm{sgn}(V_4(v))$. We emphasize that these frame vectors $U_e(v)$ are constructed from spinfoam configuration $(j_f,g_{ve},\xi_{ef},z_{vf})$ that solves the critical point equations. Note that the oriented 4-volume $V_4(v)$ in general can be either positive or negative for different 4-simplices. However for a nondegenerate critical configuration $(j_f,g_{ve},\xi_{ef},z_{vf})$, we can always make a subdivision of the triangulation, such that $\mathrm{sgn}(V_4(v))$ is a constant within each sub-triangulation. 

Fix an edge $e_1$ at the vertex $v$, we construct the inverse of the nondegenerate matrix $\Big(U^{e_2}(v),U^{e_3}(v), U^{e_4}(v),U^{e_5}(v)\Big)^t$,  denoted by $E^I_{e_i e_1}(v)$ such that
\be
U^{e_i}_I (v)E^I_{e_j e_1}(v)=\delta^i_j\ \ \ \ \ i,j=2,3,4,5
\ee
Explicitly, for example
\be
E_{e_2e_1}^{I}(v)= V_4(v)\eps^{IJKL}U^{e_3}_J(v)U^{e_4}_K(v)U^{e_5}_L(v)
\ee
Note that $E^I_{e_i e_1}(v)$ is determined only up to a sign from the data $N_e(v)$ since Eq.\Ref{U}. However if we fix $e_2$ instead of $e_1$, and find the inverse of $\Big(U^{e_1}(v),U^{e_3}(v), U^{e_4}(v),U^{e_5}(v)\Big)^t$, denoted by $E^I_{e_i e_2}(v)$, then
\be
U^{e_i}_I(v) E^I_{e_j e_2}(v)=\delta^i_j\ \ \ \ \ i,j=1,3,4,5
\ee
and
\be
E_{e_1e_2}^{I}(v)= -V_4(v)\eps^{IJKL}U^{e_3}_J(v)U^{e_4}_K(v)U^{e_5}_L(v)
\ee
where the minus sign comes from $V_4(v)$, because from the closure condition $\sum_{j=1}^5U_{e_j}(v)=0$
\be
\det\Big(U^{e_2}(v),U^{e_3}(v), U^{e_4}(v),U^{e_5}(v)\Big)=-\det\Big(U^{e_1}(v),U^{e_3}(v), U^{e_4}(v),U^{e_5}(v)\Big).
\ee
Therefore we find
\be
E_{e_1e_2}^{I}(v)=-E_{e_2e_1}^{I}(v).
\ee
Then we can fix $e_3$,  $e_4$, $e_5$, and do the same manipulation as above, to obtain $E_{e_i e_j}(v)$ $i,j=1,\cdots,5$ such that
\be
U^{e_i}_I(v) E^I_{e_j e_k}(v)=\delta^i_j-\delta^i_k\ \ \ \ \text{and}\ \ \ \ E_{e_i e_j}^{I}(v)=-E_{e_j e_i}^{I}(v)\label{1}
\ee
from which we can see that all $E^I_{e_j e_k}(v)$ are spatial vectors. One can also verify immediately that
\be
U^{e_i}_I(v) \Big(E^I_{e_j e_k}(v)+E^I_{e_k e_l}(v)+E^I_{e_l e_j}(v)\Big)=0\ \ \ \ \forall i=1,\cdots,5
\ee
By the nondegeneracy of $U^{e_i}_I(v) $, one has
\be
E^I_{e_j e_k}(v)+E^I_{e_k e_l}(v)+E^I_{e_l e_j}(v)=0\label{2}
\ee
Comparing Eqs.(\ref{1}) and (\ref{2}) with Eq.(\ref{coframe}), we see that the collection of $E_{ee'}(v)$ at $v$ is a co-frame at the vertex $v$. The bivector $X_{ee'}(v)$ can also be expressed by $E_{ee'}(v)$
\be
\eps_{e_4 e_5}(v)X^{IJ}_{e_4e_5}(v)=\eps(v)*\! \Big[E_{e_1e_2}(v)\wedge E_{e_2 e_3}(v)\Big]^{IJ}
\ee
which will also be denoted by $\eps_{ef}(v)X^{IJ}_{f}(v)=\eps(v)*\! \Big[E_{\ell_1}(v)\wedge E_{\ell_2}(v)\Big]^{IJ}$.

The above work are done essentially with in a single 4-simplex $\sig_v$. Now we consider two neighboring 4-simplices $\sig_v,\sig_{v'}$ while their center $v,v'$ are connected by the dual edge $e$. Since we only consider two simplices, we introduce a short-hand notation:
\be
U_0:=U_e(v) &\ \ \ \ \ & U'_0:=g_{vv'}U_e(v')\nonumber\\
U_i:=U_{e_i}(v) &\ \ \ \ \ & U'_i:=g_{vv'}U_{e'_i}(v')\nonumber\\
E_{ij}:=E_{e_i e_j}(v) &\ \ \ \ \ & E'_{ij}:=g_{vv'}E_{e'_i e'_j}(v')
\ee
where $i,j=1,\cdots,4$ labels the edges connecting to $v$ or $v'$ other than $e$, $E_{ij}$ and $E_{ij}'$ are orthogonal to $U_0$ and $U_0'$ respectively from Eq.\Ref{1}. Here $g_{vv'}=g_{ve}g_{ev'}$ comes from the spinfoam configuration $(j_f,g_{ve},\xi_{ef},z_{vf})$ that solves the critical point equations. From the closure condition of $U_e(v)$ we have
\be
U_0=-\sum_i U_i\ \ \ \ \text{and}\ \ \ \ U'_0=-\sum_i U'_i
\ee
By definition $N_e(v)=g_{ve}u$ and $N_e(v')=g_{v'e}u$ where $u=(1,0,0,0)^t$, thus $N_e(v)=g_{vv'}N_e(v')$ with $e=(v,v')$. Thus from the definition of $U_e(v)$ in Eq.\Ref{U}, we find
\be
\frac{U_0'}{|U_0'|}=\tilde{\eps}\frac{U_0}{|U_0|}\label{U0}
\ee
where $\tilde{\eps}=\pm$. On the other hand, from the parallel transportation relation $X_{f}(v)=g_{vv'}X_f(v')g_{v'v}$ and $\eps_{ef}(v)=-\eps_{ef}(v')$ for $e=(v,v')$, we have
\be
\eps_{0i}X^{0i}_{IJ}=\eps V(U^0\wedge U^i)_{IJ}=-\eps'V'(U'^0\wedge U'^i)_{IJ}\label{X0i}
\ee
where $X_{0i}$ is the bivector corresponds to the dual face $f$ determined by $e,e_i,e_i'$, the sign factor $\eps_{0i}=\eps_{ef}(v)$,  the sign factors $\eps$ and $\eps'$ are short-hand notations of $\eps(v)$ and $\eps(v')$ respectively, and
\be
\frac{1}{V}=\det\lt(U^1,U^2,U^3,U^4\rt)\ \ \ \ \ -\frac{1}{V'}=\det\lt(U'^1,U'^2,U'^3,U'^4\rt)
\ee
Here the minus sign for $1/V'$ is because the compatible orientations of $\sig_v$ and $\sig_{v'}$ are $[p_0,p_1,p_2,p_3,p_4]$ and $-[p_0,p_1,p_2,p_3,p_4]$. Thus we should set $\eps_{01234}(v)=-\eps_{01234}(v')=1$. Eqs.\Ref{U0} and \Ref{X0i} tell us that $U^0_I$ is proportional to $U'^0_I$ and $U'^i_I$ is a linear combination of $U^i_I$ and $U^0_I$. Explicitly
\be
U'^i_I=-\eps\eps'\tilde{\eps}\frac{|U_0|V}{|U_0'|V'}U^i_I+a_iU^0_I\label{Uprime}
\ee
where $a_i$ are the coefficients such that $\sum_iU_i'=-U_0'$. Using this expression of $U'^i$, we have
\be
-\frac{1}{V'}&=&\det\lt(U'^1,U'^2,U'^3,U'^4\rt)\ =\ \det\lt(U'^0,U'^1,U'^2,U'^3\rt)\nonumber\\
&=&\tilde{\eps}\frac{|U_0'|}{|U_0|}\lt(-\eps\eps'\tilde{\eps}\frac{|U_0|V}{|U_0'|V'}\rt)^3\det\lt(U^0,U^1,U^2,U^3\rt)\ =\ -\eps\eps'\lt(\frac{|U_0|V}{|U_0'|V'}\rt)^2\frac{1}{V'}\label{epseps}
\ee
which results in $\eps=\eps'$. Therefore $\eps(v)=\eps(v')=\eps$ is a global sign on the entire triangulation. Now for the bivectors $X_{0i}(v)$ and $X_{0i}(v')$ ($X^{ji}(v)=X^{ij}(v)$ and $\eps_{ij}(v)=-\eps_{ji}(v)$)
\be
\eps_{0i}(v)X^{0i}_{IJ}(v)&=&\eps\half\sum_{m,n}\eps^{k0imn}(v)\eps_{IJKL}E_{mk}^K(v)E_{nk}^L(v)\nonumber\\
\eps_{0i}(v')X^{0i}_{IJ}(v')&=&\eps\half\sum_{m,n}\eps^{k0imn}(v')\eps_{IJKL}E_{mk}^K(v')E_{nk}^L(v')
\ee
Since $\eps_{0i}(v)=-\eps_{0i}(v')$ and $\eps^{kijmn}(v)=-\eps^{kijmn}(v')$, we can set $\eps_{0i}(v)\eps^{k0imn}(v)=\eps_{0i}(v')\eps^{k0imn}(v')=\eps_{0i}\eps^{k0imn}$. Therefore
\be
\eps_{0i}X^{0i}_{IJ}(v)&=&\eps\half\sum_{m,n}\eps^{k0imn}\eps_{IJKL}E_{mk}^K(v)E_{nk}^L(v)\nonumber\\
\eps_{0i}X^{0i}_{IJ}(v')&=&\eps\half\sum_{m,n}\eps^{k0imn}\eps_{IJKL}E_{mk}^K(v')E_{nk}^L(v')
\ee
Given a triangle $f$, we can choose $E_{\ell_1}(v),E_{\ell_2}(v)$ (e.g. $\ell_1=(p_m,p_k)$ and $\ell_2=(p_n,p_k)$ with $\eps_{0i}=1$ and $\eps^{k0imn}=1$) such that
\be
X^{IJ}_{f}(v)=\eps\ *\! \Big[E_{\ell_1}(v)\wedge E_{\ell_2}(v)\Big]^{IJ}\ \ \ \ \text{and}\ \ \ \ X^{IJ}_{f}(v')=\eps\ *\! \Big[E_{\ell_1}(v')\wedge E_{\ell_2}(v')\Big]^{IJ}
\ee
On the other hand, Eq.\Ref{epseps} also implies that ${|U_0|V}=\pm{|U_0'|V'}$. Thus we define a sign factor $\mu:=-\tilde{\eps}{|U_0|V}\big/{|U_0'|V'}=\pm1$ such that from Eq.\Ref{Uprime}
\be
U'^i_I=\mu U^i_I+a_iU^0_I\ \ \ \ \ \mu=-\tilde{\eps}\ \mathrm{sgn}(VV')
\ee
Therefore we obtain the relation between $E_{ij}$ and $E_{ij}'$ (using $\eps_{jklm0}(v')=-\eps_{jklm0}(v)$)
\be
E'{}_{jk}^I&=&{V'}\eps_{jklm}(v')\eps^{IJKL}\ U'{}^l_J\ U'{}^m_K\ U'{}^0_L
=-\tilde{\eps}\frac{|U'_0|}{|U_0|}\mu^2{V'}\eps_{jklm}(v)\eps^{IJKL}\ U{}^l_J\ U{}^m_K\ U{}^0_L
=\mu^3{V}\eps_{jklm}(v)\eps^{IJKL}\ U{}^l_J\ U{}^m_K\ U{}^0_L\nonumber\\
&=&\mu E_{jk}^I
\ee
which means that for all tetrahedron edge $\ell$ of the tetrahedron $t_e$ dual to $e=(v,v')$, the co-frame vectors $E_\ell(v)$ and $E_{\ell}(v')$ at neighboring vertices $v$ and $v'$ are related by parallel transportation up to a sign $\mu_e$, i.e.
\be
\mu_e E_\ell(v)= g_{vv'}E_\ell(v')\ \ \ \ \forall\ \ell\subset t_e
\ee
This relation shows that the vectors $E_\ell(v)$ (constructed from spinfoam critical point configuration) satisfy the metricity condition Eq.\Ref{EE}. Therefore the collection of co-frame vectors $E_\ell(v)$ at different vertices consistently forms a discrete co-frame of the whole triangulation. At the critical configuration, we define an SO(1,3) matrix $\O_{vv'}$ relating $g_{vv'}$ (in the Spin-1 representation) by the sign $\mu_e$, i.e.
\be
g_{vv'}=\mu_e\O_{vv'}
\ee
By Lemma \ref{connection} and Definition \ref{connection1}, the SO(1,3) matrix $\O_{vv'}$ is a discrete spin connection compatible with the co-frame if $\mathrm{sgn}(V_4(v))=\mathrm{sgn}(V_4(v'))$. 

If $\mathrm{sgn}(V_4(v))=\mathrm{sgn}(V_4(v'))$, $\mu_e=-\tilde{\eps}\ \mathrm{sgn}(V_4(v)V_4(v'))=-\tilde{\eps}$. Thus from Eq.\Ref{U0},
\be
\frac{U_0'}{|U_0'|}=-\mu_e\frac{U_0}{|U_0|}
\ee
the tetrahedron normal $U_{e}(v)/|U_{e}(v)|$ is always opposite to $\O_eU_{e}(v')/|U_{e}(v')|$ when $\mathrm{sgn}(V_4(v))=\mathrm{sgn}(V_4(v'))$.

Since in Spin-1 representation $g_{vv}\in \text{SO}^+(1,3)$ and $\O\in \text{SO}(1,3)$, $\mu_e=-1$ corresponds the case that $\O_{vv'}\in \text{SO}^-(1,3)$. It means that in the case of $\mu_e=-1$ if we choose the unit vectors $\hat{U}(v),\hat{U}(v')$ orthogonal to $E_\ell(v),E_\ell(v)$ ($\ell\subset t_e$) such that
\be
\mathrm{sgn}\det\lt(E_{\ell_1}(v),E_{\ell_2}(v),E_{\ell_3}(v),\hat{U}(v)\rt)=\mathrm{sgn}\det\lt(E_{\ell_1}(v'),E_{\ell_2}(v'),E_{\ell_3}(v'),\hat{U}(v')\rt)
\ee
then one of $\hat{U}(v),\hat{U}(v')$ is future-pointing and the other is past-pointing.


\subsection{Boundary Data for Spinfoam Critical Configuration}

Given a spinfoam configuration $(j_f,g_{ve},\xi_{ef},z_{vf})$ that solves critical point equations. The boundary data of the spinfoam amplitude is given by the boundary spins and the normalized spinors $(j_f,\xi_{ef})$ for the boundary triangles $f$. Eq.\Ref{Xef0} naturally associates a bivector $X_{ef}$ to each pair $(j_f,\xi_{ef})$ for each $(e,f)$. From Eq.\Ref{Xef1},
\be
X_{ef}^{IJ}=2\g j_f\lt[\hat{n}_{ef}\wedge u\rt]
\ee
The spatial 3-vectors $j_f\hat{n}_{ef}$ satisfy the critical point equation Eq.\Ref{closure}
\be
\sum_{f}\eps_{ef}j_f\hat{n}_{ef}=0
\ee
where $v$ is the vertex connecting to the edge $e$. We define $V_3(e)$ such that
\be
\det\Big(\eps_{ef_2}j_{f_2}\hat{n}_{ef_2},\eps_{ef_3}j_{f_3}\hat{n}_{ef_3}, \eps_{ef_4}j_{f_4}\hat{n}_{ef_4}\Big)=\mathrm{sgn}(V_3(e))\lt|V_3(e)\rt|^2\label{jnV}
\ee
We rescale each vector $\eps_{ef}j_f\hat{n}_{ef}$ by 
\be
n_{ef}:=\frac{\eps_{ef}\g j_f\hat{n}_{ef}}{\lt|V_3(e)\rt|}\ \ \ \ \text{then}\ \ \ \ \sum_f n_{ef}=0\ \ \ \ \text{and}\ \ \ \ \det\Big(n_{ef_2},n_{ef_3},n_{ef_4}\Big)=\frac{1}{V_3(e)}.
\ee

We assume the nondegeneracy of the boundary data, i.e. any three of the four vectors $n_{ef}$ span the 3-dimensional spatial subspace, in another word, the following product of determinants is nonvanishing
\be
\prod_{f_1,f_2,f_3=1}^4\det\lt(n_{ef_1},n_{ef_2},n_{ef_3}\rt)\neq0.\label{nondeg3}
\ee

The nondegeneracy of the tetrahedron Eq.\Ref{nondeg3} is implied by the nondegeneracy condition in the bulk Eq.\Ref{proddet}. The reason is the following: By the parallel transportation relation $X_f(v)=g_{ve}X_{ef}g_{ev}$ and $X_{ef}=2\g j_f\hat{n}_{ef}\wedge u$, the bivector $X_f(v)$ is then given by $X_f(v)=V_{ef}(v)\wedge N_e(v)$, where $N_e(v)=g_{ve}u$ and $V_{ef}(v):=2\g j_f g_{ve}\hat{n}_{ef}$ is orthogonal to $N_e(v)$. For $f$ the triangle shared by $t_e$ and $t_{e_i}$ ($i=1,\cdots,4$), we know that $X_f(v)=\a_{e_ie}(v)N_{e_i}(v)\wedge N_e(v)$. Therefore the vector $V_{ef}(v)$ is a linear combination of $N_{e_i}$ and $N_e$. The nondegeneracy condition Eq.\Ref{proddet} in 4-dimensions implies the 4 unit vectors, say $N_e$ and any 3 out of 4 vectors $N_{e_i}$, are linear independent and span a 4-dimensional vector space. Thus any 3 out of the 4 vectors $V_{ef}(v)$ must be linear independent and span a 3-dimensional subspace orthogonal to $N_e(v)$. Then Eq.\Ref{nondeg3} is a result from parallel transporting $V_{ef}(v)$ back to the center of $t_e$.

We now denote $n_{ef}\equiv n_{p_1}(e)$, where the triangle $f$ is determined by $(p_2,p_3,p_4)$. Now we construct the spatial 3-vectors $E_{p_1p_2}(e)$, such that the matrix $\Big(E_{p_2p_1}(e),E_{p_3p_1}(e),E_{p_4p_1}(e)\Big)$ is the inverse of $\Big(n_{p_2}(e),n_{p_3}(e),n_{p_4}(e)\Big)^t$. Therefore we have
\be
n_{p_i}(e)\cdot E_{p_jp_k}(e)=\delta_{ij}-\delta_{ik}\label{nEB}
\ee
The 3-vectors $E_{p_ip_j}(e)$ are associated to the edges $\ell=(p_i,p_j)$ of the tetrahedron $t_e$, so it can be denoted by $E_\ell(e)$. Note that $E_\ell(e)$ is determined up to an overall rescaling, since the set of $n_{ef}$ is defined up to an overall scaling $\a\in\mathbb{R}$. In the following we are going to show that the vectors $E_\ell(e)$ are co-frame vectors on the boundary.

First of all, Eqs.\Ref{uE}, \Ref{EB} and \Ref{EEB} can be verified immediately from Eq.\Ref{nEB}. Since $\Big(E_{p_2p_1}(e),E_{p_3p_1}(e),E_{p_4p_1}(e)\Big)$ is the inverse of $\Big(n_{p_2}(e),n_{p_3}(e),n_{p_4}(e)\Big)^t$, we have
\be
\det\Big(E_{p_2p_1}(e),E_{p_3p_1}(e),E_{p_4p_1}(e)\Big)=V_3(e)
\ee
we also have
\be
\eps_{ef}\g j_f\hat{n}_{p_j}(e)=|V_3(e)|n_{p_j}(e)={\eps}(e)V_3(e) n_{p_j}(e)={\eps}(e)\half\sum_{k,l}\eps_{ijkl}E_{p_kp_i}(e)\times E_{p_lp_i}(e)
\ee
where we have define a sign factor $\eps(e)=\mathrm{sgn}(V_3(e))$. Equivalently for the bivector $X_{ef}$, there exists $E_{\ell_1}(e),E_{\ell_2}(e)$ such that
\be
X_{ef}^{IJ}=2\g j_f\lt[\hat{n}_{ef}\wedge u\rt]^{IJ}=\eps(e)*\!\Big[  E_{\ell_1}(e)\wedge E_{\ell_2}(e)\Big]^{IJ}.
\ee

Consider a internal vertex $v$ which connected by the edge $e$, we introduce the short-hand notation:
\be
E_{ij}:=E_{p_ip_j}(e)\ \ \ \ \ E_{ij}':=g_{ev}E_{p_ip_j}(v)\ \ \ \ \eps(e):={\eps}'\ \ \ \ \hat{n}_{j}:=\hat{n}_{p_j}(e)
\ee
Since $X_f(v)=g_{ve}X_{ef}g_{ev}$, for each triangle determined by $(p_i,p_j,p_k)$
\be
{\eps}'\half\sum_{k,l}\eps_{ijkl}*\!E_{ki}\wedge E_{li}=\eps\half\sum_{k,l}\eps_{ijkl}*\! E'_{ki}\wedge E'_{li}=2\eps_{ef}\g j_f\lt[\hat{n}_{j}\wedge u\rt]\label{EEE'E'}
\ee
We also have $N_e(v)=g_{ve}u$. So $E'_{ij}$ is orthogonal to $u^I=(1,0,0,0)$ since $E_{\ell}(v)$ ($\ell\subset t_e$) orthogonal to $N_e(v)$. Thus
\be
{\eps}'V n_{j}=2\eps_{ef}\g j_f\hat{n}_{j}=\eps\half\sum_{k,l}\eps_{ijkl} E'_{ki}\times E'_{li}
\ee
which implies that the $3\times 3$ matrix given by $E_{ki}'$ (with $i$ fixed) is the inverse of the matrix given by $n_{j}$, $j\neq i$, up to an overall constant, i.e.
\be
n_{i}\cdot E'_{jk}=\eps{\eps}'\frac{V'_3}{V_3}(\delta_{ij}-\delta_{ik})\label{nE'B}
\ee
we have used the short-hand notation
\be
V_3=V_3(e)=\det\Big(E_{21}(e),E_{31}(e),E_{41}(e)\Big)& \ \ \ \ &V'_3=V'_3(e)=\det\Big(E'_{21}(e),E'_{31}(e),E'_{41}(e)\Big)
\ee
Comparing Eq.\Ref{nE'B} and Eq.\Ref{nEB} we determine that $E_{jk}$ is proportional to $E'_{jk}$:
\be
E'_{jk}=\eps{\eps}'\frac{V'_3}{V_3}E_{jk}.
\ee
since the matrix given by $n_i$ has unique inverse. Insert this relation back into Eq.\Ref{EEE'E'}, we obtain that
\be
\eps\lt(\frac{V'_3}{V_3}\rt)^2={\eps}'
\ee
which tell us that
\be
\eps'={\eps}\ \ \ \text{and}\ \ \ \ \lt|\frac{V'_3}{V_3}\rt|=1
\ee
As a result we find the relations
\be
X_{ef}^{IJ}=\eps *\!\Big[  E_{\ell_1}(e)\wedge E_{\ell_2}(e)\Big]^{IJ}\ \ \ \ \text{and}\ \ \ \
\mu_e E_\ell(e)=g_{ev}E_\ell(v)\ \ \ \ \forall\ \ell\subset t_e\label{XEE}
\ee
where $\eps=\pm1$ is the global sign factor of the whole triangulation, and $\mu_e=\mathrm{sgn}(V_3)\mathrm{sgn}(V'_3)=\pm1$. From the second relation above, we obtain the metricity condition Eq.\Ref{EEB}. Therefore we confirm that $E_\ell(e)$ is a boundary co-frame constructed from spinfoam critical configuration.
The group element $g_{ev}$ equals to the spin connection $\O_{ev}$ up to a sign, i.e.
\be
g_{ev}=\mu_e\O_{ev}.
\ee

Since $\eps$ is a global sign of the entire triangulation and $\eps=\mathrm{sgn}(V_3(e))$ on the boundary, then prior to the construction, one has to choose a consistent orientation of the boundary triangulation such that $\mathrm{sgn}(V_3(e))=\mathrm{sgn}(V_3(e'))$ for each pair of tetrahedra $t_e,t_{e'}$.

By the following relations (we choose the orientation of the 4-simplex $\sig_v=[p_0,p_1,p_2,p_3,p_4]$):
\be
V_3=\eps_{IJK}E^I_{21}E^J_{31}E^K_{41}\ \ \ \ V'_3=\eps_{IJK}E'^I_{21}E'^J_{31}E'^K_{41}\ \ \ \ (g_{ev}U^0)_I=\frac{-1}{V_4}\eps_{IJKL}E'^I_{21}E'^J_{31}E'^K_{41}
\ee
we obtain that
\be
V'_3=-V_4U^0_I(g_{ve}u)^I=-V_4U^0_IN_0^I,\ \ \ \text{where}\ \ \ u=(1,0,0,0)^t
\ee
Then for an edge $e$ connecting to the boundary 
\be
\mu_e=-\eps\ \mathrm{sgn}(V_4(v))\mathrm{sgn}(U^0_I(v)N_0^I(v))\label{muVUN}
\ee
which implies that if we choose $\eps=\mathrm{sgn}(V_3(e))=+1$ globally on the boundary, and if $V_4(v)>0$, $\mu_e=+1$ when $U_0(v)$ is future-pointing and $\mu_e=-1$ when $U_0(v)$ is past-pointing, while $N_0(v)=g_{ve}u$ is always future-pointing.

\begin{Lemma}\label{prodmue}
Given $f$ either an internal face or a boundary face, the product $\prod_{e\subset\partial f}\mu_e$ doesn't change when $U_e(v)$ flips sign for any 4-simplex $\sig_v$, recall that the five normals $U_e(v)$ at $\sig_v$ are defined up to a overall sign. Therefore the product $\prod_{e\subset\partial f}\mu_e$ is determined by the spinfoam critical configuration.
\end{Lemma}

\startproof For a internal edge $e=(v,v')$, we have
\be
\mu_e=-\tilde{\eps}_e\mathrm{sgn}\Big(V_4(v)V_4(v')\Big)=\mathrm{sgn}\Big(U_e^I(v)(g_{vv'}U_e)^I(v')\Big)\mathrm{sgn}\Big(V_4(v)V_4(v')\Big)
\ee
where we recall that $\tilde{\eps}_e U_e(v)/|U_e(v)|= g_{vv'}U_e(v')/|U_e(v')|$. Combine with Eq.\Ref{muVUN}, it is easy to see that if we flip simultaneously the sign of all the five $U_e(v)$ at any $\sig_v$ $(v\in\partial f)$, the product $\prod_{e\subset\partial f}\mu_e$ doesn't change, for $f$ either an internal face or a boundary face. 
\finishproof

We recall FIG.\ref{boundary}, where the triangle $f_l$ is shared by two boundary tetrahedra $t_{e_0},t_{e_1}$. Because of Eq.\Ref{XEE}, we parallel transport three co-frame vectors $E_\ell(e_0)$ corresponding to the three edges of the triangle $f_l$,
\be
(\prod_{e}\mu_e) E_\ell(e_1)=G_{f_l}(e_1,e_0)E_\ell(e_0)\ \ \ \ \forall\ \ell\subset f_l
\ee
where $G_{f_l}(e_1,e_0):=\overleftarrow{\prod}_{e}g_e$ is a product of the edge holonomy $g_e$ over all the internal edges $e$ of the dual face $f_l$. Therefore the triangle formed by the three $E_\ell(e_0)$ ($\ell\subset f_l$) matches in shape with the triangle formed by $E_\ell(e_1)$ ($\ell\subset f_l$). Since both $E_\ell(e_0)$ and $E_\ell(e_1)$ are orthogonal to the unit time-like vector $u=(1,0,0,0)$. There exists an O(3) matrix $\hat{g}_l$ such that
\be
\hat{g}_l E_{\ell}(e_0)=E_{\ell}(e_1)\ \ \ \ \text{and}\ \ \ \  \hat{g}_l \hat{n}_{e_0f_l}=\hat{n}_{e_1f_l}\label{bc}
\ee
These relations give the restrictions of the boundary data for the spinfoam amplitude. We call the boundary condition given by Eq.\Ref{bc} the (nondegenerate) \emph{Regge boundary condition}. The above analysis shows that the spinfoam boundary data must satisfy the Regge boundary condition in order to have nondegenerate solutions of the critical point equations Eqs.\Ref{gluingJ}, \Ref{gluing}, \Ref{closure}.

\subsection{Summary}

Now we summarize the results in this section as a theorem:

\begin{Theorem}\label{construction} (Construction of Classical Geometry from Spinfoam Critical Configuration)

\begin{itemize}

\item Given the data $(j_f,g_{ev},\xi_{ef},z_{vf})$ be a nondegenerate spinfoam configuration that solves the critical point equations Eqs.\Ref{gluingJ}, \Ref{gluing}, and \Ref{closure}, there exists a discrete classical Lorentzian geometry on $\cm$, represented by a set of spatial co-frame vectors $E_\ell(v)$ satisfying Eqs.\Ref{E}, \Ref{edgeclose} and \Ref{EE} in the bulk, and $E_\ell(e)$ satisfying Eqs.\Ref{uE}, \Ref{EB}, \Ref{edgecloseB} and \Ref{EEB} on the boundary, such that the bivectors $X_f(v)$ and $X_{ef}$ in Proposition \ref{fromcritical} is written by
\be
X_f^{IJ}(v)=\eps\ *\!\Big[E_{\ell_1}(v)\wedge E_{\ell_2}(v)\Big]^{IJ},\ \ \ \ \ X_{ef}^{IJ}=\eps\ *\!\Big[  E_{\ell_1}(e)\wedge E_{\ell_2}(e)\Big]^{IJ}\label{XEE0}
\ee
where $\ell_1,\ell_2$ are edges of the triangle $f$. The above equation is a relation between the spinfoam data $X_f(v),X_{ef}$ and a classical geometric data $E_\ell(v)$. Such a relation is determined up to a global sign $\eps$ on the whole triangulation. Moreover the above co-frame is unique up to inversion $E_\ell\mapsto -E_\ell$ at each $v$ or $t_e$. With the co-frame vectors $E_\ell(v), E_{\ell}(e)$, we can construct a discrete metric $g_{\ell_1\ell_2}(v),g_{\ell_1\ell_2}(e)$ in the bulk and on the boundary
\be
g_{\ell_1\ell_2}(v)=\eta_{IJ}E^I_{\ell_1}(v)E^J_{\ell_2}(v)\ \ \ \ \ g_{\ell_1\ell_2}(e)=\eta_{IJ}E^I_{\ell_1}(e)E^J_{\ell_2}(e).
\ee

\item The norm of the bivector $|X_f(v)|=\lt|E_{\ell_1}(v)\wedge E_{\ell_2}(v)\rt|=2\g j_f$. Thus $\g j_f$ is understood as the area of the triangle $f$\footnote{$\lt|E_{1}\wedge E_{2}\rt|^2=\half(E_1^I E_2^J-E_1^J E_2^I)(E^1_I E^2_J-E^1_J E^2_I)=|E_1|^2|E_2|^2(1-\cos^2\theta)=(2A_f)^2$ where $E_1\cdot E_2=|E_1||E_2|\cos\theta$. $\lt|E_{1}\wedge E_{2}\rt|$ corresponds to the area of a parallelogram (two times the area of the triangle) determined by $E_1$ and $E_2$.}.

\item If the triangulation has boundary, one has to choose a consistent orientation of the boundary triangulation such that $\mathrm{sgn}(V_3(e))=\mathrm{sgn}(V_3(e'))$ for each pair of tetrahedra $t_e,t_{e'}$ (recall Eq.\Ref{jnV}). Then the global sign $\eps$ is specified by the orientation of the boundary, i.e. $\eps=\mathrm{sgn}(V_3(e))$.

\item Equivalently the bivectors in the bulk can be expressed by the frame $U_e(v)$ associated with $E_\ell(v)$
\be
X_{f}^{IJ}(v)=\eps\ V_4(v)\Big[U_{e}(v)\wedge U_{e'}(v)\Big]^{IJ}
\ee
where $e,e'$ are the dual edges of the dual face $f$, and $V_4(v)^{-1}$ is the determinant of the matrix defined by the frame co-vectors $U^{e_i}_I(v)$, $i=2,3,4,5$, i.e.
\be
\frac{1}{V_4(v)}=\det\Big(U^{e_2}(v),U^{e_3}(v), U^{e_4}(v),U^{e_5}(v)\Big).
\ee
For the bivector on the boundary, from Eq.\Ref{Xef1}
\be
X_{ef}^{IJ}=2\g j\lt[\hat{n}_{ef}\wedge u\rt]^{IJ}
\ee
where $u=(1,0,0,0)$ and $j\hat{n}_{ef}$ is the oriented area of the boundary triangle.

\item Given a dual edge $e$, for all tetrahedron edge $\ell$ of the tetrahedron $t_e$ dual to $e=(v,v')$, the associated co-frame vectors $E_\ell(v)$ and $E_{\ell}(v')$ at neighboring vertices $v$ and $v'$ are related by parallel transportation up to a sign $\mu_e$, i.e.
\be
\mu_e E_\ell(v)= g_{vv'}E_\ell(v')\ \ \ \ \forall\ \ell\subset t_e
\ee
If the dual edge $e$ connects the boundary, we have similarly
\be
\mu_e E_\ell(v)= g_{ve}E_\ell(e)\ \ \ \ \forall\ \ell\subset t_e.
\ee
We define the SO(1,3) matrices $\O_{vv'},\O_{ve}$ by
\be
\O_{vv'}=\mu_eg_{vv'}\ \ \ \ \O_{ve}=\mu_eg_{ve}.
\ee
The simplicial complex $\ck$ can be subdivided into sub-complexes $\ck_1,\cdots,\ck_n$ such that (1) each $\ck_i$ is a simplicial complex with boundary, (2) within each sub-complex $\ck_i$, $\mathrm{sgn}(V_4(v))$ is a constant. Then within each sub-complex $\ck_i$, the SO(1,3) matrices $\O_{vv'},\O_{ve}$ are the discrete spin connection compatible with the co-frame $E_\ell(v)$ and $E_{\ell}(v')$.

\item Given the boundary triangles $f$ and boundary tetrahedra $t_e$, in order to have nondegenerate solutions of the critical point equations Eqs.\Ref{gluingJ}, \Ref{gluing}, \Ref{closure}, the spinfoam boundary data $(j_f,\xi_{ef})$ must satisfy the (nondegenerate) Regge boundary condition: (1) For each boundary tetrahedron $t_e$ and its triangles $f$, $(j_f,\xi_{ef})$ determines 4 triangle normals $\hat{n}_{ef}$ that spans a 3-dimensional spatial subspace. (2) Given the tetrahedra $t_{e_0},t_{e_1}$ sharing the triangle $f$, the triangle normals $\hat{n}_{e_0f}$ and $\hat{n}_{e_1f}$ are related by an O(3) matrix $g_{l}$ ($l$ the link dual to $f$ on the boundary)
\be
\hat{g}_l \hat{n}_{e_0f}=\hat{n}_{e_1f}.
\ee
(3) The boundary triangulation is consistently oriented such that the orientation $\mathrm{sgn}(V_3(e))$ (recall Eq.\Ref{jnV}) is a constant on the boundary. If the Regge boundary condition is satisfied, there are nondegenerate solutions of the critical point equations, and the solutions implies the shape-matching of the triangle $f$ shared by the tetrahedra $t_{e_0}$ and $t_{e_1}$. If the Regge boundary condition is not satisfied, there is no nondegenerate critical configuration.

\end{itemize}
\end{Theorem}

\section{Spinfoam Amplitude at Nondegenerate Critical Configuration}

Given a nondegenerate critical configuration $(j_f,g_{ev},\xi_{ef},z_{vf})$, the previous discussions show us that we can construct a discrete classical geometry from the critical configuration. Moreover we can make a subdivision of the triangulation into sub-triangulations $\ck_1,\cdots,\ck_n$, such that (1) each $\ck_i$ is a simplicial complex with boundary, (2) within each sub-complex $\ck_i$, $\mathrm{sgn}(V_4(v))$ is a constant. To study the spinfoam (partial-)amplitude $A_j(\ck)$ at a nondegenerate critical configuration, we only need to study the amplitude $A_j(\ck_i)$ on the sub-triangulation $\ck_i$ where $\mathrm{sgn}(V_4(v))$ is a constant. Then the behavior of $A_j(\ck)$ can be expressed as a product
\be
A_j(\ck)\Big|_\text{critical}=\prod_iA_j(\ck_i)\Big|_\text{critical}
\ee
Therefore in the following analysis of this section we always assume the triangulation has a boundary and $\mathrm{sgn}(V_4)$ is a constant on the triangulation.

\subsection{Internal Faces}

We have shown previously that the action $S$ of the spinfoam amplitude can be written as a sum $S=\sum_f S_f$. We first consider the internal faces whose edges are not contained in the boundary of the triangulation. Each internal ``face action'' $S_f$ evaluated at the critical point defined by Eqs.\Ref{gluingJ}, \Ref{gluing}, and \Ref{closure} takes the form
\be
S_f=2i\g j_f\sum_{v\in\partial f}\ln\frac{||Z_{ve'f}||}{||Z_{vef}||}-2i j_f\sum_{v\in\partial f}\phi_{eve'}=-2i j_f\lt(\g\sum_{v\in\partial f}\theta_{eve'} +\sum_{v\in\partial f}\phi_{eve'}\rt)
\ee
where we have denoted
\be
\frac{||Z_{vef}||}{||Z_{ve'f}||}:=e^{\theta_{eve'}}
\ee
Recall Eqs.\Ref{gluingJ} and \Ref{gluing}, and consider the following successive actions on $\xi_{ef}$ of $g_{e'v}g_{ve}$ around the entire boundary of the face $f$
\be
\overleftarrow{\prod_{v\in\partial f}}g_{e'v}g_{ve}J\xi_{ef}&=&e^{-\sum_{v}\theta_{eve'}-i\sum_v\phi_{eve'}}J\xi_{ef}\nonumber\\
\overleftarrow{\prod_{v\in\partial f}}g_{e'v}g_{ve}\xi_{ef}&=&e^{\sum_{v}\theta_{eve'}+i\sum_v\phi_{eve'}}\xi_{ef}
\ee
Thus $\xi_{ef}$ is a eign-vector of the loop holonomy $\overleftarrow{\prod}_{v\in\partial f}g_{e'v}g_{ve}$. Since $\xi_{ef},J\xi_{ef}$ are normalized spinors and $\lag J\xi_{ef},\xi_{ef}\rag=0$, thus we represent them by
\be
\xi_{ef}=\left(\begin{array}{c}1  \\ 0 \end{array}\right)\ \ \ \ \text{and}\ \ \ \ J\xi_{ef}=\left(\begin{array}{c}0  \\ 1 \end{array}\right)
\ee
We express this loop holonomy by an arbitrary $\Slc$ matrix
\be
G_f(e):=\overleftarrow{\prod_{v\in\partial f}}g_{e'v}g_{ve}=\left(\begin{array}{cc}a & b  \\ c & d\end{array}\right)
\ee
Thus the eigenvalue equations for arbitrary complex number $\a$
\be
\left(\begin{array}{cc}a & b  \\ c & d\end{array}\right)\left(\begin{array}{c}1  \\ 0 \end{array}\right)=e^\a\left(\begin{array}{c}1  \\ 0 \end{array}\right)\ \ \ \ \text{and}\ \ \ \ \left(\begin{array}{cc}a & b  \\ c & d\end{array}\right)\left(\begin{array}{c}0  \\ 1 \end{array}\right)=e^{-\a}\left(\begin{array}{c}0  \\ 1 \end{array}\right)
\ee
implies that
\be
\left(\begin{array}{cc}a & b  \\ c & d\end{array}\right)=\left(\begin{array}{cc}e^\a & 0  \\ 0 & \ e^{-\a}\end{array}\right)=e^{\a\vec{\sig}\cdot\hat{z}}
\ee
By rotating $\hat{z}$ to the unit 3-vector $\hat{n}_{ef}$, we obtain a representation-independent expression of the loop holonomy $G_f(e)$
\be
G_f(e)=\exp\lt[\sum_{v\in\partial f}\lt(\theta_{eve'}+i\phi_{eve'}\rt)\vec{\sig}\cdot \hat{n}_{ef}\rt].
\ee
which is an exponential map from Lie algebra variable\footnote{Note that not all the elements in $\Slc$ can be written in an exponential form, because of the noncompactness.}.

Consider the following identity: for any complex number $\a$ and unit vector $\hat{n}$,
\be
\tr\lt[\half\lt(1+\vec{\sig}\cdot\hat{n}\rt)e^{\a\vec{\sig}\cdot \hat{n}}\rt]=e^\a \label{lntr}
\ee
which can be proved by the identities of Pauli matrices: $(\vec{\sig}\cdot\hat{n})^{2k}=1_{2\times 2}$ and $(\vec{\sig}\cdot\hat{n})^{2k+1}=\vec{\sig}\cdot\hat{n}$. Using this identity, we have
\be
\ln\tr\lt[\half\lt(1+\vec{\sig}\cdot\hat{n}_{ef}\rt)G_f(e)\rt]&=&\sum_{v\in\partial f}\theta_{eve'}+i\sum_{v\in\partial f}\phi_{eve'}\nonumber\\
\ln\tr\lt[\half\lt(1+\vec{\sig}\cdot\hat{n}_{ef}\rt)G_f^\dagger(e)\rt]&=&\sum_{v\in\partial f}\theta_{eve'}-i\sum_{v\in\partial f}\phi_{eve'}
\ee
where we use the fact that $\vec{\sig}$ are Hermitian matrices. Insert these into the expression of the face action $S_f$
\be
S_f
&=&-i j_f\g\lt\{\ln\tr\lt[\half\lt(1+\vec{\sig}\cdot\hat{n}_{ef}\rt)G_f(e)\rt]+\ln\tr\lt[\half\lt(1+\vec{\sig}\cdot\hat{n}_{ef}\rt)G^\dagger_f(e)\rt]\rt\} \nonumber\\
&&\ -j_f\lt\{\ln\tr\lt[\half\lt(1+\vec{\sig}\cdot\hat{n}_{ef}\rt)G_f(e)\rt]-\ln\tr\lt[\half\lt(1+\vec{\sig}\cdot\hat{n}_{ef}\rt)G^\dagger_f(e)\rt]\rt\}\nonumber\\
&=&-(i\g+1)j_f\ln\tr\lt[\half\lt(1+\vec{\sig}\cdot\hat{n}_{ef}\rt)G_f(e)\rt]-(i\g-1)j_f\ln\tr\lt[\half\lt(1+\vec{\sig}\cdot\hat{n}_{ef}\rt)G^\dagger_f(e)\rt]
\ee
We define the following variables by making a parallel transport to a vertex $v$
\be
\hat{X}_f(v):=g_{ve}\vec{\sig}\cdot\hat{n}_{ef}g_{ev}&\ \ \ \ &\hat{X}^\dagger_f(v):=g_{ev}^\dagger\vec{\sig}\cdot\hat{n}_{ef}g_{ve}^\dagger\nonumber\\
G_f(v):=g_{ve} G_f(e) g_{ev}\ &\ \ \ \ &G_f^\dagger(v):=g_{ev}^\dagger G_f(e) g_{ve}^\dagger
\ee
where one can see that $\hat{X}_f(v)$ is related to the bivector in Proposition \ref{fromcritical} by $\hat{X}_f(v)={X}_f(v)/\g j_f$. In terms of these new variables at the vertex $v$, the face action is written as
\be
S_f=-(i\g+1)j_f\ln\tr\lt[\half\lt(1+\hat{X}_f(v)\rt)G_f(e)\rt]-(i\g-1)j_f\ln\tr\lt[\half\lt(1+\hat{X}_f^\dagger(v)\rt)G^\dagger_f(e)\rt]\label{Sf0}
\ee
According to Theorem \ref{construction}, at the critical point, the bivector $\hat{X}_f(v)$ is written as
\be
\hat{X}_f(v)=2\eps\frac{*E_{\ell_1}(v)\wedge E_{\ell_2}(v)}{|*E_{\ell_1}(v)\wedge E_{\ell_2}(v)|}
\ee
and the spinfoam edge holonomy $g_{vv'}$ equals to the spin connection $\O_{vv'}$ up to a sign $\mu_e=e^{i\pi n_e}$, i.e.
\be
g_{vv'}=e^{i\pi n_e} \O_{vv'}.\label{gvv}
\ee


The spinfoam loop holonomy (in its Spin-1 representation) at the critical point satisfies
\be
G_f(v)E_{\ell}(v)=e^{i\pi\sum_{e\subset f} n_e}E_{\ell}(v)=\cos\Big(\pi\sum_{e\subset f} n_e\Big)E_{\ell}(v)\label{GE}
\ee
We pick out a $E_{\ell}(v)$ as one of the edge of the triangle dual to $f$ and construct $E_{\ell'}(v)$ as a linear combination of the edge vectors $E_{\ell_1}(v),E_{\ell_2}(v)$ and orthogonal to $E_\ell(v)$. We normalize $E_{\ell}(v),E_{\ell'}(v)$ and represented them by
\be
\hat{E}_{\ell}(v)=\left(\begin{array}{c}0 \\0 \\1 \\0\end{array}\right)\ \ \ \ \text{and}\ \ \ \ \hat{E}_{\ell'}(v)=\left(\begin{array}{c}0 \\0 \\0 \\1\end{array}\right)
\ee
We have shown that the loop holonomy $G_f(v)$ can be written as an exponential form, i.e. $G_f(v)=e^{Y_f(v)}$. If we represent $Y_f(v)$ by a $4\times4$ matrix, from Eq.\Ref{GE}, $Y_f(v)$ must be given by
\be
Y_f(v)=\left(\begin{array}{cccc}D_{11} & D_{12} & 0 & 0 \\D_{21} & D_{22} & 0 & 0 \\0 & 0 & 0 & -\pi\sum_{e} n_e \\0 & 0 & \pi\sum_{e} n_e & 0\end{array}\right)
\ee
where $D_{ij}$ is a pure boost leaving the 2-plane spaned by $E_{\ell}(v),E_{\ell'}(v)$ invariant. Then the spin-1 representation of the loop holonomy $G_f(v)$ can be expressed as 
\be
G_f(v)&=&\exp\lt[\frac{* E_{\ell_1}(v)\wedge E_{\ell_2}(v)}{|*E_{\ell_1}(v)\wedge E_{\ell_2}(v)|}\vartheta_f+\frac{E_{\ell_1}(v)\wedge E_{\ell_2}(v)}{|E_{\ell_1}(v)\wedge E_{\ell_2}(v)|}\pi\sum_{e\subset f}n_e\rt]\nonumber\\
&=&e^{\eps\frac{1}{2}\vartheta_f\hat{X}_f(v)+\frac{1}{2}\pi\sum_{e\subset f}n_e *\hat{X}_f(v)}
\ee
where $\vartheta_f$ is an arbitrary number. Since the duality map $*=i$ in the spin-$\half$ representation, thus
\be
G_f(v)=e^{\eps\frac{1}{2}\vartheta_f\hat{X}_f(v)+i\frac{1}{2}\pi\sum_{e\subset f}n_e \hat{X}_f(v)}
\ee
in the spin-$\half$ representation, where $G_f(v)\in\Slc$.

We now determine the physical meaning of the parameter $\vartheta_f$. $\mathrm{sgn}(V_4(v))$ is a constant on the triangulation for the oriented 4-volumes of the 4-simplices. By the relation between spinfoam variable $g_{vv'}$ and the spin connection: $g_{vv'}=\mu_e\O_{vv'}$, we have for the spin connection
\be
\O_f(v)=e^{i\pi\sum_en_e}G_f(v)=e^{i\pi\sum_en_e}e^{\frac{* E_{\ell_1}(v)\wedge E_{\ell_2}(v)}{|*E_{\ell_1}(v)\wedge E_{\ell_2}(v)|}\vartheta_f+\frac{E_{\ell_1}(v)\wedge E_{\ell_2}(v)}{|E_{\ell_1}(v)\wedge E_{\ell_2}(v)|}\pi\sum_{e}n_e}\in \text{SO}(1,3)\label{Ofv}
\ee
We consider a discretization of classical Einstein-Hilbert action $\int R\sqrt{-g}\rmd^4x$: 
For each dual face $f$
\be
\tr\lt[\int_{\Delta_f}\mathrm{sgn}\det(e_\mu^I)*\![e\wedge e]\int_f R\rt]\simeq \mathrm{sgn}(V_4)\half\tr\lt[*\!\Big(E_{\ell_1}(v)\wedge E_{\ell_2}(v)\Big)\ln \O^{\text{boost}}_f(v)\rt]= \mathrm{sgn}(V_4)A_f\vartheta_f\label{Regge}
\ee
This formula should be understood by ignoring the higher order correction in the continuum limit. Here we use $\Delta_f$ to denote the triangle dual to $f$. $e_\mu^I$ is a co-tetrad in the continuum. $R$ is the local curvature from the $\slc$-valued local spin connection compatible with $e_\mu^I$. Only the pure boost part $\O^{\text{boost}}_f(v)=e^{\frac{* E_{\ell_1}(v)\wedge E_{\ell_2}(v)}{|*E_{\ell_1}(v)\wedge E_{\ell_2}(v)|}\vartheta_f}$ of the spin connection $\O_f(v)$ contributes the curvature $R$ in the discrete context. When $e^{i\pi\sum_en_e}=-1$, the factor $e^{i\pi\sum_en_e}e^{\frac{E_{\ell_1}(v)\wedge E_{\ell_2}(v)}{|E_{\ell_1}(v)\wedge E_{\ell_2}(v)|}\pi\sum_{e}n_e}$ flips the overall sign of the reference frame at $v$ and rotates $\pi$ on the 2-plane spanned by $E_{\ell_1}(v), E_{\ell_2}(v)$. It serves for the case that the time-orientation of the reference frame is flipped by $\O_f$, while the triangle spanned by $E_{\ell_1}(v), E_{\ell_2}(v)$ is kept unchange. Such an operation doesn't change the quantity\footnote{$\O_f(v)\in\text{SO}^-(1,3)$ comes from an oriented but time-unoriented orthonormal frame boundle, where the co-tetrad $e_\mu^I$ can flip sign. However, the local spin connection $\G^{IJ}_\a=e^I_\mu\nabla_\a e^{\mu J}$ doesn't change as $e_\mu^I\mapsto -e_\mu^I$ and coincides with the spin connection on the oriented and time-oriented orthonormal frame bundle. The same holds also for the curvature $R$ from the spin connection. }
\be
\tr\lt[\int_{\Delta_f}\mathrm{sgn}\det(e_\mu^I)*\![e\wedge e]\int_f R\rt]
\ee
 $A_f=\half|*E_{\ell_1}(v)\wedge E_{\ell_2}(v)|$ is the area of the triangle dual to $f$. Compare Eq.\Ref{Regge} with the Regge action of discrete GR, we identify that $\mathrm{sgn}(V_4)\vartheta_f$ is the deficit angle $\Theta_f$ of $f$ responsible to the curvature $R$ from the spin connection (see also \cite{deficit}). 
\be
\Theta_f=\mathrm{sgn}(V_4)\vartheta_f
\ee
where we keep in mind that $\mathrm{sgn}(V_4)$ is a constant sign on the (sub-)triangulation.

Insert the expression of $G_f(v)$ into Eq.(\ref{Sf0}), we obtain for a internal face $f$
\be
S_f&=&-\frac{(i\g+1)}{2}j_f\lt[\eps\mathrm{sgn}(V_4)\Theta_f+i\pi\sum_{e\subset f}n_e \rt]-\frac{(i\g-1)}{2}j_f\lt[\eps\mathrm{sgn}(V_4)\Theta_f-i\pi\sum_{e\subset f}n_e \rt]\nonumber\\
&=&-i\ \eps\ \mathrm{sgn}(V_4)\ \g j_f\Theta_f-i\pi j_f\sum_{e\subset f}n_e
\ee
where we have used again the relations of Pauli matrices $(\vec{\sig}\cdot\hat{n})^{2k}=1_{2\times 2}$ and $(\vec{\sig}\cdot\hat{n})^{2k+1}=\vec{\sig}\cdot\hat{n}$, as well as the following relation
\be
\tr\lt(\hat{X}_f(v)\cdots \hat{X}_f(v)\rt)=\tr\Big(g_{ve}\vec{\sig}\cdot\hat{n}_{ef}g_{ev}\cdots g_{ve}\vec{\sig}\cdot\hat{n}_{ef}g_{ev}\Big)=\tr\Big(\vec{\sig}\cdot\hat{n}_{ef}\cdots\vec{\sig}\cdot\hat{n}_{ef}\Big).
\ee
Finally we sum over all the internal faces and construct the total internal action $S_{\text{int}}=\sum_{f\ \text{internal}} S_f$
\be
S_{\text{internal}}=-i\ \eps\ \mathrm{sgn}(V_4) \sum_{f\ \text{internal}} \g j_f\Theta_f-i\pi \sum_{f\ \text{internal}}j_f\sum_{e\in\partial f}n_e .
\ee
where $\g j_f$ is understood as the area of the triangle $f$, and $\sum_f\g j_f\Theta_f$ is the Regge action for discrete GR.

\subsection{Boundary Faces}

Let's consider a face $f$ dual to a boundary triangle (see FIG.\ref{boundary}). The corresponding face action $S_f$ reads
\be
S_f=2i\g j_f\sum_{v}\ln\frac{||Z_{ve'f}||}{||Z_{vef}||}-2i j_f\sum_{v}\phi_{eve'}=-2i j_f\lt(\g\sum_{v}\theta_{eve'} +\sum_{v}\phi_{eve'}\rt)
\ee
where the sum is over all the internal verices $v$ around the face $f$, and we have also used the notation ${||Z_{vef}||}/{||Z_{ve'f}||}:=e^{\theta_{eve'}}$.

On the boundary of the face $f$, there are at least two edges connecting to the nodes on the boundary of the triangulation. We suppose there is an edge $e_0$ of the face $f$ connecting a boundary node, associated with a boundary spinor $\xi_{e_0 f}$. Recall Eqs.\Ref{gluingJ} and \Ref{gluing}, and consider the following successive action on $\xi_{e_0 f}$ of $g_{e'v}g_{ve}$ along the boundary of the face $f$, until reaching another edge $e_1$ connecting to another boundary node. We denote by $p_{e_1 e_0}$ the path from $e_0$ to $e_1$
\be
g_{e_1v'}g_{v'e'}\cdots g_{ev}g_{ve_0}J\xi_{e_0f}&=&J\xi_{e_1f}\exp\lt[-\sum_{v\in p_{e_1 e_0}}\theta_{eve'}-i\sum_{v\in p_{e_1 e_0}}\phi_{eve'}\rt]\nonumber\\
g_{e_1v'}g_{v'e'}\cdots g_{ev}g_{ve_0}\xi_{e_0f}&=&\xi_{e_1f}\exp\lt[\sum_{v\in p_{e_1 e_0}}\theta_{eve'}+i\sum_{v\in p_{e_1 e_0}}\phi_{eve'}\rt]\label{e0e1}
\ee
We denote the holonomy along the path $p_{e_1e_0}$ by
\be
G_f(e_1,e_0):=g_{e_1v'}g_{v'e'}\cdots g_{ev}g_{ve_0}
\ee
and construct a SU(2) matrix from the normalized spinor $\xi$ by
\be
g(\xi)=(\xi,J\xi)\in\text{SU(2)}
\ee
If we denote by
\be
\a=\sum_{v\in p_{e_1 e_0}}\theta_{eve'}+i\sum_{v\in p_{e_1 e_0}}\phi_{eve'}
\ee
Eq.\Ref{e0e1} can be expressed as a matrix equation
\be
G_f(e_1,e_0)\ g(\xi_{e_0f})=g(\xi_{e_1f})\left(\begin{array}{cc}e^\a & 0  \\ 0 & e^{-\a}\end{array}\right)
\ee
Therefore $G_f(e_1,e_0)$ can be solved immediately
\be
G_f(e_1,e_0)=g(\xi_{e_1f})\ e^{\sum_{v}\lt(\theta_{eve'}+i\phi_{eve'}\rt)\vec{\sig}\cdot\hat{z}}\ g(\xi_{e_0f})^{-1}
\ee
We again employ the identity Eq.\Ref{lntr} to obtain
\be
\ln\tr\lt[\half\lt(1+\vec{\sig}\cdot\hat{z}\rt)g(\xi_{e_1f})^{-1}G_f(e_1,e_0)g(\xi_{e_0f})\rt]&=&\sum_{v}\lt(\theta_{eve'}+i\phi_{eve'}\rt)\nonumber\\
\ln\tr\lt[\half\lt(1+\vec{\sig}\cdot\hat{z}\rt)g(\xi_{e_0f})^{-1}G^\dagger_f(e_1,e_0)g(\xi_{e_1f})\rt]&=&\sum_{v}\lt(\theta_{eve'}-i\phi_{eve'}\rt)
\ee
Insert these relations into the face action $S_f$
\be
S_f
&=&-(i\g+1)j_f\ln\tr\lt[\half\lt(1+\vec{\sig}\cdot\hat{z}\rt)g(\xi_{e_1f})^{-1}G_f(e_1,e_0)g(\xi_{e_0f})\rt]\nonumber\\
&&-(i\g-1)j_f\ln\tr\lt[\half\lt(1+\vec{\sig}\cdot\hat{z}\rt)g(\xi_{e_0f})^{-1}G^\dagger_f(e_1,e_0)g(\xi_{e_1f})\rt]\label{boundarySf}
\ee
Recall that at the critical configuration $G_f(e_1,e_0)$ coincides with the spin connection $\O_f(e_1,e_0)$ up to a sign. Given the co-frame vectors $E_\ell(e_0)$ and $E_\ell(e_1)$ with $\ell$ the edges of the triangle $f$.
\be
(\prod_{e}\mu_e) E_\ell(e_1)&=&G_{f}(e_1,e_0)E_\ell(e_0)\ \ \ \ \forall\ \ell\subset f\nonumber\\
G_{f}(e_1,e_0)&=&(\prod_{e}\mu_e) \O_f(e_1,e_0)
\label{GOmega}
\ee
where the product $\prod_e$ is over all the edges along the path $p_{e_1e_0}$.

Here we are going to give an explicit expression for $G_f(e_1,e_0)$ from Eq.\Ref{GOmega}. We first define three new vectors $\tilde{E}_\ell(e_i)$ for the three $\ell$'s of the triangle $f$
\be
\tilde{E}_\ell(e_i)=\hat{g}(\xi_{e_if})^{-1}{E}_\ell(e_i)\ \ \ \ i=0,1
\ee
where $\hat{g}(\xi_{e_if})$ is the spin-1 representation of ${g}(\xi_{e_if})\in\text{SU(2)}$. Thus
\be
\hat{g}(\xi_{e_1 f})^{-1}G_{f}(e_1,e_0)\hat{g}(\xi_{e_0f})\tilde{E}_\ell(e_0)=(\prod_{e}\mu_e)\tilde{E}_\ell(e_1)
\ee
The co-frame vectors $E_\ell(e)$ of a triangle $f$ is orthogonal to $\hat{n}_{ef}$, which is given by $\hat{n}_{ef}=\hat{g}(\xi_{ef})\hat{z}$. Thus the triangles formed by $\tilde{E}_\ell(e_i)$ ($i=0,1$) are both on the 2-plane (the $xy$-plane) orthogonal to $u=(1,0,0,0)$ and $\hat{z}=(0,0,0,1)$, then they are related by a rotation $e^{\zeta_f J_3}$ on the $xy$-plane 
\be
\tilde{E}_\ell(e_1)=e^{\zeta_f J_3}\tilde{E}_\ell(e_0)\ \ \ \ \forall\ \ell\subset f.
\ee
Therefore $\hat{g}(\xi_{e_1 f})^{-1}G_{f}(e_1,e_0)\hat{g}(\xi_{e_0f})$ is the above rotation plus a pure boost along the $z$-direction and a rotation taking care the sign factor $\prod_{e}\mu_e$, both of which leaves the vector on $xy$-plane invariant. Hence
\be
G_{f}(e_1,e_0)=\hat{g}(\xi_{e_1 f})e^{\varth^B_f K_3}e^{\pi\sum_e n_e J_3}e^{\zeta_f J_3}\hat{g}(\xi_{e_0f})^{-1}
\ee
where $\varth^B_f$ is an arbitrary number. The rotation $e^{\zeta_f J_3}$ corresponds to a gauge transformation in the context of twisted geometry \cite{twisted}. Here we can always absorb $e^{\zeta_f J_3}$ into one of $\hat{g}(\xi_{e_i f})$, which leads to a redefinition of the boundary data $\xi_{e_i f}$. Such a redefinition doesn't change the triangle normal $\hat{n}_{ef}$ thus doesn't change the bivector $X_{ef}$. Then all the above analysis about constructing discrete geometry is unaffected. The boundary data after this redefinition is the Regge boundary data employed in \cite{semiclassical}. With this setting, we obtain
\be
G_{f}(e_1,e_0)= \hat{g}(\xi_{e_1 f})e^{\varth^B_f K_3}e^{\pi\sum_e n_e J_3}\hat{g}(\xi_{e_0f})^{-1}.\label{OmegaK3}
\ee
for an explicit expression of $G_{f}(e_1,e_0)$, and
\be
\tilde{E}_\ell(e_0)=\tilde{E}_\ell(e_1)=\tilde{E}_\ell
\ee
for the edges of triangle $\ell$. The three vectors $\tilde{E}_\ell$ determines the triangle geometry of $f$ in the frame at $f$. From Eq.\Ref{GOmega}, we obtain the spin connection compatible with the co-frame
\be
\O_{f}(e_1,e_0)= e^{i\pi \sum_e n_e}\hat{g}(\xi_{e_1 f})e^{\varth^B_f K_3}e^{\pi\sum_e n_e J_3}\hat{g}(\xi_{e_0f})^{-1}.\label{Ofee}
\ee
When $e^{i\pi \sum_e n_e}=1$, the spin connection $\O_{f}(e_1,e_0)\in\text{SO}^+(1,3)$, and when $e^{i\pi \sum_e n_e}=-1$, $\O_{f}(e_1,e_0)\in\text{SO}^-(1,3)$.

We now determine the physical meaning of the parameter $\varth^B_f$ in the expression of $G_f(e_1,e_0)$. It is related to the dihedral angle $\Theta_f^B$ of the two boundary tetrahedra $t_{e_0},t_{e_1}$ at the triangle $f$ sheared by them. The two tetrahedra $t_{e_0},t_{e_1}$ belongs to different 4-simplicies $\sig_{v_0},\sig_{v_1}$, while the curvature from spin connection between $\sig_{v_0},\sig_{v_1}$ are given by the pure boost part of $\O_f({v_1,v_0})$ along the internal edges of the face $f$. This curvature is responsible to the dihedral angle between $t_{e_0},t_{e_1}$. The dihedral boost between the normals of $t_{e_0},t_{e_1}$ at the triangle $f$ is given by the pure boost part of
\be
\hat{g}(\xi_{e_1 f})^{-1}\O_{f}(e_1,e_0)\hat{g}(\xi_{e_0f})=e^{i\pi \sum_e n_e}e^{\varth^B_f K_3}e^{\pi\sum_e n_e J_3}\label{gOg}
\ee
The above transformation leaves the triangle geometry $\tilde{E}_\ell$ invariant in both case of $e^{i\pi \sum_e n_e}=\pm1$. We consider the unit normal of the tetrahedron $t_{e_0}$ (viewed in its own frame) $u^I=(1,0,0,0)^t$, parallel transported by $G_{f}(e_1,e_0)$ (from the frame of $t_{e_0}$ to the frame of $t_{e_1}$)
\be
G_{f}(e_1,e_0)^I_{\ J} u^J= e^{\varth^B_f K_3}u=(\cosh\varth^B_f,0,0,\sinh\varth^B_f)^t
\ee
Contract this equation with the unit normal $u^I=(1,0,0,0)^t$ viewed in the frame of $t_{e_1}$, we obtain that for the dihedral angle $\Theta_f^B$
\be
\cosh\Theta^B_f=-u_IG_{f}(e_1,e_0)^I_{\ J} u^J=\cosh\varth^B_f\label{uOu}
\ee
which implies that $\Theta_f^B=\pm\varth_f^B$. By a generalization of the analysis in \cite{semiclassical}, we can conclude that

\begin{Lemma}
The dihedral angle $\Theta_f^B$ at the triangle $f$ relates to the parameter $\varth_f^B$ by
\be
\Theta_f^B=\eps\ \mathrm{sgn}(V_4)\varth_f^B
\ee
\end{Lemma}

\startproof In the tetrahedra $t_{e_0}$ and $t_{e_1}$, both pairs of the vectors $E_{\ell_1}(e_0),E_{\ell_2}(e_0)$ and $E_{\ell_1}(e_1),E_{\ell_2}(e_1)$ are orthogonal to $u=(1,0,0,0)^t$. Thus at the vertex $v$, both ${E}_{\ell_1}(v)$ and ${E}_{\ell_2}(v)$ are orthogonal to 
\be
F_{e_0}(v)=G_f(v,e_0)\rhd u\ \ \ \ F_{e_1}(v)=G_f(v,e_1)\rhd u
\ee
Thus both $F_{e_0}(v)$ and $F_{e_1}(v)$ are future-pointing since $G_f(v,e)\in\Slc$. Eq.\Ref{uOu} implies that 
\be
\lt|\eta_{IJ}F_{e_0}^I(v) F_{e_1}^J(v)\rt|=\cosh\Theta_f^B.
\ee
We define a dihedral boost from the dihedral angle $\Theta_f^B$ by
\be
D(e_1,e_0)&=&\exp\lt[|\Theta_f^B|\frac{F_{e_0}(v)\wedge F_{e_1}(v)}{\lt|F_{e_0}(v)\wedge F_{e_1}(v)\rt|}\rt]\nonumber\\
&=&\exp\lt[\Theta_f^B\frac{U_{e}(v)\wedge U_{e'}(v)}{| U_{e}(v)\wedge U_{e'}(v)|}\rt]
\ee
where we have chosen the sign of the dihedral angle such that \cite{semiclassical}
\be
&\text{If}&\frac{F_{e_1}(v)\wedge F_{e_0}(v)}{\lt|F_{e_1}(v)\wedge F_{e_0}(v)\rt|}=\frac{U_{e}(v)\wedge U_{e'}(v)}{| U_{e}(v)\wedge U_{e'}(v)|}: \ \ \ \ |\Theta_f^B|=-\Theta_f^B\nonumber\\
&\text{If}&\frac{F_{e_1}(v)\wedge F_{e_0}(v)}{\lt|F_{e_1}(v)\wedge F_{e_0}(v)\rt|}=-\frac{U_{e}(v)\wedge U_{e'}(v)}{| U_{e}(v)\wedge U_{e'}(v)|}: \ \ \ \ |\Theta_f^B|=\Theta_f^B
\ee
with $V_4(v){U_{e}(v)\wedge U_{e'}(v)}=G_f(v,e_0)\rhd{* E_{\ell_1}(e_0)\wedge E_{\ell_2}(e_0)}$. 

On the other hand, the boost generator $K_3$ can be related to the bivector $X^{IJ}_{ef}=2\g j_f(\hat{n}_{ef}\wedge u)^{IJ}$
\be
K_3=-\hat{z}\wedge u=-g(\xi_{ef})^{-1}\otimes g(\xi_{ef})^{-1}(\hat{n}_{ef}\wedge u)=-g(\xi_{ef})^{-1}\otimes g(\xi_{ef})^{-1}\ \frac{1}{2\g j}X_{ef}
\ee
At the critical configuration the bivector $X_{ef}$ is given by Eq.\Ref{XEE0}, which results in that
\be
K_3=-\eps g(\xi_{ef})^{-1}\otimes g(\xi_{ef})^{-1}\frac{* E_{\ell_1}(e)\wedge E_{\ell_2}(e)}{\lt| E_{\ell_1}(e)\wedge E_{\ell_2}(e)\rt|}
=-\eps \frac{* \tilde{E}_{\ell_1}\wedge \tilde{E}_{\ell_2}}{| \tilde{E}_{\ell_1}\wedge \tilde{E}_{\ell_2}|}
\ee
where $\frac{* \tilde{E}_{\ell_1}\wedge \tilde{E}_{\ell_2}}{| \tilde{E}_{\ell_1}\wedge \tilde{E}_{\ell_2}|}$ is the (unit) bivector corresponding to the triangule $f$. Therefore for the bivector at the vertex $v$
\be
\frac{U_{e}(v)\wedge U_{e'}(v)}{| U_{e}(v)\wedge U_{e'}(v)|}&=&\mathrm{sgn}(V_4)G_f(v,e_0)\rhd\frac{* E_{\ell_1}(e_0)\wedge E_{\ell_2}(e_0)}{\lt|E_{\ell_1}(e_0)\wedge E_{\ell_2}(e_0)\rt|}\nonumber\\
&=&-\mathrm{sgn}(V_4)\ \eps\ G_f(v,e_0)g(\xi_{e_0f})K_3g(\xi_{e_0f})^{-1}G_f(v,e_0)^{-1}
\ee
Then we obtain the following expression of $D(e_1,e_0)$:
\be
D(e_1,e_0)=G_f(v,e_0)g(\xi_{e_0f})e^{-\eps\ \mathrm{sgn}(V_4)\Theta^B_f K_3}g(\xi_{e_0f})^{-1}G_f(v,e_0)^{-1}\label{Dee}.
\ee

One can check that $D(e_1,e_0)$ gives a dihedral boost from $F_{e_0}(v)$ to $F_{e_1}(v)$, i.e.
\be
D(e_1,e_0)F_{e_0}(v)=F_{e_1}(v)\label{DF}
\ee
If we represent the vector $F_e(v)$ by the $2\times 2$ matrix $F_e=F_e^I\sig_{I}$, we have $F_{e}(v)=G_f(v,e)G_f(v,e)^\dagger$, Eq.\Ref{DF} can be expressed as
\be
D(e_1,e_0)G_f(v,e_0)G_f(v,e_0)^\dagger D(e_1,e_0)^\dagger=G_f(v,e_1)G_f(v,e_1)^\dagger
\ee
By using Eq.\Ref{Dee}, we obtain that ($J_3^\dagger=-J_3$)
\be
G_f(v,e_0)g(\xi_{e_0f})e^{-2\eps\ \mathrm{sgn}(V_4)\Theta^B_f K_3}g(\xi_{e_0f})^\dagger G_f(v,e_0)^\dagger=G_f(v,e_1)G_f(v,e_1)^\dagger\label{comb1}
\ee
From the expression Eq.\Ref{OmegaK3} of $G_f(e_1,e_0)=G_f(v,e_1)^{-1}G_f(v,e_0)$ in terms of $\varth_f^B$, we obtain
\be
G_f(e_0,e_1)G_f(e_0,e_1)^\dagger=\hat{g}(\xi_{e_0 f})e^{-2\varth^B_f K_3}\hat{g}(\xi_{e_0 f})^{-1}\label{comb2}
\ee
Combining Eqs.\Ref{comb1} and \Ref{comb2}, we obtain
\be
e^{-2\eps\ \mathrm{sgn}(V_4)\Theta^B_f K_3}=e^{-2\varth^B_f K_3}
\ee
which results in 
\be
\varth^B_f=\eps\ \mathrm{sgn}(V_4)\Theta^B_f.
\ee
\finishproof

The Eq.\Ref{OmegaK3} is now related to the dihedral angle $\Theta_f^B$
\be
\hat{g}(\xi_{e_1 f})^{-1}G_{f}(e_1,e_0)\hat{g}(\xi_{e_0f})= e^{\eps\ \mathrm{sgn}(V_4)\Theta_f^B K_3}e^{\pi\sum_e n_e J_3}.
\ee
Recall that in Spin-$\half$ representation $\vec{J}=\frac{i}{2}\vec{\sig}$ and $\vec{K}=\frac{1}{2}\vec{\sig}$, thus in Spin-$\half$ representation:
\be
{g}(\xi_{e_1 f})^{-1}G_{f}(e_1,e_0){g}(\xi_{e_0f})= e^{\half\eps\ \mathrm{sgn}(V_4)\Theta_f^B \sig_3}e^{\frac{i}{2}\pi\sum_e n_e \sig_3}
\ee
Insert this relation back into Eq.\Ref{boundarySf},
\be
S_f&=&-\frac{(i\g+1)}{2}j_f\lt[\eps\ \mathrm{sgn}(V_4)\Theta_f^B+i\pi\sum_{e\subset p_{e_1e_0}} n_e\rt]-\frac{(i\g-1)}{2}j_f\lt[\eps\ \mathrm{sgn}(V_4)\Theta_f^B-i\pi\sum_{e\subset p_{e_1e_0}} n_e\rt]\nonumber\\
&=&-i\eps\ \mathrm{sgn}(V_4) \g j_f\Theta_f^B-ij_f\pi\sum_{e\subset p_{e_1e_0}} n_e
\ee
Then the total boundary action $S_{\text{boundary}}=\sum_{\text{boundary }f}S_f$:
\be
S_{\text{boundary}}=-i\ \eps\ \mathrm{sgn}(V_4)\sum_{\text{boundary }f} \g j_f\Theta^B_f-i\pi \sum_{\text{boundary }f}j_f\sum_{e\subset p_{e_1e_0}}n_e.
\ee


\subsection{Spinfoam Amplitude at Nondegenerate Critical Configuration}

In this subsection we summarize our result and give spinfoam amplitude at a general nondegenerate critical configuration. First of all, we say a spin configuration $j_f$ is Regge-like, if with $j_f$ on each face the critical point equations Eqs.\Ref{gluingJ}, \Ref{gluing}, and \Ref{closure} have nondegenerate solution $(j_f,g_{ve},\xi_{ef},z_{vf})$. For a non-Regge-like spin configuration $j_f$, the critical point equations have no nondegenerate solutions.

Given a Regge-like spin configuration $j_f$ and find a solution $(j_f,g_{ve},\xi_{ef},z_{vf})$ of the critical point equations, we construct the following variables as in Section \ref{solution}:

\begin{itemize}

\item A co-frame $E_\ell(v),E_\ell(e)$ of the triangulation (bulk and boundary) can be constructed from the solution $(j_f,g_{ve},\xi_{ef},z_{vf})$, unique up to a simultaneously sign flipping $E_\ell\to -E_\ell$ within a 4-simplex, such that the Regge-like spin configuration $j_f$ satisfies
\be
2\g j_f=\lt|E_{\ell_1}(v)\wedge E_{\ell_2}(v)\rt|.
\ee
From the co-frame we can construct a unique discrete metric on the whole triangulation (bulk and boundary)
\be
g_{\ell_1\ell_2}(v)=\eta_{IJ}E^I_{\ell_1}(v)E^J_{\ell_2}(v)\ \ \ \ \ g_{\ell_1\ell_2}(e)=\eta_{IJ}E^I_{\ell_1}(e)E^J_{\ell_2}(e).
\ee
So $\g j_f$ is the triangle area from the discrete metric $g_{\ell_1\ell_2}$

\item For each dual edge $e$ we specify a sign factor $\mu_e=e^{i\pi n_e}$ that equals $1$ or $-1$ with $n_e$ equals $0$ or $1$, such that the spinfoam group element $g_{vv'}$ (in the Spin-1 representation) is related to an SO(1,3) matrix $\O_{vv'}$ by this sign factor, i.e.
\be
g_{vv'}=e^{i\pi n_e}\O_{vv'}
\ee
where $\O_{vv'}$ is compatible with the co-frame $E_\ell(v)$, i.e.
\be
(\O_{vv'})^I_{\ J}E_\ell^J(v')=E_\ell^I(v)
\ee
If $\mathrm{sgn}(V_4(v))=\mathrm{sgn}(V_4(v'))$, $\O_{vv'}$ is the unique discrete spin connection compatible with the co-frame. In addition, we note that each $\mu_e$ is not invariant under the sign flipping $E_\ell\to -E_\ell$, but the product $\prod_{e\subset\partial f}\mu_e$ is invariant for any (internal or boundary) face $f$ (see Lemma.\ref{prodmue}).

\item There is a global sign factor $\eps$ that equals $1$ or $-1$, to relate the bivectors $X_f(v)$ in the bulk and $X_{ef}$ on the boundary to the co-frame:
\be
X_f^{IJ}(v)=\eps\ *\!\Big[E_{\ell_1}(v)\wedge E_{\ell_2}(v)\Big]^{IJ},\ \ \ \ \ X_{ef}^{IJ}=\eps\ *\!\Big[  E_{\ell_1}(e)\wedge E_{\ell_2}(e)\Big]^{IJ}.
\ee
If the triangulation $\ck$ has boundary, the global sign factor $\eps=\pm1$ is specified by the orientation of the boundary triangulation, i.e. $\eps=\mathrm{sgn}(V_3)$ for the boundary tetrahedra. 

\end{itemize}
Therefore a nondegenerate solution $(j_f,g_{ve},\xi_{ef},z_{vf})$ of the spinfoam critical point equations specifies uniquely a set of variables $(g_{\ell_1\ell_2},n_e,\eps)$, which include a discrete metric and two types of sign factors.

The previous analysis shows that, given a general critical configuration $(j_f,g_{ve},\xi_{ef},z_{vf})$, we can divide the triangulation $\ck$ into sub-triangulations $\ck_1,\cdots,\ck_n$, where each of the sub-triangulations is a triangulation with boundary, with a constant $\mathrm{sgn}(V_4(v))$. On each of the sub-triangulation $\ck_i$, the spinfoam action $S$ evaluated at $(j_f,g_{ve},\xi_{ef},z_{vf})_{\ck_i}$ is a function of the variables $(g_{\ell_1\ell_2},n_e,\eps)$ and behaves mainly as a Regge action:
\be
S(g_{\ell_1\ell_2},n_e,\eps)\big|_{\ck_i}&=&S_{\text{internal}}(g_{\ell_1\ell_2},n_e,\eps)+S_{\text{boundary}}(g_{\ell_1\ell_2},n_e,\eps)\nonumber\\
&=&-i\ \eps\ \mathrm{sgn}(V_4) \sum_{\text{internal}\ f} \g j_f\Theta_f-i\pi \sum_{\text{internal}\ f}j_f\sum_{e\subset\partial f}n_e \nonumber\\
&&-i\ \eps\ \mathrm{sgn}(V_4)\sum_{\text{boundary }f} \g j_f\Theta^B_f-i\pi \sum_{\text{boundary }f} j_f\sum_{e\subset\partial f}n_e\nonumber\\
&=&-i\ \eps\ \mathrm{sgn}(V_4)\sum_{\text{internal}\ f} \g j_f\Theta_f-i\ \eps\ \mathrm{sgn}(V_4) \sum_{\text{boundary }f} \g j_f\Theta^B_f-i\pi \sum_{e}n_e\sum_{f\subset t_e} j_f
\ee
where we note that the areas $\g j_f$, deficit angles $\Theta_f$, and dihedral angles $\Theta^B_f$ are uniquely determined by the discrete metric $g_{\ell_1\ell_2}$. Moreover for each tetrahedron $t$, the sum of face spins $\sum_{f\subset t}j_f$ is an integer. If the spins $j_f$ are integers, $\sum_{f\subset t} j_f$ then is an even integer, so $e^{-i\pi \sum_{e}n_e\sum_{f\subset t_e} j_f}=1$ so the second term in the above formula doesn't contribute the exponential $e^{\l S_{\text{int}}}$. For half-integer spins, $e^{-i\pi \sum_{e}n_e\sum_{f\subset t_e}j_f}=\pm1$ gives an overall sign factor. Therefore in general at a nondegenerate spinfoam configuration $(j_f,g_{ve},\xi_{ef},z_{vf})$ that solves the critical point equations,
\be
e^{\l S}\big|_{\ck_i}=\pm\exp\l\lt[-i\ \eps\ \mathrm{sgn}(V_4) \sum_{\text{internal}\ f} \g j_f\Theta_f-i\ \eps\ \mathrm{sgn}(V_4) \sum_{\text{boundary }f} \g j_f\Theta^B_f\rt].
\ee
There exists two ways to make the overall sign factor disappear: (1) only consider integer spins $j_f$, or (2) modify the embedding from SU(2) unitary irreps to $\Slc$ unitary irreps by $j_f\mapsto (p_f,k_f):=(2\g j_f,2j_f)$, then the spinfoam action $S$ is replaced by $2S$. In these two cases the exponential $e^{\l S}$ at the critical configuration is independent of the variable $n_e$.

On the triangulation $\ck=\cup_{i=1}^n\ck_i$, $e^{\l S}$ is given by a product over all the sub-triangulations:
\be
e^{\l S}&=&\prod_{i=1}^n e^{\l S}\big|_{\ck_i}\nonumber\\
&=&\prod_{i=1}^n\exp\l\lt[-i\ \eps\ \mathrm{sgn}(V_4)\sum_{\text{internal}\ f} \g j_f\Theta_f-i\ \eps\ \mathrm{sgn}(V_4) \sum_{\text{boundary }f} \g j_f\Theta^B_f-i\pi \sum_{e}n_e\sum_{f\subset t_e} j_f\rt]_{\ck_i}
\ee
Suppose the oriented 4-volumes are different between two sub-triangulation $\ck_i$ and $\ck_j$ sharing a boundary, the spinfoam amplitude at this critical configuration exhibits a transition between two different spacetime regions with different spacetime orientation. The spacetime orientation is not continuous on the boundary between $\ck_i$ and $\ck_j$. 

We recall the difference between Einstein-Hilbert action and Palatini action
\be
\cl_{EH}=R\ \underline{\eps}=\mathrm{sgn}\det(e_\mu^I)*\![e\wedge e]_{IJ}\wedge R^{IJ}=\mathrm{sgn}\det(e_\mu^I)\cl_{Pl}
\ee
where $\cl_{EH}$ and $\cl_{Pl}$ denote the Lagrangian densities of Einstein-Hilbert action and Palatini action respectively, and $\underline{\eps}$ is a chosen volume form compatible with the metric $g_{\mu\nu}=\eta_{IJ}e_\mu^I e_\nu^J$. Since the Regge action is a discretization of the Einstein-Hilbert action, we may consider the resulting action 
\be
-i\ \eps\sum_{i=1}^n\lt[\mathrm{sgn}(V_4)\sum_{\text{internal}\ f} \g j_f\Theta_f+\mathrm{sgn}(V_4) \sum_{\text{boundary }f} \g j_f\Theta^B_f\rt]_{\ck_i}
\ee
as a discretized Palatini action with on-shell connection, where the on-shell connection means that the discrete connection is the spin connection compatible with the co-frame.

According to the properties of Regge geometry, given a collection of Regge-like areas $\g j_f$, the discrete metric $g_{\ell_1\ell_2}(v)$ is uniquely determined at each vertex $v$. Furthermore since the areas $\g j_f$ are Regge-like, There exists a discrete metric $g_{\ell_1\ell_2}$ in the entire bulk of the triangulation, such that the neighboring 4-simplicies are consistently glued together, as we constructed previously. This discrete metric $g_{\ell_1\ell_2}$ is obviously unique by the uniqueness of $g_{\ell_1\ell_2}(v)$ at each vertex. Therefore given the partial-amplitude $A_{j_f}(\ck)$ in Eq.\Ref{Aj} with a specified Regge-like $j_f$, all the critical configurations $(j_f,g_{ve},\xi_{ef},z_{vf})$ of $A_{j_f}(\ck)$ corresponds to the same discrete metric $g_{\ell_1\ell_2}$, provided a Regge boundary data. The critical configurations from the same Regge-like $j_f$ is classified in the next section.

As a result, given a Regge-like spin configurations $j_f$ and a Regge boundary data, the partial amplitude $A_{j_f}(\ck)$ has the following asymptotics
\be
A_{j_f}(\ck)\big|_{\text{Nondeg}}&\sim&\sum_{x_c}a(x_c)\lt(\frac{2\pi}{\l}\rt)^{\frac{r(x_c)}{2}-N(v,f)}\frac{e^{i\mathrm{Ind}H'(x_c)}}{\sqrt{|\det_r H'(x_c)|}}\lt[1+o\lt(\frac{1}{\l}\rt)\rt]\times\nonumber\\
&&\times\ \exp-i\l\sum_{i=1}^{n(x_c)}\lt[ \eps\ \mathrm{sgn}(V_4)\sum_{\text{internal}\ f} \g j_f\Theta_f+\eps\ \mathrm{sgn}(V_4) \sum_{\text{boundary }f} \g j_f\Theta^B_f+\pi \sum_{e}n_e\sum_{f\subset t_e} j_f\rt]_{\ck_i(x_c)}
\ee 
where $x_c\equiv (j_f,g_{ve},\xi_{ef},z_{vf})$ labels the nondegenerate critical configurations, $r(x_c)$ is the rank of the Hessian matrix at $x_c$, and $N(v,f)$ is the number of the pair $(v,f)$ with $v\in\partial f$ (recall Eq.\Ref{Aj}, there is a factor of $\dim(j_f)$ for each pair of $(v,f)$). $a(x_c)$ is the evaluation of the integration measures at $x_c$, which doesn't scale with $\l$. Here $\Theta_f$ and $\Theta^B_f$ only depend on the metric $g_{\ell_1\ell_2}$, which is uniquely determined by the Regge-like spin configuration $j_f$ and the Regge boundary data. Note that different critical configurations $x_c$ may have different subdivisions of the triangulation into sub-triangulations $\ck_1(x_c),\cdots,\ck_{n(x_c)}(x_c)$.

\section{Parity Inversion}\label{parity}

We consider a tetrahedron $t_e$ associated with spins $j_{f_1},\cdots,j_{f_4}$, we know that the set of four spinors $\xi_{ef_1},\cdots,\xi_{ef_4}$, modulo diagonal SU(2) gauge transformation, is equivalent to the shape of the tetrahedron, if the closure condition is satisfied \cite{shape}. Given a nondegenerate critical configuration $(j_f,g_{ve},\xi_{ef},z_{vf})$, as we discussed previously, the Regge-like spin configuration $j_f$ determines a discrete metric $g_{\ell_1\ell_2}$, which determines the shape of all the tetrahedra in the triangulation. At the critical configuration the closure condition of tetrahedron is always satisfied, so the spinors $\xi_{ef_1},\cdots,\xi_{ef_4}$ for each tetrahedron are determined by the Regge-like spins $j_f$, up to a diagonal SU(2) action on the spinors $\xi_{ef_1},\cdots,\xi_{ef_4}$, which is a gauge transformation of the spinfoam action\footnote{The SU(2) transformation $\xi_{ef}\mapsto h_e\xi_{ef}$ and $g_{ve}\mapsto g_{ve}h_e^{-1}$ $(h_e\in\text{SU(2)})$ is a gauge transformation of the spinfoam action $S$. }. Therefore the gauge equivalence class of the critical configurations $(j_f,g_{ve},\xi_{ef},z_{vf})$ with the same Regge-like spins $j_f$ must have the same set of spinors $\xi_{ef}$. Thus with a given Regge-like spin configuration $j_f$, the degrees of freedom of the nondegenerate critical configurations are the variables $g_{ve}$ and $z_{vf}$. The degrees of freedom of $g_{ve}$ and $z_{vf}$ are factorized into the 4-simplices. Given the Regge-like spins $j_f$ and spinors $\xi_{ef}$, within each 4-simplex, the solutions of $g_{ve}$ and $z_{vf}$ from critical point equations are completely classified in \cite{semiclassical}, which are the two solutions related by a \emph{parity} transformation.

Given a nondegenerate critical configuration $(j_f,g_{ve},\xi_{ef},z_{vf})$, it generates many other critical configurations $(j_f,\tilde{g}_{ve},\xi_{ef},\tilde{z}_{vf})$ which are the solutions of the critical point equations Eqs.\Ref{gluingJ}, \Ref{gluing}, and \Ref{closure}. In at least one simplex or some 4-simplices $\tilde{\sig}_v$  
\be
\tilde{g}_{ve}=J g_{ve} J^{-1}=(g_{ve}^\dagger)^{-1}\ \ \ \ \text{and}\ \ \ \ \frac{||\tilde{Z}_{ve'f}||}{||\tilde{Z}_{vef}||}=\frac{||{Z}_{vef}||}{||{Z}_{ve'f}||} 
\ee
while in the other 4-simplices $\tilde{g}_{ve}=g_{ve} $ and $\tilde{z}_{vf}=z_{vf}$. In \cite{semiclassical}, such a solution-generating map $g_{ve}\mapsto\tilde{g}_{ve}$ and $z_{vf}\mapsto\tilde{z}_{vf}$ is called a \emph{parity}, because $N_e(v)=g_{ve}\rhd(1,0,0,0)^t$ and $\tilde{N}_e(v)=\tilde{g}_{ve}\rhd(1,0,0,0)^t$ are different by a parity inversion. The parity inversion between $N_e(v)$ and $\tilde{N}_e(v)$ can be shown by using the Hermitian matrix representation of the vectors $V=V^0\mathbf{1}+V^j\sig_j$, thus
\be
\tilde{N}_e(v)=\tilde{g}_{ve}\tilde{g}_{ve}^\dagger=Jg_{vf}g_{vf}^\dagger J^{-1}=JN_e(v)J^{-1}=N_e^0(v)\mathbf{1}-N^j_e(v)\sig_j
\ee
since $J\vec{\sig} J^{-1}=-\vec{\sig}$. We denote the parity inversion in $(\mathbb{R}^4,\eta_{IJ})$ by $\bp=\mathrm{diag}(1,-1,-1,-1)$ then we have $\tilde{N}_e(v)=\bp N_e(v)$ in the simplices $\tilde{\sig}_v$ where $\tilde{g}_{ve}\neq g_{ve} $.

Within a single 4-simplex there are in total 2 parity-related solutions of $(g_{ve},z_{vf})$ in the nondegenereate case \cite{semiclassical}. Therefore in a general simplicial complex with $N$ simplices, given a Regge-like spin configuration $j_f$, there are in total $2^N$ nondegenerate critical configurations $(j_f,g_{ve},\xi_{ef},z_{vf})$ that solve the critical point equations. Any two critical configurations are related by the parity transformation in one 4-simplex or many 4-simplices.

We define the bivectors $\tilde{X}_f(v)=\tilde{g}_{ve}\otimes\tilde{g}_{ve}\rhd X_{ef}$ within the 4-simplices $\tilde{\sig}_v$, where
\be
X_{ef}^{IJ}=2\g j_f\lt[\hat{n}_{ef}\wedge u\rt]\ \ \ \ \ u=(1,0,0,0)^t
\ee
Consider the Hermitian matrix representation of $\hat{n}_{ef}$, the action $\tilde{g}_{ve}\rhd\hat{n}_{ef}$ is given by (note that $J^2=-1$)
\be
\tilde{g}_{ve}(\hat{n}_{ef}\cdot\vec{\sig})\tilde{g}_{ve}^\dagger=J{g}_{ve}J^{-1}(\hat{n}_{ef}\cdot\vec{\sig})J{g}_{ve}^\dagger J^{-1}=-J{g}_{ve}(\hat{n}_{ef}\cdot\vec{\sig}){g}_{ve}^\dagger J^{-1}=-\bp{g}_{ve}(\hat{n}_{ef}\cdot\vec{\sig}){g}_{ve}^\dagger
\ee
while we have shown $\tilde{g}_{ve}\rhd u=\bp\lt({g}_{ve}\rhd u\rt)$, thus we obtain that 
\be
\tilde{X}_f(v)=-(\bp\otimes\bp){X}_f(v)
\ee
Recall the construction in Section \ref{solution} and Eq.\Ref{NN}
\be
{X}_f(v)={\a}_{ee'}(v){N}_e(v)\wedge {N}_{e'}(v)
\ee 
Following the same argument towards Eq.\Ref{NN}, we obtain that for the bivectors and normals constructed from $\tilde{g}_{ve}$
\be
\tilde{X}_f(v)=\tilde{\a}_{ee'}(v)\tilde{N}_e(v)\wedge \tilde{N}_{e'}(v)\ \ \Rightarrow\ \ -(\bp\otimes\bp){X}_f(v)=\tilde{\a}_{ee'}(v)\bp{N}_e(v)\wedge \bp{N}_{e'}(v)
\ee
Then we have the relation
\be
\tilde{\a}_{ee'}(v)=-\a_{ee'}(v)\ \ \ \ \text{and}\ \ \ \ \tilde{\b}_{ee'}(v)=-\b_{ee'}(v)
\ee
where $\b_{ee'}(v)=\a_{ee'}(v)\eps_{ee'}(v)$. Following the same procedure as in Section \ref{solution}, we denote $\tilde{\b}_{e_ie_j}$ by $\tilde{\b}_{ij}$ and construct the closure condition for the 4-simplex $\tilde{\sig}_v$
\be
\sum_{j=1}^5\tilde{\b}_{ij}(v)\tilde{N}_{e_j}(v)=0
\ee
by choosing the nonvanishing diagonal elements $\tilde{\b}_{ii}$. Since we have the closure condition $\sum_{j=1}^5{\b}_{ij}{N}_{e_j}(v)=0$, the parity inversion $\tilde{N}_e(v)=\bp N_e(v)$, and $\tilde{\b}_{ij}(v)=-\b_{ij}(v)$ for $i\neq j$, we obtain that the diagonal elements $\tilde{\b}_{ii}(v)=-{\b}_{ii}(v)$. Furthermore we can show that $\tilde{\b}_{ij}$ can be factorized in the same way as in Section \ref{solution}
\be
\tilde{\b}_{ij}(v)=\mathrm{sgn}(\tilde{\b}_{j_0j_0}(v))\tilde{\b}_i(v)\tilde{\b}_j(v)\ \ \ \ \ \ \tilde{\b}_j(v)={\tilde{\b}_{jj_0}(v)}\big/{\sqrt{|\tilde{\b}_{j_0j_0}(v)|}}
\ee
which results in that
\be
\mathrm{sgn}(\tilde{\b}_{j_0j_0}(v))=-\mathrm{sgn}({\b}_{j_0j_0}(v))\ \ \ \ \text{and}\ \ \ \ \tilde{\b}_j(v)=-{\b}_j(v)
\ee
We construct the 4-volume for $\tilde{\b}_j(v)\tilde{N}_{e_j}(v)$
\be
\tilde{V}_4(v):=\det\Big(\tilde{\b}_2(v)\tilde{N}^{e_2}(v),\tilde{\b}_3(v)\tilde{N}^{e_3}(v),\tilde{\b}_4(v)\tilde{N}^{e_4}(v),\tilde{\b}_5(v)\tilde{N}^{e_5}(v)\Big)=-V_4(v)		
\ee
by the parity inversion. Since in Section.\ref{solution} we define the sign factor $\eps(v)=\mathrm{sgn}({\b}_{j_0j_0}(v))\mathrm{sgn}(V_4(v))$, then we have for the parity inversion
\be
\tilde{\eps}(v)=\mathrm{sgn}(\tilde{\b}_{j_0j_0}(v))\mathrm{sgn}(\tilde{V}_4(v))=\eps(v)
\ee
Note that one should not confuse the $\tilde{\eps}$ here with the $\tilde{\eps}$ appeared in section \ref{solution}. This result shows that the parity configuration $(j_f, \tilde{g}_{ve}, \xi_{ef},\tilde{z}_{vf})$ results in an identical global sign factor $\eps$ for the bivector (recall the proof of Theorem \ref{construction}).

The fact that the parity flips the sign of the oriented 4-volume, $\tilde{V}_4(v)=-V_4(v)$, has some interesting consequences: First of all, we mentioned that given a set of Regge-like spins, different nondegenerate critical configurations $x_c=(j_f,g_{ve},\xi_{ef},z_{vf})$ may lead to different subdivisions of the triangulation $\ck$ into sub-triangulation $\ck_1(x_c),\cdots,\ck_{n(x_c)}(x_c)$, where on each sub-triangulation $\mathrm{sgn}(V_4(v))$ is a constant. Now we understand that the difference of the subdivisions comes from a local parity transformation, which flips the sign of the oriented 4-volume. On the other hand, given a nondegenerate critical configuration $x_c=(j_f,g_{ve},\xi_{ef},z_{vf})$, there exists another nondegenerate critical configuration $\tilde{x}_c=(j_f,\tilde{g}_{ve},\xi_{ef},\tilde{z}_{vf})$, naturally associated with $x_c$, obtained by a global parity (parity transformation in all simplices) on the triangulation. The global parity flips the sign of the oriented volume $V_4(v)$ everywhere, thus flip the sign of the spinfoam action at the nondegenerate critical configuration (the deficit angle, dihedral angle, and $\sum_{e\subset\partial f}n_e$ are unchanged under the global parity, which is shown in the following), i.e.\footnote{The sign in front of the term $i\pi \sum_{e}n_e\sum_{f\subset t_e} j_f$ is unimportant.}
\be
S(\tilde{x}_c)=-S(x_c)
\ee
if $\tilde{x}_c$ and $x_c$ are related by a global parity transformation.

Since the frame vectors $U_e(v)=\pm\frac{\b_e(v)N_e(v)}{\sqrt{|V_4(v)|}}$ are defined up to a sign, the frame $\tilde{U}_e(v)$ constructed from parity configuration relates $U_e(v)$ only by a parity inversion
\be
\tilde{U}_e(v)=\bp{U}_e(v)
\ee
The same relation holds for the co-frame $\tilde{E}_{\ell}(v)$
\be
\tilde{E}_{\ell}(v)=\bp{E}_{\ell}(v)
\ee
from the relation
\be
\tilde{U}^{e_j}_I(v)\tilde{E}^I_{e_ke_l}(v)=\delta^j_k-\delta^j_l
\ee
We then obtain the same relation relating the bivector and co-frame/frame as in Theorem \ref{construction}
\be
\tilde{X}_f(v)=\eps\ \tilde{V}_4\lt[\tilde{U}_e(v)\wedge \tilde{U}_{e'}(v)\rt]\ \ \ \ \text{and}\ \ \ \ \tilde{X}_f(v)=\eps\ *\lt[\tilde{E}_{\ell_1}(v)\wedge\tilde{E}_{\ell_2}(v)\rt]
\ee
which is consistent because of the relations $\tilde{X}_f(v)=-(\bp\otimes\bp){X}_f(v)$, $\tilde{U}_e(v)=\bp{U}_e(v)$, $\tilde{E}_{\ell}(v)=\bp{E}_{\ell}(v)$, $\tilde{V}_4(v)=-V_4(v)$, and $\eps_{IJKL}\bp^I_{\ M}\bp^J_{\ N}\bp^K_{\ P}\bp^L_{\ Q}=-\eps_{MNPQ}$. Here we emphasize that the sign factor $\eps$ for the parity configuration $(j_f, \tilde{g}_{ve}, \xi_{ef},\tilde{z}_{vf})$ is the same as the original configuration $(j_f, g_{ve}, \xi_{ef},{z}_{vf})$, thus is consistent with the fact that $\eps$ is a global sign factor on the entire triangulation, i.e. the local/global parity inversion of the critical configuration doesn't change the global sign $\eps$.

The local/global parity inversion $\tilde{E}_{\ell}(v)=\bp{E}_{\ell}(v)$ doesn't change the discrete metric $g_{\ell_1\ell_2}(v)=\eta_{IJ}{E}^I_{\ell_1}(v){E}_{\ell_2}^J(v)$, so the parity configuration $(j_f, \tilde{g}_{ve}, \xi_{ef},\tilde{z}_{vf})$ leads to the same discrete metric as $(j_f, g_{ve}, \xi_{ef},{z}_{vf})$, but gives an O(1,3) gauge transformation (parity inversion) for the co-frame ${E}_{\ell}(v)$. The SO(1,3) matrix $\O_{vv'}\in\text{SO(1,3)}$ is uniquely compatible with the co-frame $E_\ell(v)$ and is a discrete spin connection when $\mathrm{sgn}(V_4(v))=\mathrm{sgn}(V_4(v'))$, as it was shown in Section \ref{geometry}. Given a nondegenerate critical configuration with a subdivision of the triangulation into sub-triangulations, in each of which $\mathrm{sgn}(V_4(v))$ is a constant, we consider a global parity transformation which doesn't change the subdivision but flip the signs of $\mathrm{sgn}(V_4(v))$ in all sub-triangulations. Given a spin connection $\O_{vv'}$ with $\sig_v,\sig_{v'}$ are both in the same sub-triangulation, i.e. $\mathrm{sgn}(V_4(v))=\mathrm{sgn}(V_4(v'))$, the spin connection $\tilde{\O}_{vv'}\in\text{SO(1,3)}$ after a parity transformation in both $\sig_v,\sig_{v'}$ is given by 
\be
\tilde{\O}_{vv'}=\mathbf{P}\O_{vv'}\mathbf{P}
\ee
since $\tilde{\O}_{vv'}$ is uniquely determined by 
\be
\tilde{\O}_{vv'}\tilde{E}_\ell(v')=\tilde{E}_\ell(v)\ \ \ \ \ \ell\subset t_e,\ e=(v,v')
\ee
On the other hand we can check from
\be
\tilde{g}=JgJ^{-1}\ \ \ \ \tilde{g}(-\vec{\sig})\tilde{g}^\dagger=\mathbf{P}\rhd g\vec{\sig}g^\dagger
\ee
that given a 4-vector $V^I$
\be
\tilde{g}\mathbf{P}(V^I\sig_I)\tilde{g}^\dagger=\mathbf{P}({g}V^I\sig_I{g}^\dagger)\ \ \ \ \text{i.e.}\ \ \ \ \tilde{g}\mathbf{P}V=\mathbf{P}gV\ \ \ \text{in Spin-1 representation}
\ee
Let $V={E}_\ell(v')$, using $g_{vv'}=\mu_e\O_{vv'}$,
\be
\tilde{g}_{vv'}\tilde{E}_\ell(v')=\tilde{g}_{vv'}\mathbf{P}{E}_\ell(v')=\mathbf{P}{g}_{vv'}{E}_\ell(v')=\mu_e\mathbf{P}{E}_\ell(v)=\mu_e\tilde{E}_\ell(v)
\ee
Therefore we obtain from $\tilde{g}_{vv'}=\tilde{\mu}_e\tilde{\O}_{vv'}$ that the sign $\mu_e$ is invariant under the parity transformation:
\be
\mu_e=\tilde{\mu}_e
\ee
when $e$ is a internal edge. In case $t_e$ is a boundary tetrahedron, the parity transformation changes the co-frame $E_\ell(v)\mapsto \tilde{E}_{\ell}(v)=\mathbf{P}E_\ell(v)$ at the vertex $v$, while leaves the boundary co-frame $E_\ell(e)$ invariant. Therefore the spin connection $\tilde{\O}_{ve}\in\text{SO(1,3)}$ is uniquely determined by
\be
\tilde{\O}_{ve}{E}_\ell(e)=\tilde{E}_\ell(v)\ \ \ \ \ \ell\subset t_e,
\ee 
Before the parity transformation, ${\O}_{ve}{E}_\ell(e)={E}_\ell(v)$ determines uniquely the spin connection ${\O}_{ve}$. Then the relation between $\tilde{\O}_{ve}$ and ${\O}_{ve}$ is given by
\be
\tilde{\O}_{ve}=\mathbf{P}{\O}_{e}\mathbf{T}\ \ \ \ \text{where}\ \ \ \mathbf{T}=\mathrm{diag}(-1,1,1,1)
\ee
by the fact that the co-frame vectors ${E}_\ell(e)$ are orthogonal to $(1,0,0,0)^t$ and both $\tilde{\O}_e$ and ${\O}_e$ belong to SO(1,3). Here the matrix $\mathbf{T}$ is a time-reversal in the Minkowski space, which leaves $E_\ell(e)$ invariant. Given a spatial vector $V^I$ orthogonal to $(1,0,0,0)^t$
\be
\tilde{g}(V^i\sig_i)\tilde{g}^\dagger=-\mathbf{P}g(V^i\sig_i)g^\dagger\ \ \ \ \text{i.e.}\ \ \ \ \tilde{g}V=-\mathbf{P}g V\ \ \ \text{in Spin-1 representation}
\ee
Let $V={E}_\ell(e)$, using $g_{ve}=\mu_e\O_{ve}$ in Spin-1 representation
\be
\tilde{g}_{ve}{E}_\ell(e)=-\mathbf{P}g_{ve}{E}_\ell(e)=-\mu_e\mathbf{P}{E}_\ell(v)=-\mu_e\tilde{E}_\ell(v)
\ee
Therefore we obtain from $\tilde{g}_{ve}{E}_\ell(e)=\tilde{\mu}_e\tilde{E}_\ell(v)$ that
\be
\mu_e=-\tilde{\mu}_e
\ee
for an edge connecting to the boundary. A boundary triangle is shared by exactly two boundary tetrahedra, in the dual language, a boundary face has exactly two edges connecting to the boundary. Thus the product $\prod_{e\subset\partial f}\mu_e$ is invariant under the parity transformaiton, i.e.
\be
\prod_{e\subset\partial f}\mu_e=\prod_{e\subset\partial f}\tilde{\mu}_e
\ee
for either a boundary face or an internal face. If we write $\mu_e=e^{i\pi n_e}$ and $\tilde{\mu}_e=e^{i\pi \tilde{n}_e}$, then we have
\be
\sum_{e\subset\partial f}n_e=\sum_{e\subset\partial f}\tilde{n}_e
\ee

We consider $\tilde{\O}_{f}(v)$ a loop holonomy of the spin connection along the boundary of an internal face $f$, based at the vertex $v$, which is constructed from a global parity configuration $(j_f, \tilde{g}_{ve}, \xi_{ef},\tilde{z}_{vf})$ with $\tilde{g}_{ve}\neq g_{ve}$ at all the vertices. It is different from the original $\O_f(v)$ by 
\be
\tilde{\O}_{f}(v)=\mathbf{P}^{}{\O}_{f}(v)\mathbf{P}
\ee
From Eq.\Ref{Ofv}, $\O_f(v)$ can be expressed in terms of the co-frame vectors $E_{\ell_1}(v), E_{\ell_2}(v)$ for the edges $\ell_1,\ell_2$ of the triangle $f$
\be
\O_f(v)&=&e^{i\pi\sum_en_e}e^{\frac{* E_{\ell_1}(v)\wedge E_{\ell_2}(v)}{|*E_{\ell_1}(v)\wedge E_{\ell_2}(v)|}\mathrm{sgn}(V_4)\Theta_f+\frac{E_{\ell_1}(v)\wedge E_{\ell_2}(v)}{|E_{\ell_1}(v)\wedge E_{\ell_2}(v)|}\pi\sum_{e}n_e}\nonumber\\
\tilde{\O}_f(v)&=&e^{i\pi\sum_e\tilde{n}_e}e^{\frac{* \tilde{E}_{\ell_1}(v)\wedge \tilde{E}_{\ell_2}(v)}{|*\tilde{E}_{\ell_1}(v)\wedge \tilde{E}_{\ell_2}(v)|}\mathrm{sgn}(\tilde{V}_4)\tilde{\Theta}_f+\frac{\tilde{E}_{\ell_1}(v)\wedge \tilde{E}_{\ell_2}(v)}{|\tilde{E}_{\ell_1}(v)\wedge \tilde{E}_{\ell_2}(v)|}\pi\sum_{e}\tilde{n}_e}
\ee
From the previous results $\mathrm{sgn}(\tilde{V}_4)=-\mathrm{sgn}({V}_4)$, $\sum_{e}n_e=\sum_{e}\tilde{n}_e$ and the relation $\mathbf{P}\otimes\mathbf{P}(* E_1\wedge E_2)=-*\mathbf{P}E_1\wedge\mathbf{P} E_2$, we obtain that 
\be
\Theta_f=\tilde{\Theta}_f
\ee
which is consistent with the fact that the deficit angle $\Theta_f$ is determined by the metric $g_{\ell_1\ell_2}$ which is invariant under the parity transformation.

For the holonomy $\O_f(e_1,e_0)$ for a boundary face $f$, under a global parity 
\be
\tilde{\O}_f(e_1,e_0)=\mathbf{T}\O_f(e_1,e_0)\mathbf{T}
\ee
Recall Eq.\Ref{Ofee}, we have for both $\tilde{\O}_f(e_1,e_0)$ and $\O_f(e_1,e_0)$
\be
\hat{g}(\xi_{e_1 f})^{-1}\O_{f}(e_1,e_0)\hat{g}(\xi_{e_0f})&=&e^{i\pi \sum_e n_e}e^{\eps\ \mathrm{sgn}({V}_4)\Theta^B_f K_3}e^{\pi\sum_e n_e J_3}\nonumber\\
\hat{g}(\xi_{e_1 f})^{-1}\tilde{\O}_{f}(e_1,e_0)\hat{g}(\xi_{e_0f})&=&e^{i\pi \sum_e \tilde{n}_e}e^{\eps\ \mathrm{sgn}(\tilde{V}_4)\tilde{\Theta}^B_f K_3}e^{\pi\sum_e \tilde{n}_e J_3}
\ee
Since $\mathbf{T}$ commutes with $\hat{g}(\xi_{ef})\in \text{SU(2)}$ and $\mathbf{T}K_3\mathbf{T}=-K_3,\mathbf{T}J_3\mathbf{T}=J_3$, we obtain that
\be
\Theta_f^B=\tilde{\Theta}_f^B
\ee
and consistent with the fact that the dihedral angle $\Theta^B_f$ is determined by the metric $g_{\ell_1\ell_2}$ which is invariant under the parity transformation. 

Before we come to the next section, we emphasize that given a Regge-like spin configuration $j_f$, there exists only two nondegenerate critical configurations $(j_f, {g}^c_{ve}, \xi_{ef},{z}^c_{vf})$ such that the oriented 4-volume has a constant sign on the triangulation, i.e. $\mathrm{sgn}(V_4(v))$ is a constant for all $\sig_v$. The existence can be shown in the following way: given a nondegenerate critical configuration $(j_f, {g}_{ve}, \xi_{ef},{z}_{vf})$, it determines a subdivision of the triangulation into sub-triangulations $\ck_1,\cdots,\ck_n$, where on each $\ck_i$, $\mathrm{sgn}(V_4(v))$ is a constant, but $\mathrm{sgn}(V_4(v))$ is not a constant for neighboring $\ck_i$ and $\ck_j$. However we can always make a parity transformation for all the simplices within some sub-triangulations, to flip the sign of the oriented 4-volume, such that $\mathrm{sgn}(V_4(v))$ is a constant on the entire triangulation. Any two nondegenerate solutions $(j_f, {g}_{ve}, \xi_{ef},{z}_{vf})$ are related by a (local) parity tranformation, which flips the sign of $V_4(v)$ at least locally. There exists two nondegenerate critical configurations $(j_f, {g}^c_{ve}, \xi_{ef},{z}^c_{vf})$ such that the oriented 4-volume has a constant sign on the entire triangulation, while the two configurations are related by a global parity transformation. If there was another nondegenerate critical configurations such that the oriented 4-volume has a constant sign on the entire triangulation, it must relate the existed two configurations by a local parity transformation, which flips $\mathrm{sgn}(V_4(v))$ only locally thus breaks the constancy of $\mathrm{sgn}(V_4(v))$.

\section{Asymptotics of Degenerate Amplitudes}

\subsection{Degenerate Critical Configurations}

The previous discussions of the critical configuration and asymptotic formula are under the nondegenerate assumption:
\be
\prod_{e_1,e_2,e_3,e_4=1}^5\det\Big(N_{e_1}(v),N_{e_2}(v),N_{e_3}(v),N_{e_4}(v)\Big)\neq0
\ee
where $N_e(v)=g_{ve}(1,0,0,0)^t$, i.e. any four of the five normal vectors $N_e(v)$ form a linearly independent set and span the 4-dimensional Minkowski space.

Now we consider a degenerate critical configuration $(j_f,g_{ve},\xi_{ef},z_{vf})$ that solves the critical equations Eqs.\Ref{gluingJ}, \Ref{gluing}, and \Ref{closure}, but violates the above nondegenerate assumption at \emph{all} vertices on a triangulation (with boundary). if we assume the nondegeneracy of the tetrahedra, i.e. given a tetrahedron $t_e$, the 4 vectors $\hat{n}_{ef}$ obtained from the spinors $\xi_{ef}$ span a 3-dimensional subspace, then the Lemma 3 in the first reference of \cite{semiclassical} shows that within each 4-simplex, all five normals $N_e(v)$ from the degenerate critical configuration $(j_f,g_{ve},\xi_{ef},z_{vf})$ are parallel and more precisely $N_e(v)=u=(1,0,0,0)$\footnote{Recall that we have fixed $g_{ve_5}=1$ to make the vertex amplitude finite. }. By definition $N_e(v)=g_{ve}(1,0,0,0)^t$, we find that all the group variables $g_{ve}\in\text{SU(2)}$ for a degenerate critical configuration $(j_f,g_{ve},\xi_{ef},z_{vf})$. For the bivectors $*\!X_{f}(v)$, they are all orthogonal to the same unit vector $u=(1,0,0,0)$.

From $*\!X_{f}(v)\cdot u=0$, we can write the bivector $X_f(v)=V_f(v)\wedge u$ for a vector $V_f(v)$ orthogonal to $u$. The vector $V_f(v)$ can be determined by the parallel transportation $X_f(v)=g_{ve}X_{ef}g_{ev}$ and $X_{ef}=2 \g j_f\hat{n}_{ef}\wedge u$, thus
\be
V_f(e)= 2\g j_f\hat{n}_{ef}\ \ \ \ \ V_f(v)=2\g j_f\ g_{ve}\hat{n}_{ef}
\ee
The above relation doesn't depend on the choice of $e$ (recall Proposition \ref{fromcritical}). From the closure condition Eq.\Ref{closure}, we have
\be
\sum_{f\subset t_e}\eps_{ef}(v)V_f(v)=0\label{closureV}
\ee
Therefore a degenerate critical configuration $(j_f,g_{ve},\xi_{ef})$ assign uniquely a spatial vector $V_f(v)\bot u$ at the vertex $v$ for each triangle $f$, satisfying the closure condition Eq.\Ref{closureV}. The collection of the vectors $V_f(v)$ is referred as a \emph{vector geometry} in \cite{semiclassical}.

Since $g_{ev}\in\text{SU(2)}$ in the degenerate critical configuration $(j_f,g_{ve},\xi_{ef})$, we have immediately $\frac{||Z_{ve'f}||}{||Z_{vef}||}=1$. Then for each face action $S_f$ (internal face or boundary face)
\be
S_f=2i\g j_f\sum_{v}\ln\frac{||Z_{ve'f}||}{||Z_{vef}||}-2i j_f\sum_{v}\phi_{eve'}=-2i\ j_f\sum_{v}\phi_{eve'}
\ee
In the same way as we did for the nondegenerate amplitude, we make use of Eqs.\Ref{gluingJ} and \Ref{gluing}, which now take the following forms
\be
g_{ve}\left( J\xi_{ef}\right)  &=&e^{-i\phi _{eve^{\prime
}}}g_{ve^{\prime }}\left( J\xi_{e^{\prime }f}\right)  \nonumber
\\
g_{ve}\xi_{ef} &=&e^{i\phi _{eve^{\prime }}}g_{ve^{\prime
}}\xi_{e^{\prime }f}  \label{degcritical}
\ee
First of all, for a internal face $f$, we again consider the successive actions on $\xi_{ef}$ of $g_{e'v}g_{ve}$ around the entire boundary of the face $f$,
\be
\overleftarrow{\prod_{v\in\partial f}}g_{e'v}g_{ve}J\xi_{ef}&=&e^{-i\sum_v\phi_{eve'}}J\xi_{ef}\nonumber\\
\overleftarrow{\prod_{v\in\partial f}}g_{e'v}g_{ve}\xi_{ef}&=&e^{+i\sum_v\phi_{eve'}}\xi_{ef}
\ee
where $g_{ve}\in\text{SU(2)}$. In the same way as we did for the nondegenerate case, the above equations imply that for the loop holonomy $G_f(e)=\overleftarrow{\prod}_{v\in\partial f}g_{e'v}g_{ve}$,
\be
G_f(e)=\exp\lt[i\sum_{v\in\partial f}\phi_{eve'}\vec{\sig}\cdot \hat{n}_{ef}\rt].\label{Gf1}
\ee
For a boundary face $f$, again in the same way as we did for the nondegenerate case, we obtain
\be
G_f(e_1,e_0)=g(\xi_{e_1f})\ e^{i\sum_{v}\phi_{eve'}\vec{\sig}\cdot\hat{z}}\ g(\xi_{e_0f})^{-1}.\label{Gf2}
\ee
We then need to determine the physical interpretation of the angle $\sum_{v\in\partial f}\phi_{eve'}$ in different cases.

Recall the degenerate critical equations Eq.\Ref{degcritical} together with the closure condtion Eq.\Ref{closure}, we find they are essentially the same as the critical equations in \cite{HZ} for a Euclidean spinfoam amplitude:
\be
g_{ve}^\pm\left( J\xi_{ef}\right)  &=&e^{-i\phi^\pm_{eve^{\prime
}}}g_{ve^{\prime }}^\pm\left( J\xi_{e^{\prime }f}\right)  \nonumber
\\
g_{ve}^\pm\xi_{ef} &=&e^{i\phi^\pm_{eve^{\prime }}}g^\pm_{ve^{\prime
}}\xi_{e^{\prime }f} \nonumber\\
 0&=&\sum_{f\subset t_e}^{4}\eps_{ef}(v)j_{f}\hat{n}_{ef}\label{Eucl}
\ee
where the equations for self-dual or anti-self-dual sector are essentially the same, and both of them are the same as the above degenerate critical equation for Lorentzian amplitude. Therefore given a degenerate critical configuration $(j_f,g_{ve},\xi_{ef},z_{vf})$ for the Lorentzian amplitude, there exists a critical configuration $(j_f,g^{\pm}_{ve},\xi_{ef})$ for the Euclidean amplitude in \cite{HZ}, such that $g_{ve}=g^+_{ve}$. In the following, we classify the degenerate Lorentzian critical configurations into two type (type A and type B) and discuss the uniqueness of the corresponding Euclidean critical configurations:

\begin{description}

\item[Type-A configuration:] A degenerate Lorentzian critical configuration $(j_f,g_{ve},\xi_{ef},z_{vf})$ corresponds to an Euclidean critical configuration $(j_f,g^{\pm}_{ve},\xi_{ef})$, which is nondegenerate at each 4-simplex $\sig_v$ of the triangulation, i.e. any four of the five normals $N_e(v)=(g_{ve}^+,g_{ve}^-)\vartriangleright(1,0,0,0)^t$ span a 4-dimensional vector space. Since the Euclidean spins $j_f$ and spinors $\xi_{ef}$ are uniquely specified by the Lorentzian configuration $(j_f,g_{ve},\xi_{ef},z_{vf})$, we only need to consider how many solutions $(g^+_{ve},g^-_{ve})$ in Eq.\Ref{Eucl} if the variables $j_f$ and $\xi_{ef}$ are fixed. It is shown in \cite{semiclassical} that for a 4-simplex $\sig_v$, there are only two solutions in the nondegenerate case\footnote{The notion of nondegenercy here is different from the notion in \cite{semiclassical}. In the Lemma 4 of the first reference of \cite{semiclassical}, there are 4 solutions in a 4-simplex $(g_{ve}^1,g^2_{ve}),\ (g_{ve}^2,g^1_{ve}),\ (g_{ve}^1,g^1_{ve}),\ (g_{ve}^2,g^2_{ve})$ for the nondegenerate case (in the sense of \cite{semiclassical}). However the two solutions $(g_{ve}^1,g^1_{ve}),\ (g_{ve}^2,g^2_{ve})$ are degenerate in our notion of degeneracy. }
\be
(g_{ve}^+,g^-_{ve})=(g_{ve}^1,g^2_{ve}) \ \ \ \ \text{and}\ \ \ \ (g_{ve}^+,g^-_{ve})=(g_{ve}^2,g^1_{ve})
\ee
Then the correspondence $g_{ve}=g^+_{ve}$ fix uniquely a solution $(g_{ve}^+,g^-_{ve})$ for the Euclidean critical configuration $(j_f,g^{\pm}_{ve},\xi_{ef})$.

\item[Type-A configuration:] The degenerate Lorentzian critical configuration $(j_f,g_{ve},\xi_{ef},z_{vf})$ could always correspond to a degenerate Euclidean critical configuration $(j_f,g^{\pm}_{ve},\xi_{ef})$ with $g^+_{ve}=g^-_{ve}$ by $(g_{ve}^+,g^-_{ve})=(g_{ve},g_{ve})$, even the data $j_f$ and $\xi_{ef}$ can have two nondegenerate solutions as above. Then in this case, we alway make the above nondegenerate choice as the canonical choice.

\item[Type-B configuration:] The data $j_f$ and $\xi_{ef}$ in a degenerate Lorentzian critical configuration $(j_f,g_{ve},\xi_{ef},z_{vf})$ lead to only one Euclidean solutions $(g_{ve},g_{ve})\in \text{SO(4)}$ for Eq.\Ref{Eucl} in each 4-simplex $\sig_v$. Then the Euclidean configuration $(j_f,g_{ve}^\pm,\xi_{ef})$ is degenerate in $\sig_v$ in the sense of \cite{semiclassical}. Then obviously the correspondence is unique by $g_{ve}\mapsto (g_{ve},g_{ve})$.

\end{description}

\subsection{Type-A Degenerate Critical Configuration: Euclidean Geometry}

First of all, we consider a type A degenerate Lorentzian critical configuration $(j_f,g_{ve},\xi_{ef},z_{vf})$ on the triangulation (with boundary). The corresponding Euclidean critical configuration $(j_f,g^{\pm}_{ve},\xi_{ef})$ is nondegenerate everywhere. We can construct a nondegenerate discrete Euclidean geometry on the triangulation such that (see \cite{HZ}, see also \cite{CF})

\begin{itemize}

\item An Euclidean co-tetrad $E_\ell(v),E_\ell(e)$ of the triangulation (bulk and boundary) can be constructed from $(j_f,g^{\pm}_{ve},\xi_{ef})$, unique up to a sign fliping $E_\ell\to -E_\ell$, such that the spins $j_f$ satisfies
\be
2\g j_f=\lt|E_{\ell_1}(v)\wedge E_{\ell_2}(v)\rt|.
\ee
From the co-tetrad we can construct a unique discrete metric with Euclidean signature on the whole triangulation (bulk and boundary)
\be
{}^E\! g_{\ell_1\ell_2}(v)=\delta_{IJ}E^I_{\ell_1}(v)E^J_{\ell_2}(v)\ \ \ \ \ {}^E\!g_{\ell_1\ell_2}(e)=\delta_{IJ}E^I_{\ell_1}(e)E^J_{\ell_2}(e).
\ee
So $\g j_f$ is the triangle area from the discrete metric ${}^E\!g_{\ell_1\ell_2}$.

\item For the bivectors in the bulk,
\be
 j_f(g_{ve}^+,g_{ve}^-)(\hat{n}_{ef},\hat{n}_{ef})=\eps*\!E_{\ell_1}(v)\wedge E_{\ell_2}(v)
\ee
For the bivector on the boundary
\be
 j_f(\hat{n}_{ef},\hat{n}_{ef})=\eps*\!E_{\ell_1}(e)\wedge E_{\ell_2}(e)
\ee
where $\eps$ is a global sign on the entire triangulation. If the triangulation has boundary, the sign factor $\eps$ is specified by the orientation of the boundary triangulation, i.e. $\eps=\mathrm{sgn}(V_3)$ for the boundary tetrahedra. 

\item The SO(4) group variable $(g_{e}^+,g_{e}^-)$ equals to the Euclidean spin connection ${}^E\!\O_{e}$ compatible with $E_\ell(v)$, up to a sign $\mu_e=e^{i\pi n_e}$ ($n_e=0,1$), i.e.
\be
(g_{e}^+,g_{e}^-)=\mu_e{}^E\!\O_{e}
\ee
in the Spin-1 representation. Here ${}^E\!\O_{e}\in\text{SO(4)}$ is compatible with the co-frame $E_\ell(v), E_\ell(e)$
\be
({}^E\!\O_{vv'})^I_{\ J}E_\ell^J(v')=E_\ell^I(v)\ \ \ \ \text{and}\ \ \ \ ({}^E\!\O_{ve})^I_{\ J}E_\ell^J(e)=E_\ell^I(v)
\ee
If $\mathrm{sgn}(V_4(v))=\mathrm{sgn}(V_4(v'))$, $\O_{vv'}$ is the unique discrete spin connection compatible with the co-frame. In addition, we note that each $\mu_e$ is not invariant under the sign flipping $E_\ell\to -E_\ell$, but the product $\prod_{e\subset\partial f}\mu_e$ is invariant for any (internal or boundary) face $f$ (see Lemma.\ref{prodmue}).

\end{itemize}
Therefore in this way, a type-A degnerate \emph{Lorentzian} critical configuration determines uniquely a triple of (Euclidean) variables $({}^E\!g_{\ell_1\ell_2},n_e,\eps)$ corresponding to a \emph{Euclidean} Geometry and two types of sign factors.

Given a nondegenerate Euclidean critical configuration $(j_f,g^{\pm}_{ve},\xi_{ef})$, in the same way as the nondegenerate Lorentzian critical configuration, it determines a subdivision of the triangulation into sub-triangulations (with boundaries) $\ck_1,\cdots,\ck_n$, on each of the sub-triangulation, the sign of the oriented 4-volume $\mathrm{sgn}(V_4(v))$ is a constant.

Now we discuss the spinfoam amplitude at a Type-A degenerate configuration, while we restrict our attention into a sub-triangulation $\ck_i$ where $\mathrm{sgn}(V_4(v))$ is a constant. For a internal face $f$, it is shown in \cite{HZ} that the loop holonomy along the boundary of $f$ is given by
\be
\lt(G_f^+(e),G_f^-(e)\rt)=\lt(e^{\frac{i}{2}\lt[\eps\ \mathrm{sgn}(V_4){}^E\!\Theta_f+\pi\sum_e n_e\rt]\vec{\sig}\cdot \hat{n}_{ef}},e^{-\frac{i}{2}\lt[\eps\ \mathrm{sgn}(V_4){}^E\!\Theta_f-\pi\sum_e n_e\rt]\vec{\sig}\cdot \hat{n}_{ef}}\rt)
\ee
where ${}^E\!\Theta_f$ is the deficit angle from the Euclidean spin connection compatible with the metric ${}^E\!g_{\ell_1\ell_2}$. By the above identification $g_{ve}=g^+_{ve}$ between the degenerate Lorentzian critical configuration $(j_f,g_{ve},\xi_{ef},z_{vf})$ and a nondegenerate Euclidean critical configuration $(j_f,g^{\pm}_{ve},\xi_{ef})$. We obtain that for the degenerate Lorentzian critical configuration, the loop holonomy $G_f(e)=G^+_f(e)$. Comparing with Eq.\Ref{Gf1},
\be
\sum_{v\in\partial f}\phi_{eve'}=\frac{1}{2}\lt[\eps\ \mathrm{sgn}(V_4){}^E\!\Theta_f+\pi\sum_e n_e\rt]
\ee
Therefore the angle $\sum_{v\in\partial f}\phi_{eve'}$ has the physical meaning as a deficit angle in a corresponding Euclidean geometry. Then the face action (as a function of $({}^E\!g_{\ell_1\ell_2},n_e,\eps)$) reads
\be
S_f({}^E\!g_{\ell_1\ell_2},n_e,\eps)=-i\eps\ \mathrm{sgn}(V_4)\ j_f{}^E\!\Theta_f-i\pi\sum_e n_e j_f
\ee
for a internal face $f$.

For a boundary face $f$, we have the path holonomy along its internal boundary $p_{e_1e_0}$ is given by
\be
&&\lt(G_f^+(e_1,e_0),G_f^-(e_1,e_0)\rt)\nonumber\\
&=&\lt(g(\xi_{e_1f})\ e^{\frac{i}{2}\lt[\eps\ \mathrm{sgn}(V_4){}^E\!\Theta^B_f+\pi\sum_e n_e\rt]{\sig}_3}\ g(\xi_{e_0f})^{-1},g(\xi_{e_1f})\ e^{-\frac{i}{2}\lt[\eps\ \mathrm{sgn}(V_4){}^E\!\Theta^B_f-\pi\sum_e n_e\rt]{\sig}_3}\ g(\xi_{e_0f})^{-1}\rt)
\ee
where ${}^E\!\Theta^B_f$ is the dihedral angle (determined by the metric ${}^E\!g_{\ell_1\ell_2}$) between two boundary tetrahedra $t_{e_0},t_{e_1}$ at the triangle $f$ shared by them. The degenerate Lorentzian critical configuration $G_f(e_1,e_0)$ is identify with $G_f^+(e_1,e_0)$ here. Comparing to Eq.\Ref{Gf2} we obtain that
\be
\sum_{v\in p_{e_1e_0}}\phi_{eve'}=\frac{1}{2}\lt[\eps\ \mathrm{sgn}(V_4){}^E\!\Theta^B_f+\pi\sum_e n_e\rt]
\ee
Therefore the face action $S_f$ for a boundary face $f$ is given by
\be
S_f({}^E\!g_{\ell_1\ell_2},n_e,\eps)=-i\eps\ \mathrm{sgn}(V_4)\ j_f{}^E\!\Theta^B_f-i\pi\sum_e n_e j_f.
\ee

As a result, at a type-A degenerate critical configuration (restricted to a sub-triangulation $\ck_i$), the Lorentzian spinfoam action $S$ is a function of the variables $({}^E\!g_{\ell_1\ell_2},n_e,\eps)$ and behaves mainly as an Euclidean Regge action:
\be
S({}^E\!g_{\ell_1\ell_2},n_e,\eps)\Big|_{\ck_i}&=&\sum_{f\ \text{internal}}S_{f}({}^E\!g_{\ell_1\ell_2},n_e,\eps)+\sum_{f\ \text{boundary}}S_{f}({}^E\!g_{\ell_1\ell_2},n_e,\eps)\nonumber\\
&=&\lt[-i\ \eps\ \mathrm{sgn}(V_4)\sum_{\text{internal}\ f} j_f{}^E\!\Theta_f-i\ \eps\ \mathrm{sgn}(V_4) \sum_{\text{boundary }f} j_f{}^E\!\Theta^B_f-i\pi \sum_{e}n_e\sum_{f\subset t_e} j_f\rt]_{\ck_i}
\ee
where we note that the areas $\g j_f$, deficit angles ${}^E\!\Theta_f$, and dihedral angles ${}^E\!\Theta^B_f$ are uniquely determined by the discrete metric $g_{\ell_1\ell_2}$. Moreover for each tetrahedron $t$, the sum of face spins $\sum_{f\subset t}j_f$ is an integer. For half-integer spins, $e^{-i\pi \sum_{e}n_e\sum_{f\subset t_e}j_f}=\pm1$ gives an overall sign factor. Therefore in general at a type-A degenerate critical configuration $(j_f,g_{ve},\xi_{ef},z_{vf})$ for Lorentzian amplitude,
\be
e^{\l S}\Big|_{\ck_i}=\pm\exp\l\lt[-i\ \eps\ \mathrm{sgn}(V_4)\sum_{\text{internal}\ f}  j_f{}^E\!\Theta_f-i\ \eps\ \mathrm{sgn}(V_4)\sum_{\text{boundary }f} j_f{}^E\!\Theta^B_f\rt]_{\ck_i}.
\ee
Again there exists two ways to make the overall sign factor disappear: (1) only consider integer spins $j_f$, or (2) modify the embedding from SU(2) unitary irreps to $\Slc$ unitary irreps by $j_f\mapsto (p_f,k_f):=(2\g j_f,2j_f)$, then the spinfoam action $S$ is replaced by $2S$. In these two cases the exponential $e^{\l S}$ at the critical configuration is independent of the variable $n_e$.

According to the properties of Euclidean Regge geometry, given a collection of (Euclidean) Regge-like areas $\g j_f$, the discrete Euclidean metric ${}^E\!g_{\ell_1\ell_2}(v)$ is uniquely determined at each vertex $v$. Furthermore since the areas $\g j_f$ are Regge-like, There exists a discrete Euclidean metric ${}^E\!g_{\ell_1\ell_2}$ in the entire bulk of the triangulation, such that the neighboring 4-simplicies are consistently glued together, as we constructed in \cite{HZ}. This discrete metric ${}^E\!g_{\ell_1\ell_2}$ is obviously unique by the uniqueness of $g_{\ell_1\ell_2}(v)$. Therefore given the partial-amplitude $A_{j_f}(\ck)$ in Eq.\Ref{Aj} with a specified Euclidean Regge-like $j_f$, all the degenerate critical configurations $(j_f,g_{ve},\xi_{ef},z_{vf})$ of type-A corresponds to the same discrete Euclidean metric ${}^E\!g_{\ell_1\ell_2}$, provided a Regge boundary data. Any two type-A critical configurations $(j_f,g_{ve},\xi_{ef},z_{vf})=(j_f,g^\pm_{ve},\xi_{ef})$ with the same $j_f$ are related by local or global parity transformation in the Euclidean theory, see \cite{HZ}, similar to the Lorentzian nondegenerate case.

As a result, given an Euclidean Regge-like spin configurations $j_f$ and a Regge boundary data, the degenerate critical configurations of type-A give the following asymptotics
\be
A_{j_f}(\ck)\big|_{\text{Deg-A}}&\sim&\sum_{x_c}a(x_c)\lt(\frac{2\pi}{\l}\rt)^{\frac{r(x_c)}{2}-N(v,f)}\frac{e^{i\mathrm{Ind}H'(x_c)}}{\sqrt{|\det_r H'(x_c)|}}\lt[1+o\lt(\frac{1}{\l}\rt)\rt]\times\nonumber\\
&&\times\prod_{i=1}^{n(x_c)} \exp-i\l\lt[\eps\ \mathrm{sgn}(V_4)\sum_{\text{internal}\ f} j_f{}^E\!\Theta_f+\eps\ \mathrm{sgn}(V_4)\sum_{\text{boundary }f} j_f{}^E\!\Theta^B_f+\pi \sum_{e}n_e\sum_{f\subset t_e} j_f\rt]_{\ck_{i}(x_c)}
\ee 
where $x_c= (j_f,g_{ve},\xi_{ef},z_{vf})=(j_f,g^\pm_{ve},\xi_{ef})$ labels the degenerate critical configurations of type-A,  $r(x_c)$ is the rank of the Hessian matrix at $x_c$, and $N(v,f)$ is the number of the pair $(v,f)$ with $v\in\partial f$ (recall Eq.\Ref{Aj}, there is a factor of $\dim(j_f)$ for each pair of $(v,f)$). $a(x_c)$ is the evaluation of the integration measures at $x_c$, which doesn't scale with $\l$. Here ${}^E\!\Theta_f$ and ${}^E\!\Theta^B_f$ only depend on the Euclidean metric ${}^E\!g_{\ell_1\ell_2}$, which is uniquely determined by the Euclidean Regge-like spin configuration $j_f$ and the Regge boundary data.

\subsection{Type-B Degenerate Critical Configuration: Vector Geometry}

Given a type-B degenerate Lorentzian critical configuration $(j_f,g_{ve},\xi_{ef},z_{vf})$, the data $\xi_{ef}$ lead to only one Euclidean solution $(g_{ve},g_{ve})\in \text{SU(2)}\times \text{SU(2)}$ for Eq.\Ref{Eucl} in each 4-simplex $\sig_v$. Then the Euclidean configuration $(j_f,g_{ve}^\pm,\xi_{ef})$ is degenerate in $\sig_v$ in the sense of \cite{semiclassical}. Therefore there is no nondegenerate geometric interpretation of a type-B degenerate Lorentzian critical configuration $(j_f,g_{ve},\xi_{ef},z_{vf})$. It can only be interpreted as a vector geometry in terms of $V_f(v),V_f(e)$ on the triangulation (bulk and boundary), where all the vectors $V_f(v),V_f(e)$ are orthogonal to the unit time-like vector $u=(1,0,0,0)^t$, and $|V_f(v)|=|V_f(e)|=2\g j_f$.  The vectors $V_f(v),V_f(e)$ are uniquely determined by $j_f$ and $\xi_{ef}$ by $V_f(e)= 2\g j_f\hat{n}_{ef}$ and $V_f(v)=2\g j_f\ g_{ve}\hat{n}_{ef}$, since the group variable $g_{ve}$ is uniquely determined by $\xi_{ef}$.  We have the parallel transportation using the Spin-1 representation of $g_{ve}$
\be
g_{vv'}\rhd V_f(v')=V_f(v)\ \ \ \ \text{and}\ \ \ \ g_{ve}\rhd V_f(e)=V_f(v)
\ee
for all triangles $f$ in the tetrahedron $t_e$ (shared by $v,v'$ if not a boundary tetrahedron). Then the unique group variables $g_{vv'},g_{ve}\in \text{SU(2)}$ are said to be compatible with the vector geometry $V_f(v),V_f(e)$. Therefore a type-B degenerate Lorentzian critical configuration $(j_f,g_{ve},\xi_{ef})$ determine uniquely a vector geometry $V_f(v),V_f(e)$. Conversely, given a vector geometry $V_f(v),V_f(e)$, it uniquely determine the SU(2) group variables $g_{ve}$ up to a sign $e^{i\pi n_e}$, due to the 2-to-1 correspondence between SU(2) and SO(3). 

Since we have shown from the critical point equations that
\be
G_f(e)=e^{i\sum_{v}\phi_{eve'}\vec{\sig}\cdot \hat{n}_{ef}}\ \ \ \ \
G_f(e_1,e_0)=g(\xi_{e_1f})\ e^{i\sum_{v}\phi_{eve'}\vec{\sig}\cdot\hat{z}}\ g(\xi_{e_0f})^{-1},
\ee
the above SU(2) angle $\sum_{v}\phi_{eve'}$ is determined uniquely by the group variables $g_{ve}$ (which is uniquely compatible with the vector geometry $V_f(v),V_f(e)$ up to a sign $e^{i\pi n_e}$)
\be
\sum_{v\in\partial f}\phi_{eve'}=\half\Phi_f+\pi\sum_{e\subset\partial f}n_e\ \ \ \ \text{and}\ \ \ \ \sum_{v\in p_{e_1e_0}}\phi_{eve'}=\half\Phi^B_f+\pi\sum_{e\subset p_{e_1e_0}}n_e
\ee
respectively for a internal face and a boundary face, where the SO(3) angle $\Phi_f$ is uniquely determined by the vector geometry $V_f$ only (the factor $\frac{1}{2}$ shows the relation between an SU(2) angle and SO(3) angle). Therefore for the face action (internal face and boundary face)
\be
S_f(V_f,n_e)=i\ j_f\Phi_f-2i\pi\sum_{e\subset\partial f}n_ej_f\ \ \ \ \text{and}\ \ \ \ S_f(V_f,n_e)=i\ j_f\Phi_f^B-2i\pi\sum_{e\subset\partial f}n_ej_f
\ee

As a result, at a type-B degenerate critical configuration, the Lorentzian spinfoam action $S$ is a function of the variables $(V_f,n_e)$:
\be
S(V_f,n_e)
=-i \sum_{\text{internal}\ f} j_f\Phi_f-i\sum_{\text{boundary }f} j_f\Phi^B_f-2\pi i \sum_{e\subset\partial f}n_e\sum_{f\subset t_e} j_f
\ee
Moreover for each tetrahedron $t$, the sum of face spins $\sum_{f\subset t}j_f$ is an integer. Therefore in general at a type-B degenerate critical configuration $(j_f,g_{ve},\xi_{ef},z_{vf})$ for Lorentzian amplitude, $e^{\l S}$ is a function of vector geometry $V_f$ only:
\be
e^{\l S}=\exp\l\lt[-i \sum_{\text{internal}\ f} j_f\Phi_f-i\sum_{\text{boundary }f} j_f\Phi^B_f\rt].
\ee
where the area $\g j_f=\half|V_f|$ and the angle $\Phi_f$ is uniquely determined by the vector geometry $V_f$.

As a result, given an spin configurations $j_f$ and a boundary data that admit a vector geometry on the triangulation, the degenerate critical configurations of type-B give the following asymptotics
\be
A_{j_f}(\ck)\big|_{\text{Deg-B}}&\sim&\sum_{x_c}a(x_c)\lt(\frac{2\pi}{\l}\rt)^{\frac{r(x_c)}{2}-N(v,f)}\frac{e^{i\mathrm{Ind}H'(x_c)}}{\sqrt{|\det_r H'(x_c)|}}\lt[1+o\lt(\frac{1}{\l}\rt)\rt]\times\nonumber\\
&&\times\ \exp\l\lt[-i \sum_{\text{internal}\ f} j_f\Phi_f-i\sum_{\text{boundary }f} j_f\Phi^B_f\rt]
\ee 
where $x_c\equiv (j_f,g_{ve},\xi_{ef},z_{vf})$ labels the degenerate critical configurations of type-B. Note that if we make a suitable gauge fixing for the boundary data, we can always set $\Phi_f^B=0$ \cite{semiclassical}.

\section{Transition between Lorentzian, Euclidean and Vector Geometry}

\begin{figure}[h]
\begin{center}
\includegraphics[width=9cm]{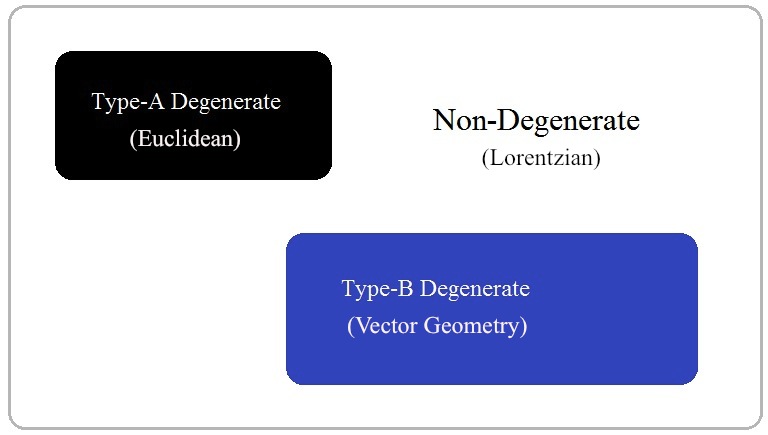}
\caption{A generic critical configuration with both nondegenerate and degenerate configurations. The critical configuration is non-degenerate in the white region, thus corresponds to a non-degenerate Lorentzian discrete geometry. The critical configuration is degenerate in type-A in the black region, thus corresponds to a non-degenerate Euclidean discrete geometry. The critical configuration is degenerate in type-B in the blue region, thus corresponds to a vector geometry. All the three regions are the triangulations with boundaries. }
\label{transition}
\end{center}
\end{figure}

All the previous analysis assume that on the \emph{entire} triangulation, the critical configuration $(j_f,g_{ve},\xi_{ef},z_{vf})$ is one of the three types: nondegenerate, degenerate of type-A or degenerate of type-B. However they are not the most general case. In principle one should admit the critical configuration that mixes the three types on the triangulation: Given a most general critical configuration $(j_f,g_{ve},\xi_{ef},z_{vf})$ that mixes the three types, one can always make a partition of the triangulation into three regions (maybe disconnected regions) $\calr_{\text{Nondeg}},\calr_{\text{Deg-A}},\calr_{\text{Deg-B}}$. Each of the three regions $\calr_{*}$, $*=\text{Nondeg},\text{Deg-A},{\text{Deg-B}}$ is a triangulation with boundary, on which the critical configuration $(j_f,g_{ve},\xi_{ef},z_{vf})_{\calr_{*}}$ is of single type $*=\text{Nondeg},\text{Deg-A},{\text{Deg-B}}$. See FIG.\ref{transition} for an illustration.

Therefore for a generic spin configuration $j_f$, the asymptotics of the partial amplitude $A_{j_f}(\ck)$ is given by
\be
A_{j_f}(\ck)\sim\sum_{x_c}a(x_c)\lt(\frac{2\pi}{\l}\rt)^{\frac{r(x_c)}{2}-N(v,f)}\frac{e^{i\mathrm{Ind}H'(x_c)}}{\sqrt{|\det_r H'(x_c)|}}\lt[1+o\lt(\frac{1}{\l}\rt)\rt]\ca_{j_f}(\calr_{\text{Nondeg}})\ca_{j_f}(\calr_{\text{Deg-A}})\ca_{j_f}(\calr_{\text{Deg-B}})\label{AAA}
\ee 
where $x_c$ labels the general critical configuration $(j_f,g_{ve},\xi_{ef},z_{vf})$ admitted by the spin configuration $j_f$ and boundary data, and $(j_f,g_{ve},\xi_{ef},z_{vf})$ determines the regions $\calr_{*}$, $*=\text{Nondeg},\text{Deg-A},{\text{Deg-B}}$ such that $(j_f,g_{ve},\xi_{ef},z_{vf})_{\calr_{*}}$ is of single type. The amplitudes $\ca_{j_f}(\calr_{\text{Nondeg}}), \ca_{j_f}(\calr_{\text{Deg-A}}), \ca_{j_f}(\calr_{\text{Deg-B}})$ are given respectively by
\be
\ca_{j_f}(\calr_{\text{Nondeg}})&=&\prod_{i=1}^{n(x_c)}\exp-i\l\lt[\eps\ \mathrm{sgn}(V_4) \sum_{\text{internal}\ f} \g j_f\Theta_f+\eps\ \mathrm{sgn}(V_4) \sum_{\text{boundary }f} \g j_f\Theta^B_f+\pi \sum_{e}n_e\sum_{f\subset t_e} j_f\rt]_{\calr_{\text{Nondeg}},\ck_i(x_c)}\nonumber\\
\ca_{j_f}(\calr_{\text{Deg-A}})&=&\prod_{j=1}^{n'(x_c)}\exp-i\l\lt[\eps\ \mathrm{sgn}(V_4) \sum_{\text{internal}\ f} j_f{}^E\!\Theta_f+\eps\ \mathrm{sgn}(V_4) \sum_{\text{boundary }f} j_f{}^E\!\Theta^B_f+\pi \sum_{e}n_e\sum_{f\subset t_e} j_f\rt]_{\calr_{\text{Deg-A}},\ck'_j(x_c)}\nonumber\\
\ca_{j_f}(\calr_{\text{Deg-B}})&=&\exp-i\l\lt[\sum_{\text{internal}\ f} j_f\Phi_f+\sum_{\text{boundary }f} j_f\Phi^B_f\rt]_{\calr_{\text{Deg-B}}}
\ee
As we discussed previously, given a general critical configuration $(j_f,g_{ve},\xi_{ef},z_{vf})$, the regions $\calr_{\text{Nondeg}}$ and $\calr_{\text{Deg-A}}$ should be respectively divided into sub-triangulations $\ck_1,\cdots,\ck_{n(x_c)}$ and $\ck'_1,\cdots,\ck'_{n(x_c)}$, such that in each $\ck_i$ or $\ck_i'$, $\mathrm{sgn}(V_4)$ is a constant.

Interestingly, from Eq.\Ref{AAA} we find an transition between a nondegenerate Lorentzian geometry and a nondegenerate Euclidean geometry through the boundary shared by $\calr_{\text{Nondeg}}$ and $\calr_{\text{Deg-A}}$. In $\calr_{\text{Nondeg}}$ the asymptotics gives a Regge action in Lorentzian signature (plus an additional term):
\be
S_{\text{Nondeg}}=-i\ \eps\ \mathrm{sgn}(V_4) \sum_{\text{internal}\ f} A_f\Theta_f-i\ \eps\ \mathrm{sgn}(V_4) \sum_{\text{boundary }f} A_f\Theta^B_f-\frac{i\pi}{\g} \sum_{e}n_e\sum_{f\subset t_e} A_f
\ee
where we set the physical area $A_f=\g j_f$ (in Planck unit). In $\calr_{\text{Deg-A}}$ the asymptotics gives a Euclidean Regge action divided by the Barbero-Immirzi parameter (plus an additional term)
\be
S_{\text{Deg-A}}=-\frac{i}{\g}\ \eps\ \mathrm{sgn}(V_4) \sum_{\text{internal}\ f} A_f{}^E\!\Theta_f-\frac{i}{\g}\ \eps\ \mathrm{sgn}(V_4) \sum_{\text{boundary }f} A_f{}^E\!\Theta^B_f-\frac{i\pi}{\g} \sum_{e}n_e\sum_{f\subset t_e} A_f
\ee
In the case of a single simplex, this asymptotics has been presented in \cite{semiclassical}. One might expect the transition between Lorentzian and Euclidean geometry is a quantum tunneling effect. But surprisingly in the large-j regime $e^{ S_{\text{Deg-A}}}$ is not damping exponentially but oscillatory. Similarly there is also a transition between a nondegenerate Lorentzian/Euclidean geometry and a vector geometry through the boundary of $\calr_{\text{Deg-B}}$, and in the region $\calr_{\text{Deg-B}}$, the asymptotics give 
\be
S_{\text{Deg-B}}=-\frac{i}{\g} \sum_{\text{internal}\ f} A_f\Phi_f-\frac{i}{\g}\sum_{\text{boundary }f} A_f\Phi^B_f
\ee
Thus $e^{ S_{\text{Deg-B}}}$ is also oscillatory and gives nontrivial transition in the large-j regime. However there are some specialities for the phases $e^{ S_{\text{Deg-A}}}$, $e^{S_{\text{Deg-B}}}$. These phases oscillates much more violently than the Regge action part in $e^{S_{\text{Nondeg}}}$ when the Barbero-Immirzi parameter $\g$ is small, unless ${}^E\!\Theta_f,{}^E\!\Theta^B_f,\Phi_f,\Phi^B_f$ are all vanishing\footnote{The term $\frac{i\pi}{\g} \sum_{e}n_e\sum_{f\subset t_e} A_f$ in both $S_{\text{Nondeg}}$ and $S_{\text{Deg-A}}$ may need special treatment by imposing the boundary semiclassical state carefully.}. We expect that when we take into account the sum over spins $j_f$, the violently oscillating phases $e^{ S_{\text{Deg-A}}}$ and $e^{S_{\text{Deg-B}}}$ may only have relatively small contribution to the total amplitude $A(\ck)=\sum_j A_j(\ck)$, as is suggested by the the Riemann-Lebesgue lemma\footnote{The Riemann-Lebesgue lemma states that for all complex $L^1$-function $f(x)$ on $\mathbb{R}$,  
\be
\int_{-\infty}^{\infty}f(x)e^{i\a x}\rmd x=0\ \ \ \ \text{as}\ \ \ \ \a\to\pm\infty.
\ee }. But surely the nontrivial transition between different types of geometries is a interesting phenomena exhibiting in the semiclassical analysis of Lorentzian spinfoam amplitude, thus requires further investigation and clarification.

\section{Conclusion and Discussion}

The present work studies the large-j asymptotics of the Lorentzian EPRL spinfoam amplitude on a 4d simplicial complex with an arbitrary number of simplices. The asymptotics of the spinfoam amplitude is determined by the critical configurations of the spinfoam action. Here we show that, given a critical configuration $(j_f, g_{ve},\xi_{ef},z_{vf})$ in general, there exists a partition of the simplicial complex $\ck$ into three types of regions $\calr_{\text{Nondeg}},\calr_{\text{Deg-A}},\calr_{\text{Deg-B}}$, where the three regions are simplicial sub-complexes with boundaries. The critical configuration implies different types of geometries in different types of regions, i.e. (1) the critical configuration restricted into $\calr_{\text{Nondeg}}$ implies a nondegenerate discrete Lorentzian geometry in $\calr_{\text{Nondeg}}$. (2) the critical configuration restricted into $\calr_{\text{Deg-A}}$ is degenerate of type-A in our definition of degeneracy, but implies a nondegenerate discrete Euclidean geometry in $\calr_{\text{Deg-A}}$, (3) the critical configuration restricted into $\calr_{\text{Deg-B}}$ is degenerate of type-B, and implies a vector geometry in $\calr_{\text{Deg-B}}$.

With the critical configuration $(j_f, g_{ve},\xi_{ef},z_{vf})$, we further make a subdivision of the regions $\calr_{\text{Nondeg}}$ and $\calr_{\text{Deg-A}}$ into sub-complexes (with boundary) $\ck_{1}(\calr_{*}),\cdots,\ck_n(\calr_{*})$ ($*$=Nondeg,Deg-A) according to their Lorentzian/Euclidean oriented 4-volume $V_4(v)$ of the 4-simplices, such that $\mathrm{sgn}(V_4(v))$ is a constant sign on each $\ck_i(\calr_{*})$. Then in the each sub-complex $\ck_i(\calr_{\text{Nondeg}})$ or $\ck_i(\calr_{\text{Deg-A}})$, the spinfoam amplitude at the critical configuration gives an exponential of Regge action in Lorentzian or Euclidean signature respectively. However we should note that the Regge action reproduced here contains a sign prefactor $\mathrm{sgn}(V_4(v))$ related to the oriented 4-volume of the 4-simplices. Therefore the Regge action reproduced here is actually a discretized Palatini action with on-shell connection.

Finally the asymptotic formula of the spinfoam amplitude is given by a sum of the amplitudes evaluated at all possible critical configurations, which are the products of the amplitudes associated to different type of geometries. 

The present work gives explicitly the critical configurations of the spinfoam amplitude and their geometrical interpretations. However we didn't answer the question such as whether or not the nondegenerate critical configurations are dominating the large-j asymptotic behavior, although we expect the Lorentzian nondegenerate configurations are dominating when the Barbero-Immirzi parameter $\g$ is small. To answer this question in general requires a detailed investigation about the rank of the Hessian matrix in general circumstances. In the appendix, we compute the Hessian matrix of the spinfoam action. However we leave the detailed study about its rank to the future research. 

In this work we show that given a Regge-like spin configuration $j_f$ on the simplicial complex, the critical configurations $(j_f, g_{ve},\xi_{ef},z_{vf})$ with the Regge-like $j_f$ are nondegenerate, and there is a unique critical configuration $(j_f, g_{ve}^c,\xi_{ef},z_{vf}^c)$ with the oriented 4-volume $V_4(v)>0$ (or $V_4(v)<0$) everywhere. We can regard the critical configuration $(j_f, g_{ve}^c,\xi_{ef},z_{vf}^c)$ with $V_4(v)>0$ as a classical background geometry, and define the perturbation theory with the background field method. Thus with the background field method, the n-point functions in spinfoam formulation should be investigated as a generalization of \cite{propagator} to the context with arbitrary simplicial complex, which is a research undergoing.

\section*{Acknowledgements}

The authors would like to thank E. Bianchi, L. Freidel, T. Krajewski, S. Speziale, and C. Rovelli for discussions and communications. M.Z. is supported by CSC scholarship No.2010601003.

\section{Appendix A: Hessian Matrix}\label{Hessian}

In this section we compute the Hessian matrix of the spinfoam action
\be
S_{f}=\sum_{v\in f}S_{vf}=\sum_{v\in f}\left( j_{f}\ln \frac{\left\langle
\xi_{ef},Z_{vef}\right\rangle ^{2}\left\langle Z_{ve^{\prime }f},\xi_{e^{\prime
}f}\right\rangle ^{2}}{\left\langle Z_{vef},Z_{vef}\right\rangle
\left\langle Z_{ve^{\prime }f},Z_{ve^{\prime }f}\right\rangle }+i\gamma
j_{f} \ln \frac{\left\langle Z_{ve^{\prime }f},Z_{ve^{\prime
}f}\right\rangle }{\left\langle Z_{vef},Z_{vef}\right\rangle }\right)
\ee

First of all, we compute the double variation $\delta_{\xi_{ef}}\delta_{\xi_{e'f}}S$, by use $\delta\xi_{ef}=\o_{ef}( J\xi_{ef})+i\eta_{ef}\xi_{ef}$ for complex infinitesimal parameter $\o_{ef}\in\bbc$ and $\eta_{ef}\in\mathbb{R}$. We see immediately from the variation in Eq.\Ref{deltaxi} that $\delta_{\xi_{ef}}\delta_{\xi_{e'f}}S=0$ if $e\neq e'$. Then for the double variation $\delta _{\xi_{ef}}^{2}S$ for the same $\xi_{ef}$
\begin{eqnarray}
\delta _{\xi_{ef}}^{2}S &=&j_{f}\delta _{\xi_{ef}}\left( 2\frac{\delta
_{\xi_{ef}}\left\langle \xi _{ef},Z_{vef}\right\rangle }{\left\langle \xi
_{ef},Z_{vef}\right\rangle }+2\frac{\delta _{\xi_{ef}}\left\langle
Z_{v^{\prime }ef},\xi _{ef}\right\rangle }{\left\langle Z_{v^{\prime
}ef},\xi _{ef}\right\rangle }\right) \nonumber \\
&=&2j_{f}\Bigg(-\frac{\delta _{\xi_{ef}}\left\langle \xi _{ef},Z_{vef}\right\rangle
\delta _{\xi_{ef}}\left\langle \xi _{ef},Z_{vef}\right\rangle }{\left\langle
\xi _{ef},Z_{vef}\right\rangle ^{2}}+\frac{\delta _{\xi_{ef}}^{2}\left\langle
\xi _{ef},Z_{vef}\right\rangle }{\left\langle \xi _{ef},Z_{vef}\right\rangle 
}\nonumber \\
&&-\frac{\delta _{\xi_{ef}}\left\langle Z_{v^{\prime }ef},\xi
_{ef}\right\rangle \delta _{\xi_{ef}}\left\langle Z_{v^{\prime }ef},\xi
_{ef}\right\rangle }{\left\langle Z_{v^{\prime }ef},\xi _{ef}\right\rangle
^{2}}+\frac{\delta _{\xi_{ef}}^{2}\left\langle Z_{v^{\prime }ef},\xi
_{ef}\right\rangle }{\left\langle Z_{v^{\prime }ef},\xi _{ef}\right\rangle }\Bigg)
\end{eqnarray}%
where we use the relation for double variation $\delta(X^{-1}\delta X)=-X^{-2}(\delta X)^2+X^{-1}\delta^2 X$. We compute the above double variation term by term, by using the following relations: 
\begin{eqnarray*}
\delta _{\xi_{ef}}\left\langle \xi _{ef},Z_{vef}\right\rangle  &=&\bar{%
\o}\left\langle J\xi _{ef},Z_{vef}\right\rangle -i\eta \left\langle
\xi _{ef},Z_{vef}\right\rangle  \\
\delta _{\xi_{ef}}\left\langle J\xi _{ef},Z_{vef}\right\rangle 
&=&-\o \left\langle \xi _{ef},Z_{vef}\right\rangle +i\eta
\left\langle J\xi _{ef},Z_{vef}\right\rangle  \\
\delta _{\xi_{ef}}^{2}\left\langle \xi _{ef},Z_{vef}\right\rangle  &=&\bar{%
\o}\delta _{\xi_{ef}}\left\langle J\xi _{ef},Z_{vef}\right\rangle
-i\eta \delta _{\xi_{ef}}\left\langle \xi _{ef},Z_{vef}\right\rangle  \\
&=&-\bar{\o}\o \left\langle \xi _{ef},Z_{vef}\right\rangle
-\eta ^{2}\left\langle \xi _{ef},Z_{vef}\right\rangle  \\
\delta _{\xi_{ef}}\left\langle Z_{v^{\prime }ef},\xi _{ef}\right\rangle 
&=&\o \left\langle Z_{v^{\prime }ef},J\xi _{ef}\right\rangle +i\eta
\left\langle Z_{v^{\prime }ef},\xi _{ef}\right\rangle  \\
\delta _{\xi_{ef}}\left\langle Z_{v^{\prime }ef},J\xi _{ef}\right\rangle  &=&-%
\bar{\o}\left\langle Z_{v^{\prime }ef},\xi _{ef}\right\rangle
-i\eta \left\langle Z_{v^{\prime }ef},J\xi _{ef}\right\rangle  \\
\delta _{\xi_{ef}}^{2}\left\langle Z_{v^{\prime }ef},\xi _{ef}\right\rangle 
&=&\o \delta _{\xi_{ef}}\left\langle Z_{v^{\prime }ef},J\xi
_{ef}\right\rangle +i\eta \delta _{\xi_{ef}}\left\langle Z_{v^{\prime }ef},\xi
_{ef}\right\rangle  \\
&=&-\o \bar{\o}\left\langle Z_{v^{\prime }ef},\xi
_{ef}\right\rangle -\eta ^{2}\left\langle Z_{v^{\prime }ef},\xi
_{ef}\right\rangle 
\end{eqnarray*}%
Then each term in the above $\delta^2_{\xi_{ef}}S$ can be computed
\begin{eqnarray}
-\frac{\delta _{\xi_{ef}}\left\langle \xi _{ef},Z_{vef}\right\rangle \delta
_{n_{ef}}\left\langle \xi _{ef},Z_{vef}\right\rangle }{\left\langle \xi
_{ef},Z_{vef}\right\rangle ^{2}}
&=&-\frac{\left( \bar{\o}^2\left\langle J\xi
_{ef},Z_{vef}\right\rangle ^{2}-2i\eta \bar{\o}\left\langle J\xi
_{ef},Z_{vef}\right\rangle \left\langle \xi _{ef},Z_{vef}\right\rangle -\eta
^{2}\left\langle \xi _{ef},Z_{vef}\right\rangle ^{2}\right) }{\left\langle
\xi _{ef},Z_{vef}\right\rangle ^{2}}\nonumber \\
&=&-\frac{\bar{\o}^{2}\left\langle J\xi _{ef},Z_{vef}\right\rangle
^{2}}{\left\langle \xi _{ef},Z_{vef}\right\rangle ^{2}}+\frac{2i\eta \bar{%
\o}\left\langle J\xi _{ef},Z_{vef}\right\rangle }{\left\langle \xi
_{ef},Z_{vef}\right\rangle }+\eta ^{2}\nonumber\\
\frac{\delta _{\xi_{ef}}^{2}\left\langle \xi _{ef},Z_{vef}\right\rangle }{%
\left\langle \xi _{ef},Z_{vef}\right\rangle }&=&\frac{-\bar{\o}%
\o \left\langle \xi _{ef},Z_{vef}\right\rangle -\eta
^{2}\left\langle \xi _{ef},Z_{vef}\right\rangle }{\left\langle \xi
_{ef},Z_{vef}\right\rangle }=-\bar{\o}\o -\eta ^{2}\nonumber\\
-\frac{\delta _{\xi_{ef}}\left\langle Z_{v^{\prime }ef},\xi
_{ef}\right\rangle \delta _{\xi_{ef}}\left\langle Z_{v^{\prime }ef},\xi
_{ef}\right\rangle }{\left\langle Z_{v^{\prime }ef},\xi _{ef}\right\rangle
^{2}} 
&=&-\frac{\left( \o ^{2}\left\langle Z_{v^{\prime }ef},J\xi
_{ef}\right\rangle ^{2}+2i\eta \o \left\langle Z_{v^{\prime
}ef},J\xi _{ef}\right\rangle \left\langle Z_{v^{\prime }ef},\xi
_{ef}\right\rangle -\eta ^{2}\left\langle Z_{v^{\prime }ef},\xi
_{ef}\right\rangle ^{2}\right) }{\left\langle Z_{v^{\prime }ef},\xi
_{ef}\right\rangle ^{2}} \nonumber\\
&=&-\frac{\o^{2}\left\langle Z_{v^{\prime }ef},J\xi
_{ef}\right\rangle ^{2}}{\left\langle Z_{v^{\prime }ef},\xi
_{ef}\right\rangle ^{2}}-\frac{2i\eta \o \left\langle Z_{v^{\prime
}ef},J\xi _{ef}\right\rangle }{\left\langle Z_{v^{\prime }ef},\xi
_{ef}\right\rangle }+\eta ^{2}\nonumber\\
\frac{\delta _{xi_{ef}}^{2}\left\langle Z_{v^{\prime }ef},\xi
_{ef}\right\rangle }{\left\langle Z_{v^{\prime }ef},\xi _{ef}\right\rangle }%
&=&-\o \bar{\o}-\eta ^{2}
\ee
Therefore $\delta _{\xi_{ef}}^{2}S$ is obtained explicitly
\[
\delta _{\xi_{ef}}^{2}S=2j_{f}\left( -\frac{\bar{\o}^{2}\left\langle
J\xi _{ef},Z_{vef}\right\rangle ^{2}}{\left\langle \xi
_{ef},Z_{vef}\right\rangle ^{2}}+\frac{2i\eta \bar{\o}\left\langle
J\xi _{ef},Z_{vef}\right\rangle }{\left\langle \xi
_{ef},Z_{vef}\right\rangle }-\frac{\o ^{2}\left\langle Z_{v^{\prime
}ef},J\xi _{ef}\right\rangle ^{2}}{\left\langle Z_{v^{\prime }ef},\xi
_{ef}\right\rangle ^{2}}-\frac{2i\eta \o \left\langle Z_{v^{\prime
}ef},J\xi _{ef}\right\rangle }{\left\langle Z_{v^{\prime }ef},\xi
_{ef}\right\rangle }-2\o \bar{\o}\right) 
\]%
Because of Eq.\Ref{eq:PreGlu}, at the critical configuration $Z_{vef}\sim \xi _{ef}$. Therefore by using the relation $\left\langle J\xi _{ef},\xi
_{ef}\right\rangle =0$, we obtain the result
\be
\delta _{\xi_{ef}}^{2}S=-4j_{f}\o \bar{\o}
\ee
which means that the $H_{\o_{ef}\bar{\o}_{ef}}$ components of the Hessian matrix are the only nonvanishing components in the Hessian submatrix respect to the spinorial variables $\xi_{ef}$, and
\be
\addtolength{\fboxsep}{1pt}
\boxed{H_{\o_{ef}\bar{\o}_{ef}}=H_{\bar{\o}_{ef}\o_{ef}}=-4j_f}
\ee

Secondly we compute the double variation with respect to both the spinorial variabe $\xi_{ef}$ and the group variable $g_{ve}$, where $\delta_{g_{ve}}:={\partial }\big/{\partial \theta _{IJ}^{ve}}\big|_{\theta ^{ve}=0}$ with the parametrization $g'_{ve}={g}_{ve}e^{\theta _{IJ}^{ve}\cj^{IJ}}$ at a critical configuration $g_{ve}$
\begin{eqnarray}
\delta _{g_{ve}}\delta _{\xi_{ef}}S &=&j_{f}\delta _{g_{ve}}\left( 2\frac{%
\delta _{\xi_{ef}}\left\langle \xi _{ef},Z_{vef}\right\rangle }{\left\langle
\xi _{ef},Z_{vef}\right\rangle }+2\frac{\delta _{\xi_{ef}}\left\langle
Z_{v^{\prime }ef},\xi _{ef}\right\rangle }{\left\langle Z_{v^{\prime
}ef},\xi _{ef}\right\rangle }\right) \nonumber \\
&=&2j_{f}\left( -\frac{\delta _{g_{ve}}\left\langle \xi
_{ef},Z_{vef}\right\rangle \delta _{\xi_{ef}}\left\langle \xi
_{ef},Z_{vef}\right\rangle }{\left\langle \xi _{ef},Z_{vef}\right\rangle ^{2}%
}+\frac{\delta _{g_{ve}}\delta _{\xi_{ef}}\left\langle \xi
_{ef},Z_{vef}\right\rangle }{\left\langle \xi _{ef},Z_{vef}\right\rangle }%
\right) 
\end{eqnarray}
We use the following relations:
\begin{eqnarray}
\delta _{g_{ve}}\left\langle \xi _{ef},Z_{vef}\right\rangle  &=&\left\langle
\xi _{ef},\mathcal{J}^{\dag }Z_{vef}\right\rangle\nonumber  \\
\delta _{g_{ve}}\delta _{\xi_{ef}}\left\langle \xi _{ef},Z_{vef}\right\rangle 
&=&\bar{\o}\left\langle J\xi _{ef},\mathcal{J}^{\dag
}Z_{vef}\right\rangle -i\eta \left\langle \xi _{ef},\mathcal{J}^{\dag
}Z_{vef}\right\rangle 
\end{eqnarray}%
Thus using Eq.\Ref{eq:PreGlu} we have
\begin{eqnarray}
\delta _{g_{ve}}\delta _{\xi_{ef}}S &=&2j_{f}\Bigg[-\frac{\left\langle \xi _{ef},%
\mathcal{J}^{\dag }Z_{vef}\right\rangle \left( \bar{\o}\left\langle
J\xi _{ef},Z_{vef}\right\rangle -i\eta \left\langle \xi
_{ef},Z_{vef}\right\rangle \right) }{\left\langle \xi
_{ef},Z_{vef}\right\rangle ^{2}} \nonumber\\
&&+\frac{\bar{\o}\left\langle J\xi _{ef},\mathcal{J}^{\dag
}Z_{vef}\right\rangle -i\eta \left\langle \xi _{ef},\mathcal{J}^{\dag
}Z_{vef}\right\rangle }{\left\langle \xi _{ef},Z_{vef}\right\rangle }\Bigg] \nonumber\\
&=&2j_{f}\left( \frac{i\eta \left\langle \xi _{ef},\mathcal{J}^{\dag
}Z_{vef}\right\rangle }{\left\langle \xi _{ef},Z_{vef}\right\rangle }+\frac{%
\bar{\o}\left\langle J\xi _{ef},\mathcal{J}^{\dag
}Z_{vef}\right\rangle -i\eta \left\langle \xi _{ef},\mathcal{J}^{\dag
}Z_{vef}\right\rangle }{\left\langle \xi _{ef},Z_{vef}\right\rangle }\right) \nonumber
\\
&=&2j_{f}\frac{\bar{\o}\left\langle J\xi _{ef},\mathcal{J}^{\dag
}\xi _{ef}\right\rangle }{\left\langle \xi _{ef},\xi _{ef}\right\rangle }%
\end{eqnarray}
where explicitly for $\lag J\xi _{ef},\sigma ^{i}\xi _{ef}\rag $, $\xi =\left(
\xi_{0},\xi_{1}\right)^t $ we have 
\begin{eqnarray*}
\lag J \xi _{ef},\sigma ^{1}\xi _{ef}\rag  &=&-2\xi_{0}\xi_{1} \\
\lag J \xi _{ef},\sigma ^{2}\xi _{ef}\rag  &=&0 \\
\lag J \xi _{ef},\sigma ^{3}\xi _{ef}\rag  &=&-\left(
\xi_{0}^{2}-\xi_{1}^{2}\right) 
\end{eqnarray*}
However there are also nonvanishing components $\delta _{g_{v'e}}\delta _{\xi_{ef}}S$ with $e=(v,v')$. Similarly we obtain
\be
\delta _{g_{v'e}}\delta _{\xi_{ef}}S=2j_{f}{{\o}\left\langle \mathcal{J}^{\dag
}\xi _{ef},J\xi _{ef}\right\rangle }
\ee 
Thus the nonvanishing components of the Hessian matrix are
\be
\addtolength{\fboxsep}{1pt}
\boxed{H_{\theta_{ve}\bar{\o}_{ef}}=H_{\bar{\o}_{ef}\theta_{ve}}=2j_{f}{\left\langle J\xi _{ef},\mathcal{J}^{\dag
}\xi _{ef}\right\rangle }\ \ \ \ \text{and}\ \ \ \ H_{\theta_{v'e}{\o}_{ef}}=H_{{\o}_{ef}\theta_{v'e}}=2j_{f}{\left\langle \mathcal{J}^{\dag
}\xi _{ef},J\xi _{ef}\right\rangle }.}
\ee

Next we compute the double variation $\delta _{z_{vf}}\delta _{\xi_{ef}}S$. Here $z_{vf}$ is a $\mathbb{CP}^1$ variable so $\delta z_{vf}=\eps_{vf}Jz_{vf}$
\begin{eqnarray}
\delta _{z_{vf}}\delta _{\xi_{ef}}S &=&j_{f}\delta _{z_{vf}}\left( 2\frac{%
\delta _{\xi_{ef}}\left\langle \xi _{ef},Z_{vef}\right\rangle }{\left\langle
\xi _{ef},Z_{vef}\right\rangle }+2\frac{\delta _{\xi_{ef}}\left\langle
Z_{v^{\prime }ef},\xi _{ef}\right\rangle }{\left\langle Z_{v^{\prime
}ef},\xi _{ef}\right\rangle }\right)\nonumber \\
&=&2j_{f}\left( -\frac{\delta _{z_{vf}}\left\langle \xi
_{ef},Z_{vef}\right\rangle \delta _{\xi_{ef}}\left\langle \xi
_{ef},Z_{vef}\right\rangle }{\left\langle \xi _{ef},Z_{vef}\right\rangle ^{2}%
}+\frac{\delta _{z_{vf}}\delta _{\xi_{ef}}\left\langle \xi
_{ef},Z_{vef}\right\rangle }{\left\langle \xi _{ef},Z_{vef}\right\rangle }%
\right)
\end{eqnarray}%
We use the following relations:
\begin{eqnarray}
\delta _{z_{vf}}\left\langle \xi _{ef},Z_{vef}\right\rangle &=&\varepsilon
\left\langle \xi _{ef},g_{ve}^{\dag }Jz_{vf}\right\rangle \\
\delta _{z_{vf}}\delta _{\xi_{ef}}\left\langle \xi _{ef},Z_{vef}\right\rangle
&=&\bar{\o}\delta _{z_{vf}}\left\langle J\xi
_{ef},Z_{vef}\right\rangle -i\eta \delta _{z_{vf}}\left\langle \xi
_{ef},Z_{vef}\right\rangle \nonumber\\
&=&\bar{\o}\varepsilon \left\langle J\xi _{ef},g_{ve}^{\dag
}Jz_{vf}\right\rangle -i\eta \varepsilon\left\langle \xi _{ef},g_{ve}^{\dag
}Jz_{vf}\right\rangle 
\end{eqnarray}%
Using Eq.\Ref{eq:PreGlu} we have at a critical configuration:
\begin{eqnarray}
\delta _{z_{vf}}\delta _{\xi_{ef}}S &=&2j_{f}\Bigg[-\frac{ \varepsilon
\left\langle \xi _{ef},g_{ve}^{\dag }Jz_{vf}\right\rangle  \left( \bar{%
\o}\left\langle J\xi _{ef},Z_{vef}\right\rangle -i\eta \left\langle
\xi _{ef},Z_{vef}\right\rangle \right) }{\left\langle \xi
_{ef},Z_{vef}\right\rangle ^{2}} +\frac{\bar{\o}\varepsilon \left\langle J\xi
_{ef},g_{ve}^{\dag }Jz_{vf}\right\rangle  -i\eta \varepsilon\left\langle \xi _{ef},g_{ve}^{\dag }Jz_{vf}\right\rangle  }{\left\langle \xi
_{ef},Z_{vef}\right\rangle }\Bigg]\nonumber \\
&=&2j_{f}\frac{\bar{\o}\varepsilon \left\langle J\xi
_{ef},g_{ve}^{\dag }Jz_{vf}\right\rangle }{\left\langle \xi
_{ef},Z_{vef}\right\rangle }
=2j_{f}\bar{\o}\varepsilon e^{2i\phi _{ev}}\left\langle
g_{ve}J\xi _{ef},g_{ve}J\xi _{ef}\right\rangle
\end{eqnarray}%
Similar we also have for $e=(v,v')$ that
\be
\delta _{z_{v'f}}\delta _{\xi_{ef}}S=2j_{f}{\o}\bar{\varepsilon} e^{-2i\phi _{ev}}\left\langle
g_{ve}J\xi _{ef},g_{ve}J\xi _{ef}\right\rangle
\ee
Then the nonvanishing components of the Hessian matrix are
\be
  \addtolength{\fboxsep}{1pt} 
   \boxed{ 
   \begin{split} 
&&H_{\eps_{vf}\bar{\o}_{ef}}=H_{\bar{\o}_{ef}\eps_{vf}}=2j_{f} e^{2i\phi _{ev}}\left\langle
g_{ve}J\xi _{ef},g_{ve}J\xi _{ef}\right\rangle\nonumber\\
&&H_{\bar{\eps}_{v'f}\o_{ef}}=H_{\o_{ef}\bar{\eps}_{v'f}}=2j_{f} e^{-2i\phi _{ev}}\left\langle
g_{ve}J\xi _{ef},g_{ve}J\xi _{ef}\right\rangle
 \end{split} 
   } 
\ee
Note that in the degenerate case $g_{ve}\in\mathrm{SU(2)}$
\be
H_{\eps_{vf}\bar{\o}_{ef}}\big|_{\text{deg}}=H_{\bar{\o}_{ef}\eps_{vf}}\big|_{\text{deg}}=2j_{f} e^{2i\phi _{ev}}\ \ \ \ \ \ H_{\bar{\eps}_{v'f}\o_{ef}}\big|_{\text{deg}}=H_{\o_{ef}\bar{\eps}_{v'f}}\big|_{\text{deg}}=2j_{f} e^{-2i\phi _{ev}}.
\ee

For the double variations $\delta_{z_{vf}}\delta_{z_{v'f'}}S$, it is obvious that the nonvanishing components are $\delta^2_{z_{vf}}S$
\begin{eqnarray*}
\delta _{z_{vf}}^{2}S_{vf} &=&j_{f}\delta _{z_{vf}}\left( 2\frac{\delta
_{z_{vf}}\left\langle \xi _{ef},Z_{vef}\right\rangle }{\left\langle \xi
_{ef},Z_{vef}\right\rangle }+2\frac{\delta _{z_{vf}}\left\langle
Z_{ve^{\prime }f},\xi _{e^{\prime }f}\right\rangle }{\left\langle
Z_{ve^{\prime }f},\xi _{e^{\prime }f}\right\rangle }-\frac{\delta
_{z_{vf}}\left\langle Z_{vef},Z_{vef}\right\rangle }{\left\langle
Z_{vef},Z_{vef}\right\rangle }-\frac{\delta _{z_{vf}}\left\langle
Z_{ve^{\prime }f},Z_{ve^{\prime }f}\right\rangle }{\left\langle
Z_{ve^{\prime }f},Z_{ve^{\prime }f}\right\rangle }\right) \\
&&+i\gamma j_{f} \delta _{z_{vf}}\left( \frac{\delta
_{z_{vf}}\left\langle Z_{ve^{\prime }f},Z_{ve^{\prime }f}\right\rangle }{%
\left\langle Z_{ve^{\prime }f},Z_{ve^{\prime }f}\right\rangle }-\frac{\delta
_{z_{vf}}\left\langle Z_{vef},Z_{vef}\right\rangle }{\left\langle
Z_{vef},Z_{vef}\right\rangle }\right)
\end{eqnarray*}%
In the following we compute $\delta _{z_{vf}}^{2}S_{vf}$ term by term: 
\begin{eqnarray}
2\delta _{z_{vf}}\frac{\delta _{z_{vf}}\left\langle \xi
_{ef},Z_{vef}\right\rangle }{\left\langle \xi _{ef},Z_{vef}\right\rangle }
&=&-2\varepsilon \varepsilon e^{4i\phi _{ev}}\left\langle g_{ve}\xi
_{ef},g_{ve}J\xi _{ef}\right\rangle ^{2}-2\varepsilon \bar{\varepsilon}\nonumber\\
2\delta _{z_{vf}}\frac{\delta _{z_{vf}}\left\langle Z_{ve^{\prime }f},\xi
_{e^{\prime }f}\right\rangle }{\left\langle Z_{ve^{\prime }f},\xi
_{e^{\prime }f}\right\rangle } 
&=&-2\bar{\varepsilon}\bar{\varepsilon}e^{-4i\phi _{e^{\prime
}v}}\left\langle g_{ve^{\prime }}J\xi _{e^{\prime }f},g_{ve^{\prime }}\xi
_{e^{\prime }f}\right\rangle ^{2}-2\bar{\varepsilon}\varepsilon \nonumber\\
-\delta _{z_{vf}}\frac{\delta _{z_{vf}}\left\langle
Z_{vef},Z_{vef}\right\rangle }{\left\langle Z_{vef},Z_{vef}\right\rangle }
&=&\left( \varepsilon e^{2i\phi _{ev}}\left\langle g_{ve}\xi _{ef},g_{ve}J\xi _{ef}\right\rangle +\bar{\varepsilon}e^{-2i\phi
_{ev}}\left\langle g_{ve}J\xi _{ef},g_{ve}\xi _{ef}\right\rangle
\right) ^{2} \nonumber\\
&&2\bar{\varepsilon}\varepsilon -2\varepsilon \bar{\varepsilon}\left\langle
g_{ve}^{\dag }g_{ve}J\xi _{ef},g_{ve}^{\dag }g_{ve}J\xi _{ef}\right\rangle \nonumber\\
-\delta _{z_{vf}}\frac{\delta _{z_{vf}}\left\langle Z_{ve^{\prime
}f},Z_{ve^{\prime }f}\right\rangle }{\left\langle Z_{ve^{\prime
}f},Z_{ve^{\prime }f}\right\rangle } 
&=&\left( \varepsilon e^{2i\phi _{e^{\prime }v}}\left\langle g_{ve^{\prime
}}\xi _{e^{\prime }f},g_{ve^{\prime }}J\xi _{e^{\prime }f}\right\rangle +%
\bar{\varepsilon}e^{-2i\phi _{e^{\prime }v}}\left\langle g_{ve^{\prime
}}J\xi _{e^{\prime }f},g_{ve^{\prime }}\xi _{e^{\prime }f}\right\rangle
\right) ^{2}\nonumber \\
&&2\bar{\varepsilon}\varepsilon -2\varepsilon \bar{\varepsilon}\left\langle
g_{ve^{\prime }}^{\dag }g_{ve^{\prime }}J\xi _{e^{\prime }f},g_{ve^{\prime
}}^{\dag }g_{ve^{\prime }}J\xi _{e^{\prime }f}\right\rangle 
\end{eqnarray}%
We obtain explicitly the expression of $\delta _{z_{vf}}^{2}S_{vf}$
\begin{eqnarray}
\delta _{z_{vf}}^{2}S_{vf} &=&j_{f}\Bigg( 
2\varepsilon \bar{\varepsilon}\left\langle g_{ve}\xi _{ef},g_{ve}J\xi
_{ef}\right\rangle \left\langle g_{ve}J\xi _{ef},g_{ve}\xi
_{ef}\right\rangle+2\varepsilon \bar{\varepsilon}\left\langle g_{ve'}\xi _{e'f},g_{ve'}J\xi
_{e'f}\right\rangle \left\langle g_{ve'}J\xi _{e'f},g_{ve'}\xi
_{e'f}\right\rangle \nonumber \\ 
&&-2\varepsilon \bar{\varepsilon}\left\langle g_{ve}^{\dag }g_{ve}J\xi
_{ef},g_{ve}^{\dag }g_{ve}J\xi _{ef}\right\rangle -2\varepsilon \bar{%
\varepsilon}\left\langle g_{ve^{\prime }}^{\dag }g_{ve^{\prime }}J\xi
_{e^{\prime }f},g_{ve^{\prime }}^{\dag }g_{ve^{\prime }}J\xi _{e^{\prime
}f}\right\rangle 
\Bigg)  \nonumber\\
&&+i\gamma j_{f}\Bigg( 2\varepsilon \bar{\varepsilon}%
\left\langle g_{ve}^{\dag }g_{ve}J\xi _{ef},g_{ve}^{\dag }g_{ve}J\xi
_{ef}\right\rangle -2\varepsilon \bar{\varepsilon}\left\langle g_{ve^{\prime
}}^{\dag }g_{ve^{\prime }}J\xi _{e^{\prime }f},g_{ve^{\prime }}^{\dag
}g_{ve^{\prime }}J\xi _{e^{\prime }f}\right\rangle \Bigg) 
\end{eqnarray}
Therefore we obtain the nonvanishing components of the Hessian matrix
\be
  \addtolength{\fboxsep}{1pt} 
   \boxed{ 
   \begin{split} 
H_{\eps_{vf}\bar{\eps}_{vf}}=H_{\bar{\eps}_{vf}\eps_{vf}}&=&2j_{f}\Bigg( 
\left\langle g_{ve}\xi _{ef},g_{ve}J\xi
_{ef}\right\rangle \left\langle g_{ve}J\xi _{ef},g_{ve}\xi
_{ef}\right\rangle+\left\langle g_{ve'}\xi _{e'f},g_{ve'}J\xi
_{e'f}\right\rangle \left\langle g_{ve'}J\xi _{e'f},g_{ve'}\xi
_{e'f}\right\rangle \nonumber \\ 
&&-\left\langle g_{ve}^{\dag }g_{ve}J\xi
_{ef},g_{ve}^{\dag }g_{ve}J\xi _{ef}\right\rangle -\left\langle g_{ve^{\prime }}^{\dag }g_{ve^{\prime }}J\xi
_{e^{\prime }f},g_{ve^{\prime }}^{\dag }g_{ve^{\prime }}J\xi _{e^{\prime
}f}\right\rangle 
\Bigg)  \nonumber\\
&&+2i\gamma j_{f}\Bigg( \left\langle g_{ve}^{\dag }g_{ve}J\xi _{ef},g_{ve}^{\dag }g_{ve}J\xi
_{ef}\right\rangle -\left\langle g_{ve^{\prime
}}^{\dag }g_{ve^{\prime }}J\xi _{e^{\prime }f},g_{ve^{\prime }}^{\dag
}g_{ve^{\prime }}J\xi _{e^{\prime }f}\right\rangle \Bigg)
\end{split}
}
\ee
Note that in the degenerate case $g_{ve}\in\mathrm{SU(2)}$
\be
H_{\eps_{vf}\bar{\eps}_{vf}}\big|_{\text{deg}}=H_{\bar{\eps}_{vf}\eps_{vf}}\big|_{\text{deg}}=-4j_{f}
\ee

For the double variation $\delta _{g_{ve}}\delta _{z_{vf}}S$ we have 
\begin{eqnarray}
\delta _{g_{ve}}\delta _{z_{vf}}S_{vf} &=&j_{f}\left( 2\delta _{g_{ve}}\frac{%
\delta _{z_{vf}}\left\langle \xi _{ef},Z_{vef}\right\rangle }{\left\langle
\xi _{ef},Z_{vef}\right\rangle }-\delta _{g_{ve}}\frac{\delta
_{z_{vf}}\left\langle Z_{vef},Z_{vef}\right\rangle }{\left\langle
Z_{vef},Z_{vef}\right\rangle }\right)  -i\gamma  j_{f}\delta _{g_{ve}}\frac{\delta
_{z_{vf}}\left\langle Z_{vef},Z_{vef}\right\rangle }{\left\langle
Z_{vef},Z_{vef}\right\rangle }
\end{eqnarray}%
which can be computed term by term:
\begin{eqnarray}
\delta _{g_{ve}}\frac{\delta _{z_{vf}}\left\langle \xi
_{ef},Z_{vef}\right\rangle }{\left\langle \xi _{ef},Z_{vef}\right\rangle }
&=&-\varepsilon e^{2i\phi _{ev}}\left\langle g_{ve}\xi _{ef},g_{ve}J\xi
_{ef}\right\rangle \left\langle \xi _{ef},\mathcal{J}^{\dag }\xi
_{ef}\right\rangle +\varepsilon e^{2i\phi _{ev}}\left\langle \xi _{ef},%
\mathcal{J}^{\dag }g_{ve}^{\dag }g_{ve}J\xi _{ef}\right\rangle \nonumber\\
\delta _{g_{ve}}\frac{\delta _{z_{vf}}\left\langle Z_{vef},Z_{vef}\right\rangle }{\left\langle Z_{vef},Z_{vef}\right\rangle }
&=&-\lt(\varepsilon e^{2i\phi_{ev}}\left\langle g_{ve}\xi_{ef},g_{ve}J\xi_{ef}\right\rangle 
+\bar{\varepsilon}e^{-2i\phi_{ev}}\left\langle g_{ve}J\xi_{ef},g_{ve}\xi_{ef}\right\rangle\rt) 
\lt(\big\langle \cj^{\dagger} \xi_{ef},\xi_{ef}\big\rangle 
+\big\langle\xi_{ef},\cj^{\dagger} \xi_{ef}\big\rangle\rt)\nonumber\\
&&+\eps e^{2i\phi_{ev}}\left\langle \cj^{\dagger}\xi_{ef},g_{ve}^\dagger g_{ve}J\xi_{ef}\right\rangle  
+\bar{\eps}e^{-2i\phi_{ev}}\big\langle \cj^{\dagger}g_{ve}^\dagger g_{ve}J\xi_{ef},\xi_{ef}\big\rangle\nonumber\\
&&+\eps e^{2i\phi_{ev}}\big\langle\xi_{ef}, \cj^{\dagger}g_{ve}^\dagger g_{ve}J\xi_{ef}\big\rangle
+\bar{\eps}e^{-2i\phi_{ev}}\left\langle g_{ve}^\dagger g_{ve}J\xi_{ef}, \cj^{\dagger}\xi_{ef}\right\rangle
\end{eqnarray}%
Thus we have explicitly
\be
\delta _{g_{ve}}\delta _{z_{vf}}S_{vf} &=&2j_{f}\lt[-\varepsilon e^{2i\phi _{ev}}\left\langle g_{ve}\xi _{ef},g_{ve}J\xi
_{ef}\right\rangle \left\langle \xi _{ef},\mathcal{J}^{\dag }\xi
_{ef}\right\rangle +\varepsilon e^{2i\phi _{ev}}\left\langle \xi _{ef},%
\mathcal{J}^{\dag }g_{ve}^{\dag }g_{ve}J\xi _{ef}\right\rangle\rt]\nonumber\\
&&- (1+i\g)j_f\Bigg[-\lt(\varepsilon e^{2i\phi_{ev}}\left\langle g_{ve}\xi_{ef},g_{ve}J\xi_{ef}\right\rangle 
+\bar{\varepsilon}e^{-2i\phi_{ev}}\left\langle g_{ve}J\xi_{ef},g_{ve}\xi_{ef}\right\rangle\rt) 
\lt(\big\langle \cj^{\dagger} \xi_{ef},\xi_{ef}\big\rangle 
+\big\langle\xi_{ef},\cj^{\dagger} \xi_{ef}\big\rangle\rt)\nonumber\\
&&+\eps e^{2i\phi_{ev}}\left\langle \cj^{\dagger}\xi_{ef},g_{ve}^\dagger g_{ve}J\xi_{ef}\right\rangle  
+\bar{\eps}e^{-2i\phi_{ev}}\big\langle \cj^{\dagger}g_{ve}^\dagger g_{ve}J\xi_{ef},\xi_{ef}\big\rangle\nonumber\\
&&+\eps e^{2i\phi_{ev}}\big\langle\xi_{ef}, \cj^{\dagger}g_{ve}^\dagger g_{ve}J\xi_{ef}\big\rangle
+\bar{\eps}e^{-2i\phi_{ev}}\left\langle g_{ve}^\dagger g_{ve}J\xi_{ef}, \cj^{\dagger}\xi_{ef}\right\rangle\Bigg] 
\ee
Therefore we obtain the components of the Hessian matrix:
\be
  \addtolength{\fboxsep}{1pt} 
   \boxed{ 
   \begin{split} 
H_{\theta_{ve}\eps_{vf}}=H_{\eps_{vf}\theta_{ve}} &=&2j_{f}\lt[- e^{2i\phi _{ev}}\left\langle g_{ve}\xi _{ef},g_{ve}J\xi
_{ef}\right\rangle \left\langle \xi _{ef},\mathcal{J}^{\dag }\xi
_{ef}\right\rangle +e^{2i\phi _{ev}}\left\langle \xi _{ef},%
\mathcal{J}^{\dag }g_{ve}^{\dag }g_{ve}J\xi _{ef}\right\rangle\rt]\nonumber\\
&&- (1+i\g)j_f\Bigg[- e^{2i\phi_{ev}}\left\langle g_{ve}\xi_{ef},g_{ve}J\xi_{ef}\right\rangle 
\lt(\big\langle \cj^{\dagger} \xi_{ef},\xi_{ef}\big\rangle 
+\big\langle\xi_{ef},\cj^{\dagger} \xi_{ef}\big\rangle\rt)\nonumber\\
&&+ e^{2i\phi_{ev}}\left\langle \cj^{\dagger}\xi_{ef},g_{ve}^\dagger g_{ve}J\xi_{ef}\right\rangle  + e^{2i\phi_{ev}}\big\langle\xi_{ef}, \cj^{\dagger}g_{ve}^\dagger g_{ve}J\xi_{ef}\big\rangle\Bigg] \nonumber\\
H_{\theta_{ve}\bar{\eps}_{vf}}=H_{\bar{\eps}_{vf}\theta_{ve}} &=&- (1+i\g)j_f\Bigg[-\bar{\varepsilon}e^{-2i\phi_{ev}}\left\langle g_{ve}J\xi_{ef},g_{ve}\xi_{ef}\right\rangle
\lt(\big\langle \cj^{\dagger} \xi_{ef},\xi_{ef}\big\rangle 
+\big\langle\xi_{ef},\cj^{\dagger} \xi_{ef}\big\rangle\rt)\nonumber\\
&&+e^{-2i\phi_{ev}}\big\langle \cj^{\dagger}g_{ve}^\dagger g_{ve}J\xi_{ef},\xi_{ef}\big\rangle+e^{-2i\phi_{ev}}\left\langle g_{ve}^\dagger g_{ve}J\xi_{ef}, \cj^{\dagger}\xi_{ef}\right\rangle\Bigg] 
\end{split}
}
\ee
while in the degenerate case
\be
H_{\theta_{ve}\eps_{vf}}\big|_{\text{deg}}=H_{\eps_{vf}\theta_{ve}}\big|_{\text{deg}}
&=&2j_f e^{2i\phi _{ev}}\left\langle \xi _{ef},\mathcal{J}^{\dag }J\xi _{ef}\right\rangle\nonumber\\
&&- (1+i\g)j_f\Bigg[e^{2i\phi_{ev}}\left\langle \cj^{\dagger}\xi_{ef},J\xi_{ef}\right\rangle  + e^{2i\phi_{ev}}\big\langle\xi_{ef}, \cj^{\dagger}J\xi_{ef}\big\rangle\Bigg] \nonumber\\
H_{\theta_{ve}\bar{\eps}_{vf}}\big|_{\text{deg}}=H_{\bar{\eps}_{vf}\theta_{ve}}\big|_{\text{deg}} 
&=&- (1+i\g)j_f\Bigg[e^{-2i\phi_{ev}}\big\langle \cj^{\dagger}J\xi_{ef},\xi_{ef}\big\rangle+e^{-2i\phi_{ev}}\left\langle J\xi_{ef}, \cj^{\dagger}\xi_{ef}\right\rangle\Bigg] 
\ee
However there are nonvanishing components of Hessian matrix from $\delta _{g_{ve'}}\delta _{z_{vf}}S_{vf}$ 
\be
\delta _{g_{ve'}}\delta _{z_{vf}}S_{vf}&=&j_{f}\left( 2\delta _{g_{ve'}}\frac{%
\delta _{z_{vf}}\left\langle Z_{ve'f},\xi _{e'f}\right\rangle }{\left\langle
Z_{ve'f},\xi _{e'f}\right\rangle }-\delta _{g_{ve'}}\frac{\delta
_{z_{vf}}\left\langle Z_{ve'f},Z_{ve'f}\right\rangle }{\left\langle
Z_{ve'f},Z_{ve'f}\right\rangle }\right)  +i\gamma  j_{f}\delta _{g_{ve'}}\frac{\delta
_{z_{vf}}\left\langle Z_{ve'f},Z_{ve'f}\right\rangle }{\left\langle
Z_{ve'f},Z_{ve'f}\right\rangle }\nonumber\\
&=&2j_{f}\lt[-\bar{\varepsilon} e^{-2i\phi _{e'v}}\left\langle g_{ve'}J\xi
_{e'f},g_{ve'}\xi _{e'f}\right\rangle \left\langle\mathcal{J}^{\dag }\xi
_{e'f}, \xi _{e'f}\right\rangle +\bar{\varepsilon} e^{-2i\phi _{e'v}}\left\langle
\mathcal{J}^{\dag }g_{ve'}^{\dag }g_{ve'}J\xi _{e'f}, \xi _{e'f}\right\rangle\rt]\nonumber\\
&&- (1-i\g)j_f\Bigg[-\lt(\varepsilon e^{2i\phi_{e'v}}\left\langle g_{ve'}\xi_{ef},g_{ve'}J\xi_{e'f}\right\rangle 
+\bar{\varepsilon}e^{-2i\phi_{e'v}}\left\langle g_{ve'}J\xi_{e'f},g_{ve'}\xi_{e'f}\right\rangle\rt) 
\lt(\big\langle \cj^{\dagger} \xi_{e'f},\xi_{e'f}\big\rangle 
+\big\langle\xi_{e'f},\cj^{\dagger} \xi_{e'f}\big\rangle\rt)\nonumber\\
&&+\eps e^{2i\phi_{e'v}}\left\langle \cj^{\dagger}\xi_{e'f},g_{ve'}^\dagger g_{ve'}J\xi_{e'f}\right\rangle  
+\bar{\eps}e^{-2i\phi_{e'v}}\big\langle \cj^{\dagger}g_{ve'}^\dagger g_{ve'}J\xi_{e'f},\xi_{e'f}\big\rangle\nonumber\\
&&+\eps e^{2i\phi_{e'v}}\big\langle\xi_{e'f}, \cj^{\dagger}g_{ve'}^\dagger g_{ve'}J\xi_{e'f}\big\rangle
+\bar{\eps}e^{-2i\phi_{e'v}}\left\langle g_{ve'}^\dagger g_{ve'}J\xi_{e'f}, \cj^{\dagger}\xi_{e'f}\right\rangle\Bigg] 
\ee
Therefore we obtain the components of the Hessian matrix:
\be
  \addtolength{\fboxsep}{1pt} 
   \boxed{ 
   \begin{split} 
H_{\theta_{ve'}\eps_{vf}}=H_{\eps_{vf}\theta_{ve'}}&=&- (1-i\g)j_f\Bigg[-e^{2i\phi_{e'v}}\left\langle g_{ve'}\xi_{ef},g_{ve'}J\xi_{e'f}\right\rangle 
\lt(\big\langle \cj^{\dagger} \xi_{e'f},\xi_{e'f}\big\rangle 
+\big\langle\xi_{e'f},\cj^{\dagger} \xi_{e'f}\big\rangle\rt)\ \ \ \ \ \ \ \ \ \ \ \ \nonumber\\
&&+ e^{2i\phi_{e'v}}\left\langle \cj^{\dagger}\xi_{e'f},g_{ve'}^\dagger g_{ve'}J\xi_{e'f}\right\rangle  + e^{2i\phi_{e'v}}\big\langle\xi_{e'f}, \cj^{\dagger}g_{ve'}^\dagger g_{ve'}J\xi_{e'f}\big\rangle
\Bigg]\ \ \ \ \ \ \ \ \ \ \ \ \nonumber\\
H_{\theta_{ve'}\bar{\eps}_{vf}}=H_{\bar{\eps}_{vf}\theta_{ve'}}&=&2j_{f}\lt[-e^{-2i\phi _{e'v}}\left\langle g_{ve'}J\xi
_{e'f},g_{ve'}\xi _{e'f}\right\rangle \left\langle\mathcal{J}^{\dag }\xi
_{e'f}, \xi _{e'f}\right\rangle + e^{-2i\phi _{e'v}}\left\langle
\mathcal{J}^{\dag }g_{ve'}^{\dag }g_{ve'}J\xi _{e'f}, \xi _{e'f}\right\rangle\rt]\nonumber\\
&&- (1-i\g)j_f\Bigg[- e^{-2i\phi_{e'v}}\left\langle g_{ve'}J\xi_{e'f},g_{ve'}\xi_{e'f}\right\rangle 
\lt(\big\langle \cj^{\dagger} \xi_{e'f},\xi_{e'f}\big\rangle 
+\big\langle\xi_{e'f},\cj^{\dagger} \xi_{e'f}\big\rangle\rt)\nonumber\\
&&+e^{-2i\phi_{e'v}}\big\langle \cj^{\dagger}g_{ve'}^\dagger g_{ve'}J\xi_{e'f},\xi_{e'f}\big\rangle
+e^{-2i\phi_{e'v}}\left\langle g_{ve'}^\dagger g_{ve'}J\xi_{e'f}, \cj^{\dagger}\xi_{e'f}\right\rangle\Bigg]
\end{split}
} 
\ee
while in the degenerate case:
\be
H_{\theta_{ve'}\eps_{vf}}\big|_{\text{deg}}=H_{\eps_{vf}\theta_{ve'}}\big|_{\text{deg}}&=&- (1-i\g)j_f\Bigg[ e^{2i\phi_{e'v}}\left\langle \cj^{\dagger}\xi_{e'f},J\xi_{e'f}\right\rangle  + e^{2i\phi_{e'v}}\big\langle\xi_{e'f}, \cj^{\dagger}J\xi_{e'f}\big\rangle
\Bigg] \nonumber\\
H_{\theta_{ve'}\bar{\eps}_{vf}}\big|_{\text{deg}}=H_{\bar{\eps}_{vf}\theta_{ve'}}\big|_{\text{deg}}&=&2j_{f} e^{-2i\phi _{e'v}}\left\langle
\mathcal{J}^{\dag }J\xi _{e'f}, \xi _{e'f}\right\rangle\nonumber\\
&&- (1-i\g)j_f\Bigg[e^{-2i\phi_{e'v}}\big\langle \cj^{\dagger}J\xi_{e'f},\xi_{e'f}\big\rangle
+e^{-2i\phi_{e'v}}\left\langle J\xi_{e'f}, \cj^{\dagger}\xi_{e'f}\right\rangle\Bigg]
\ee

Finally the nonvanishing Hessian components $H_{\theta_{ve}\theta_{ve}}$ are computed in \cite{semiclassical}
\be
  \addtolength{\fboxsep}{1pt} 
   \boxed{ 
   \begin{split}
H_{\theta_{ve}\theta_{ve}}^{rr}&=&\half\sum_{f}j_{f}\lt(-\delta^{ij}+\hat{n}_{ef}^i\hat{n}_{ef}^j+i\eps^{ijk}\hat{n}_{ef}^k\rt)\ \ \ \ \ \ \ \ \ \  \nonumber\\
H_{\theta_{ve}\theta_{ve}}^{rb}&=&-\frac{i}{2}\sum_{f}j_{f}\lt(-\delta^{ij}+\hat{n}_{ef}^i\hat{n}_{ef}^j+i\eps^{ijk}\hat{n}_{ef}^k\rt)\ \ \ \ \ \ \ \  \nonumber\\
H_{\theta_{ve}\theta_{ve}}^{br}&=&-\frac{i}{2}\sum_{f}j_{f}\lt(-\delta^{ij}+\hat{n}_{ef}^i\hat{n}_{ef}^j+i\eps^{ijk}\hat{n}_{ef}^k\rt)\ \ \ \ \ \ \ \  \nonumber\\
H_{\theta_{ve}\theta_{ve}}^{rr}&=&2(1+\frac{i}{2}\g)\sum_{f}j_{f}\lt(-\delta^{ij}+\hat{n}_{ef}^i\hat{n}_{ef}^j+i\eps^{ijk}\hat{n}_{ef}^k\rt)
\end{split}
}
\ee
where $r$ and $b$ label respectively the rotation and boost parts of the generators.

\end{document}